\documentclass[preprint,12pt,authoryear,table]{elsarticle}

\date{}
\usepackage{amsthm,amssymb,amsfonts}
\usepackage{amsmath}
 \usepackage{graphicx}
\usepackage{booktabs}
\usepackage{float}
\usepackage{comment}
\usepackage{natbib}
\usepackage{listings}
\usepackage[ruled,vlined]{algorithm2e}
\usepackage{mathrsfs} 
\usepackage{tikz} 
\usepackage[skip=0.5\baselineskip]{caption}

\usepackage{marginnote,datetime,enumitem,subfigure,rotating,fancyvrb}
\usepackage{longtable, multirow, colortbl, graphicx, array}
\usepackage{xcolor}
\usepackage[margin=1.3in]{geometry} 

\usepackage{algorithmic}
\usepackage{wrapfig}
\input xy
\xyoption{all}

\newtheorem{definition}{Definition}
\newtheorem{proposition}{Proposition}
\newtheorem{corollary}{Corollary}
\newtheorem{remark}{Remark}

\newtheorem{example}{Example}

\DeclareMathOperator{\PP}{{\mathbb{P}}}

\usepackage[colorlinks,pagebackref=true]{hyperref}
\begin{document}

\begin{frontmatter}


\title{Modeling Systemic Risk: A Time-Varying Nonparametric Causal Inference Framework}





\author[tum]{Jalal Etesami}
\ead{j.etesami@tum.de}
\author[vt]{Ali Habibnia}
\ead{habibnia@vt.edu}
\author[epfl]{Negar Kiyavash}
\ead{negar.kiyavash@epfl.ch}

\address[tum]{School of Computation, Information and Technology, Technical University of Munich, Germany}
\address[vt]{Department of Economics, Virginia Tech, USA}
\address[epfl]{College of Management of Technology, Ecole Polytechnique F\'ed\'erale de Lausanne, Switzerland}

\tnotetext[t1]{}



\begin{abstract}
  We propose a nonparametric and time-varying directed information graph (TV-DIG) framework to estimate the evolving causal structure in time series networks, thereby addressing the limitations of traditional econometric models in capturing high-dimensional, nonlinear, and time-varying interconnections among series. This framework employs an information-theoretic measure rooted in a generalized version of Granger-causality, which is applicable to both linear and nonlinear dynamics. Our framework offers advancements in measuring systemic risk and establishes meaningful connections with established econometric models, including vector autoregression and switching models. We evaluate the efficacy of our proposed model through simulation experiments and empirical analysis, reporting promising results in recovering simulated time-varying networks with nonlinear and multivariate structures. We apply this framework to identify and monitor the evolution of interconnectedness and systemic risk among major assets and industrial sectors within the financial network. We focus on cryptocurrencies' potential systemic risks to financial stability, including spillover effects on other sectors during crises like the COVID-19 pandemic and the Federal Reserve's 2020 emergency response. Our findings reveals significant, previously underrecognized pre-2020 influences of cryptocurrencies on certain financial sectors, highlighting their potential systemic risks and offering a systematic approach in tracking evolving cross-sector interactions within financial networks.

\end{abstract}




\end{frontmatter}



\newpage

\section{Introduction} \label{sec:intro}
Understanding the interconnection between financial institutions, especially in the context of systemic risk, is of great importance. In principle, there are two main approaches to measuring such interconnections between institutions in the literature, which are mainly visualized by a directed graph.  One approach is based on a mathematical model of financial market participants, and the relations are derived from a combination of information extracted from financial statements, like the market value of liabilities of counterparties.
The other approach, which is also adopted in this work, is based on statistical analysis of financial series related to the institutions of interest. Nevertheless, most existing methods in the literature rely on  \textit{pairwise} comparisons or impose additional assumptions on the underlying model, such as \textit{linearity}, and often presume \textit{time-invariant} interconnections. In this work, we introduce a novel framework designed to capture causal relationships within a time series network while relaxing these assumption.  


To highlight a few relevant works, \cite{billio2012econometric} propose systemic risk measures to capture the connections between the monthly returns of various financial institutions (hedge funds, banks, brokers, and insurance companies), utilizing pairwise linear Granger-causality tests. Alternatively, in their work, \cite{diebold2014network} propose a connectedness measure based on generalized variance decomposition (GVD). This measure, again limited to linear systems, also suffers from disregarding the entire network akin to pairwise analysis commonly used in the literature, as we will discuss in Section \ref{sec:ma}. \cite{barigozzi2016network} focus on one particular network structure: the long-run variance decomposition network (LVDN). Similar to the work by \cite{diebold2014network}, the LVDN defines a weighted directed graph where the associated weight to edge \((i,j)\) represents the proportion of h-step-ahead forecast error variance of variable \(i\) which is accounted for by the innovations in variable \(j\). LVDNs are also limited to linear systems.

Building on these approaches, there are several attempts to go beyond linear models in the literature. For instance, \cite{andersen2019unified} provide unifying theory for parametric nonlinear factor models based on a panel of noisy observations. The nonlinear model 
is governed by stochastic differential equations and 
the estimation procedure is carried out via penalized least squares.
The inference theory for penalized least squares with fixed time span has been studied by \cite{andersen2015parametric}.
Later, \cite{andersen2019unified} develop an inference theory that applies in either fixed or large time span cases.

Expanding further on nonlinear methodologies, \cite{bonaccolto2019estimation} study quantile based methods of Granger causality and multiplex networks.
This work provides a possible solution to combine multiplex network, or a collection of networks existing between a set of subjects.
Ideally, the constituents of the multiplex network represent the outcomes of different approaches to the estimation or identification of links between the analyzed subjects.
Another contribution of this work is developing a quantile based method to interpret the risk of the estimated financial networks that identifies causality among quantiles of the modelled variables. It is important to realize that this method is consistent with the works by \cite{hong2009granger} and \cite{corsi2018measuring} that focus on causality among tail events.
The main idea is to capture causalities that are not in the center of their distributions, or in the mean but they are in the tails of their distributions. We discuss this method in details in Section \ref{sec:quantile}. \cite{etesami2017econometric} use a non-parametric information-theoretic measure to infer causal relations in a network however, it relies on an assumption that the causal network is time invariant. Herein, we extend this result to time-varying networks. 

Connectedness measures based on correlation remain widespread. Such measures capture only pairwise association and are mainly studied for linear Gaussian models. This makes them of limited value in finance. 
Different approaches have been developed to relax these assumptions. For example, equi-correlation approach of \cite{engle2012dynamic} uses average correlations across all pairs. 
The CoVaR approach of \cite{adrian2008covar} measures the value-at-risk of financial institutions conditional on other institutions experiencing financial distress. 
The marginal expected shortfall approach of \cite{acharya2017measuring} measures the expected loss to each financial institution conditional on the entire set of institutions’ poor performance. Although these measures rely less on linear Gaussian methods, 
a general framework that can capture non-pairwise connectedness in time-varying networks remains elusive. Introducing such measure is the main purpose of this work. 


Recently, there have been various attempts to infer time-varying causal relations among time series. These methods generally fall into three categories: rolling-window methods, change-point detection techniques, and model-based approaches. 

Rolling-window methods, like \cite{lu2014time}, use rolling sub-samples for time-varying Granger causality in oil markets, building on \cite{hong2001test}'s Granger causality test. This method, and its adaptations by \citep{hong2009granger,shi2018change,baum2021dynamics,zhang2021time}, primarily reveal causal relationships in means and variances between time series. However, absence of such causalities doesn't rule out all causality, and these tests assume an ARMA-GARCH model for conditional variance. \cite{shi2018change} introduced a recursive evolving window test for detecting causal relationship changes, while \cite{phillips2017estimating} examined nonlinear cointegration models with time-varying structural coefficients, using non-parametric kernel methods for estimation.

In statistical analysis, change point detection is crucial for identifying instances when a process's probability distribution changes. This technique is particularly useful in pinpointing alterations in causal networks, such as changes in the parent sets of network nodes or in their conditional probabilities. Detecting these shifts in high-dimensional distributions remains a challenging task \citep{aminikhanghahi2017survey,wang2021optimal,barigozzi2021time,pelger2022state}. To overcome these challenges, several model-based methods have been developed. For example, \cite{barigozzi2018simultaneous,barigozzi2021time} propose a framework for the estimation of time series factor models with multiple change-points in their second-order structure. 
As we will discuss in Section \ref{sec:time_v}, our proposed method is also capable of detecting such changes in the conditional distributions as well.


Model-based approaches like Markov switching models describe  dynamic interconnections via time-dependent or state-dependent model parameters. \cite{bianchi2019modeling} introduce a Markov switching graphical model for analyzing time-varying systemic risk using multi-factor asset pricing models. This model assumes each time series as a dynamic multi-factor linear model with normally distributed residuals. They employ a Markov chain Monte Carlo (MCMC) scheme for parameter inference and use weighted eigenvector centrality to identify latent states. However, their time-varying network, defined as a Markov random field in the error terms, fails to adequately represent causal relationships as it forms an undirected graph. Recently, \cite{billio2022bayesian} suggest a tensor-on-tensor regression for multilayer networks, combining zero-inflated logistic regression and Markov-switching coefficients for structural changes, using Bayesian inference with Gibbs sampling. Despite innovations, this approach shares limitations in causal network inference similar to \citep{bianchi2019modeling}.

 
Most of the aforementioned methods are studied in small size networks. The inference problem in high dimensional settings requires estimating a large number of parameters and thus most of the above methods may not be applicable.  
\cite{billio2019bayesian} propose a Bayesian nonparametric Lasso prior for high-dimensional VAR models that can improve estimation efficiency and prediction accuracy. 
This approach clusters the VAR coefficients into groups and shrinks the coefficients of each group toward a common location to overcome over-parametrization and overfitting issues.
Another related work is \citep{petrova2019quasi}, in which the authors propose a quasi-Bayesian local likelihood estimation methodology for multivariate models in high dimensional setting with time-varying parameters.
However, these methods are limited to linear models.
To overcome this limitation, \cite{kalli2018bayesian} propose a Bayesian non-parametric VAR model that allows for nonlinearity in the conditional mean, heteroscedasticity in the conditional variance, and non-Gaussian innovations. But, unlike the BNP-Lasso, it does not allow sparsity in the model.

\cite{iacopini2019bayesian} and  \cite{bernardi2019high} tackle the curse of dimensionality through a two-stage prior specification. In the first stage, a spike-and-slab prior distribution is used for each entry of the coefficient matrix.
In the second stage, it imposes prior dependence on the coefficients by specifying a Markov process for their random distribution. Causal network is a by-product of this estimation procedure.
Within the hierarchical conditional Bayesian prior methods, \cite{korobilis2019adaptive} develop a new estimation algorithm for VARs 
that breaks the multivariate estimation problem into a series of independent tasks and consequently 
reducing the curse of dimensionality.
Note that all aforementioned methods are limited to VAR models.

\cite{hue2019measuring} propose a new network measure of systemic risk that combines the pair-wise Granger causality approach with the leave-one-out concept. 
This methodology allows them to deal with the issue of indirect causalities, without facing the inherent curse of dimensionality that arises in the multivariate approaches.
However, since this approach only leaves out one financial institution at a time, it fails to capture  causalities with more than one indirect paths. 
In this work, we propose a framework for estimating time-varying causal relations in a general and possibly high-dimensional network of time series.



\subsection{Contributions}

In our research, we contribute both theoretically and empirically to the understanding of dynamic causal relationships within complex networks of time series, especially in the financial domain and the emerging influence of crypto assets. Theoretically, we introduce the time-varying directed information (TV-DI) measure, an information-theoretic framework specifically designed to capture the evolving nature of causal relationships in a general network of time series. Additionally, we enhance the estimation process of directed information (DI) by proposing a non-parametric method. This method leverages rolling window moving block bootstrapping in conjunction with non-parametric estimators for mutual information, such as k-nearest and kernel-based methods. The efficacy of this estimator is rigorously validated through a series of simulated experiments, showcasing its robustness and accuracy.  
We also discuss different approaches to tackle the curse of dimensionality for estimating DI in large networks. 
Furthermore, we introduce a DI-based measure for inferring non-parametric and non-pairwise aggregated causal effects between two disjoint subsets of time series. This can be used to quantify the causal effect of, for instance, a financial sector on another sector. In the context of network analysis, this translates into estimating the influence of an entire sub-network, rather than individual nodes, on another sub-network. Moreover, we quantify the contributions of each individual time series in such aggregated causal effects. This is useful to detect the most or the least influential asset in a financial sector influencing another sector, providing deeper insights into the dynamics of systemic risk. We also establish the connection between our framework with several well-known econometric models such as VAR, GARCH, switching models and others.
We show how our framework improves the measurement of systemic risk.
Empirically, to the best of our knowledge, this is the inaugural study that investigates the evolving dynamics of systemic risk posed by crypto assets to financial networks, marking a significant advancement in empirical financial network analysis.

\section{Methodology}\label{Sec1}

In order to investigating the dynamic of systemic risk, it is important to measure the causal relationship between financial institutions.
In this section, we introduce a statistical approach to measure such causal interconnections using  a generalized version
of Granger causality.
We begin by introducing some notations.
Plain capital letters denote random variables or processes, while lowercase letters denote their realizations. 
Bold lowercase and capital letters are used for column vectors and matrices, respectively. Calligraphy letters are used for denoting sets. 
We use $X_{j,t}$ to denote a time series $X_j$ at time $t$ 
and $X^{t}_j$ to denote the time series $X_j$ up to time $t$. 
For a set $\mathcal{A}=\{a_1,...,a_n\}$ and an index set $\mathcal{I}\subseteq\{1,...,n\}$, we define $\mathcal{A}_{-\mathcal{I}}:=\mathcal{A}\setminus\{a_i: i\in\mathcal{I}\}$.

\subsection{Graphical Models and Granger Causality}\label{sec:granger}

Researchers from different fields have developed various graphical models suitable for their application of interest to encode interconnections among variables or processes.
For instance, \cite{koller2009probabilistic} define Markov Networks, Bayesian networks (BNs), and \cite{murphy2002dynamic} introduces Dynamic Bayesian networks (DBNs). These are three examples of such graphical models that have been used extensively in the literature. 
Markov networks are undirected graphs that represent the conditional independence between the variables. On the other hand BNs and DBNs are directed acyclic graphs (DAGs) that encode conditional dependencies in a reduced factorization of the joint distribution. 
In DBNs, the size of the graphs depend on the time-homogeneity and the Markov order of the random processes. Therefore, in general, the size of the graphs can grow with time. 
As an example, the DBN graph of a vector autoregressive (VAR) introduced by \cite{dahlhaus2003causality} with $m$ processes each of order $L$ requires $mL$ nodes.
Hence, they are not suitable for succinct visualization of relationships between the time series. 

Similar to the works by  \cite{directed} and \cite{massey}, we use directed information graphs (DIGs) to encode interconnections among the financial institutions in which each node represents a time series. 
Below, we formally introduce this type of graphical models.
We use an information-theoretical generalization of the notion of Granger causality to determine the interconnection between time series. 
The basic idea in this framework was originally introduced by \cite{wiener1956theory} and later formalized by \cite{granger}.
The idea is as follows: ``we say that $X$ is causing $Y$ if we are better able to predict the future of $Y$ using all available information than if the information apart from the past of $X$ had been used."

Despite broad philosophical viewpoint of \cite{granger1963economic}, his formulation for practical implementation was done using autoregressive models and linear regression. This version has been widely adopted in econometrics and other disciplines. 
More precisely, in order to identify the influence of $X_t$ on $Y_t$ in a VAR comprises of three time series $\{X,Y,Z\}$, Granger's idea is to compare the performance of two linear regressions: the first one predicts $Y_t$ using $\{X^{t-1},Y^{t-1},Z^{t-1}\}$ and the second one predicts $Y_t$ given $\{Y^{t-1},Z^{t-1}\}$. 
Clearly, the performance of the second predictor is bounded by the first one. 
If they have the same performance, then we say $X$ does not Granger cause $Y$.  

Below, we introduce directed information (DI), an information-theoretical measure that generalized Granger causality beyond linear models.  
DI has been used in many applications to infer causal relationships. For example, \cite{spike} and \cite{di12} used it for analyzing neuroscience data and \cite{etesami2017econometric} applied for market data .

\subsection{Directed Information Graphs (DIGs)}\label{sec:dig}


Consider a causal\footnote{In causal systems, given the full past of the system, the present of the processes become independent. In other words, there are no simulations relationships between the processes.} dynamical system comprised of three time series $\{X,Y,Z\}$. To answer whether $X$ has influence on $Y$ or not over time horizon $[1,T]$, we compare the average performance of two particular predictors over this time horizon. 
The first predictor is non-nested that uses the full history of all three time series while the second one is nested and it uses the history of all processes excluding process $X$. 
On average, the performance of the predictor with less information (the second one) is upper bounded by the performance of the predictor with more information (the first one). 
However, when the performance of the predictors are close over time horizon $[1,T]$, then we declare that $X$ does not influence $Y$ over this time horizon.

In order to consider higher order moments, our prediction lies in the space of probability measures.  
More precisely, the first prediction at time $t$ is given by $\PP(Y_t|X^{t-1},Y^{t-1},Z^{t-1})$ that is the conditional distribution of $Y_t$ given the past of all processes and the second predictor is given by $\PP(Y_t|Y^{t-1},Z^{t-1})$.

To measure the performance of a predictor, we use a nonnegative loss function 
which defines the quality of the prediction. 
This loss function increases as the prediction deviates from the true outcome. 
Although there are many candidate loss functions, e.g., the squared error loss, absolute loss, etc, for the purpose of this work, we consider the logarithmic loss. 
More precisely, when the outcome $y_t$ is revealed for $Y_t$, the two predictors incur losses $\ell_{1,t}:=-\log\PP(Y_t=y_t|X^{t-1},Y^{t-1},Z^{t-1})$ and $\ell_{2,t}:=-\log\PP(Y_t=y_t|Y^{t-1},Z^{t-1})$, respectively.
This loss function has meaningful information-theoretical interpretations and it is related to the Shannon entropy.
The log-loss is the Shannon code length, i.e., the number of bits required to efficiently represent a symbol $y_t$. Thus, it may be thought of the description length of $y_t$.
For more justifications on this loss function, see the work by \cite{directed}.

The reduction in the loss at time $t$, known as regret is defined as
\begin{align}
r_t:=\ell_{2,t}-\ell_{1,t}=\log \frac{\PP(Y_t=y_t|X^{t-1},Y^{t-1},Z^{t-1})}{\PP(Y_t=y_t|Y^{t-1},Z^{t-1})}.
\end{align}
Note that the regrets are non-negative. The average regret over the time horizon $[1,T]$, i.e., $\frac{1}{T}\sum_{t=1}^T\mathbb{E}[r_t]$ is called \textit{directed information} (DI). This will be our measure of causation and its value determines the strength of influence.
If this quantity is close to zero, it indicates that the past values of time series $X$ contain no information that would help in predicting the future of time series $Y$ given the history of $Y$ and $Z$.
This definition can be generalized to more than three processes as follows,

 \begin{definition}\label{cdi}
Consider a network of $m$ time series $\mathcal{R}:=\{R_1,...,R_m\}$. We define the directed information from $R_i$ to $R_j$ as follows
\begin{small}
\begin{align}\label{cdif}
I(R_i\rightarrow R_j||\mathcal{R}_{-\{i,j\}}):=\frac{1}{T}\sum_{t=1}^T\mathbb{E}\left[\log\frac{\PP(R_{j,t}|\mathcal{R}^{t-1})}{\PP(R_{j,t}|\mathcal{R}^{t-1}_{-\{i\}})}\right],
\end{align}
\end{small}
where $\mathcal{R}_{-\{i,j\}}\!\!:=\!\!\mathcal{R}\setminus\{R_i,R_j\}$ and $\mathcal{R}_{-\{i\}}^{t-1}$ denotes $\{\!R_1^{t-1},...,R_m^{t-1}\!\}\setminus\{R_i^{t-1}\}$. We declare $R_i$ causes $R_j$ within time horizon $[1,T]$, if and only if 
$
I(R_i\rightarrow R_j||\mathcal{R}_{-\{i,j\}})>0.
$
\end{definition}

\begin{definition}\label{DIG1}
Directed information graph (DIG) of a set of $m$ processes $\{R_1,...,R_m\}$ is a weighted directed graph $G=(\mathcal{V},\mathcal{E},\mathcal{W})$, where nodes $\mathcal{V}$ represent processes and an arrow $(R_i,R_j)\in \mathcal{E}$ denotes that $R_i$ influences $R_j$ with weight $w_{j,i}:=I(R_i\rightarrow R_j||\mathcal{R}_{-\{i,j\}})$. 
Consequently, $(R_i,R_j)\notin \mathcal{E}$ if and only if $w_{j,i}=0$.
\end{definition} 

Pairwise comparison has been applied in the literature to identify the causal structure of time series. The works by \cite{billio2012econometric}, \cite{billio2010measuring}, and \cite{allen2010financial} are such examples.
Pairwise comparison is not correct in general and fails to capture the true causal relationships.  See \ref{sec:examples} for an example. 


\subsection{Estimating the DIs}\label{sec:estimation}
Given the definition of DI in Equation \eqref{cdif}, it is straightforward to see that DI can be written as the summation of conditional mutual information, 
\begin{small}
\begin{align}\label{estdi}
I(R_i\rightarrow R_j||\mathcal{R}_{-\{i,j\}})=\frac{1}{T}\sum_{t=1}^T I(R_{j,t};R_{i}^{t-1}|\mathcal{R}^{t-1}_{-\{i,j\}},R^{t-1}_{j}),
\end{align}
\end{small}
where $I(X;Y|Z)$ denotes the conditional mutual information between $X$ and $Y$ given $Z$. For more details see the book by \cite{cover2012elements}.
Therefore, parametric and non-parametric estimators for the conditional mutual information can be used to estimate DI. 
 There are different methods that can be used to estimate the terms in Equation \eqref{estdi} given i.i.d. samples such as plug-in empirical estimator, kernel estimator, and $k$-nearest neighbor estimator.
 For an overview of such estimators including their asymptotic behavior see the articles by \cite{paninski2003estimation}, \cite{noshad2019scalable}, and \cite{jiao2013universal}.
 For our experimental results, we used the $k$-nearest method since it shows relatively better performance compared to the other non-parametric estimators. For more details see \ref{sec:k-nearest}.  

{
\paragraph{Moving Block Bootstrap}
It is noteworthy that the aforementioned non-parametric estimators require large amount of i.i.d. data to output accurate DIs' estimations. 
However, often in practice (e.g., analysing financial market) such amount of i.i.d. data is not available for a short time window. 
To overcome this challenge, we adopt Moving block bootstrap (MBB) method. 
The MBB, introduced by \cite{kunsch1989jackknife}, is a non-parametric bootstrap procedure that can be applied to time series for replicating data. 
It obtains replicated series of data by drawing with replacement from the blocks of consecutive data. This method is well investigated in the case of strictly stationary strong mixing time series by \cite{kunsch1989jackknife,lahiri1999theoretical} and also in non-stationary case by \cite{fitzenberger1998moving,gonccalves2002bootstrap,synowiecki2007consistency}.
\begin{figure}[H]
    \centering
    \includegraphics[width=13cm,height=5cm]{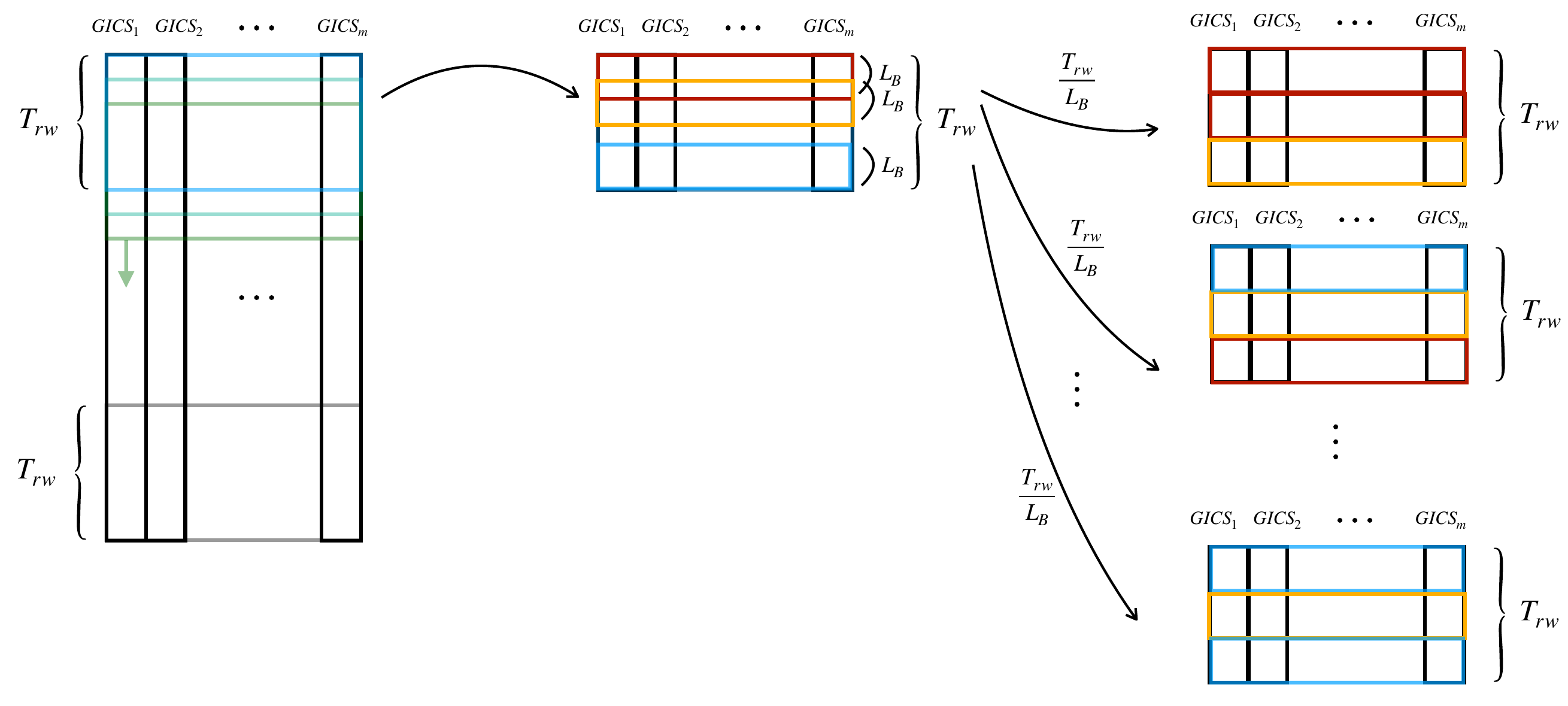}
    \caption{Rolling window MBB for constructing resamples.}
    \label{fig:rolling_moving}
\end{figure}

In this work, we combined the rolling window method and MBB to estimate the DIs in \eqref{eq:time_DI} as depicted in Figure \ref{fig:rolling_moving}. 
More specifically, a window of length $T_{rw}$ is selected (this is the rolling window and will be moved throughout the entire horizon $[1,T]$) and then batches of bootstrap resamples from the selected window are generated. 
This is done by first dividing $T_{rw}$ into overlapping blocks of length $L_B$. This leads to total $T_{rw}-L_B+1$ number of blocks. 
By randomly choosing $T_{rw}/L_B$ blocks (with replacement) and concatenating them, we create a bootstrap resample. 
Repeating this process several times constructs a batch of bootstrap resamples. 
The generated bootstrap resamples are treated as i.i.d. samples to estimate the DIs in Equation \eqref{eq:time_DI} via k-nearest estimator. 
}

\subsection{Strength of Causal Relationships}
Measuring the causal strength is important in many applications, e.g., finding which financial company has the strongest or weakest influence. 
There are several works in literature that consider different measures for causal strength. For instance, \cite{janzing2013quantifying} propose an axiomatic measure of causal strength based on a set of postulates. 
In this section, we show that the DI introduced in \eqref{cdif} can capture the strength of causal relationships as well. 
We do so using a simple linear example and then extend it to general systems.
Consider a network of three time series $\textbf{x}_t=(X_{1,t},X_{2,t},X_{3,t})^T$ with the following joint dynamics 
\begin{small}
\begin{equation}\label{eq:ex2}
\vcenter{\hbox{\begin{minipage}{5cm}
\centering
\xygraph{ !{<0cm,0cm>;<1.5cm,0cm>:<0cm,1.7cm>::} 
!{(-.7,1) }*+{X_{1}}="x1"
!{(.4,1.4) }*+{X_{2}}="x2"
!{(1.5,1)}*+{X_{3}}="x3"
"x2":"x1" "x3":"x2"  "x3":"x1" }
\captionof{figure}{The DIG of \eqref{eq:ex2}.}
\end{minipage}}}
\qquad\qquad
\begin{aligned}
\textbf{x}_t=
 \begin{pmatrix}
 0 & 0.1 & 0.3\\
 0 & 0 & -0.2\\
 0 & 0 & 0
 \end{pmatrix}\textbf{x}_{t-1}+\mathbf{\epsilon}_t,
\end{aligned}
\end{equation}
\end{small}
where $\epsilon_t$ denotes a vector of exogenous noises that has normal distribution with mean zero and identity covariance matrix. 
Figure 2 shows the corresponding DIG of this network.
Note that in this particular example where the relationships are linear, the support of the coefficient matrix encodes the corresponding DIG of the network \citep{acc2014}. 

In order to compare the strength of causal relationships $X_2\rightarrow X_1$ and $X_3\rightarrow X_1$ over a time horizon $[1,T]$, we compare the performance of two linear predictors of $X_{1,t}$. The first one predicts $X_{1,t}$ using $\{X^{t-1}_1,X^{t-1}_3\}$ and the other one uses $\{X^{t-1}_1,X^{t-1}_2\}$. If the first predictor shows better performance compared to the second one, it implies that $X_3$ contains more relevant information about $X_1$ compared to $X_2$. In other words, $X_3$ has a stronger influence on $X_1$ than $X_2$. 
To measure the performance of these two predictors over the time horizon $[1,T]$, we consider the mean squared errors. 
\begin{small}
\begin{align}
e_1:=\min_{(a,b)\in \mathbb{R}^{2}}\frac{1}{T}\sum_{t=1}^T  \mathbb{E}||X_{1,t}-(aX_{1,t-1}+bX_{3,t-1})||^2, \\
e_2:=\min_{(c,d)\in \mathbb{R}^{2}}\frac{1}{T}\sum_{t=1}^T  \mathbb{E}||X_{1,t}-(cX_{1,t-1}+dX_{2,t-1})||^2, 
\end{align}
\end{small}
It is easy to show that $e_1=1+0.1^2$ and $e_2=1+0.3^2$. Since $e_1<e_2$, we infer that $X_3$ has stronger influence on $X_1$ compared to $X_2$.

Similar to the directed information, we generalize the above framework to non-linear systems. Consider a network of $m$ time series $\{R_1,...,R_m\}$ with corresponding DIG $G=(\mathcal{V},\mathcal{E},\mathcal{W})$. Suppose that $(R_i,R_j), (R_k,R_j)\in \mathcal{E}$, i.e., $R_i$ and $R_k$ both are parents of $R_j$. 
We say $R_i$ has stronger influence on $R_j$ compared to $R_k$ within $[1,T]$ if $\PP(R_{j,t}|\mathcal{R}^{t-1}_{-\{k\}})$ is a better predictor than $\PP(R_{j,t}|\mathcal{R}^{t-1}_{-\{i\}})$ over that time horizon, i.e.,  
\begin{small}
\begin{align}
\frac{1}{T}\sum_{t=1}^T\mathbb{E}\left[-\log \PP(R_{j,t}|\mathcal{R}^{t-1}_{-\{k\}})\right]<
\frac{1}{T}\sum_{t=1}^T\mathbb{E}\left[-\log \PP(R_{j,t}|\mathcal{R}^{t-1}_{-\{i\}})\right].
\end{align}
\end{small}
The above inequality holds if and only if $I(R_i\rightarrow R_j||\mathcal{R}_{-\{i,j\}})>I(R_k\rightarrow R_j||\mathcal{R}_{-\{k,j\}})$. Thus, the DI in Equation \eqref{cdif} quantifies the causal relationships in a network.
For instance, in the system of (\ref{eq:ex2}), we have
\begin{small}
\begin{equation}
I(X_2\rightarrow X_1||X_3)=\frac{1}{2}\log(1.01)< \frac{1}{2}\log(1.09)=I(X_3\rightarrow X_1||X_2). 
\end{equation}
\end{small}
This means that $X_3$ has stronger influence on $X_1$ compared to $X_2$, which is consistent with the result of linear predictors.

\subsection{DIG in High-dimensional Settings}
For large networks with thousands nodes or millions of edges, DIGs become too complex to infer (i.e., it requires large amount of data) and also too complicated for direct humans analysis. 
There are several approaches for inferring the DIG of large networks. 
To mention a few: approximate large networks with smaller and simpler networks, use side information (e.g., model class) to develop alternative simpler inferring algorithms, and reduce the complexity of DI by reducing the size of the conditioning set. Below, we will briefly discuss each of these approaches. 

\begin{figure}
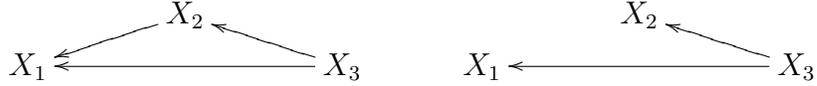

\hspace{2.5cm}
\xygraph{ !{<0cm,0cm>;<1.9cm,0cm>:<0cm,1.7cm>::} 
!{(-.7,1) }*+{X_{1}}="x1"
!{(.4,1.4) }*+{X_{2}}="x2"
!{(1.5,1)}*+{X_{3}}="x3"
"x2":"x1" "x3":"x2"  "x3":"x1" }
\hspace{1cm}
\xygraph{ !{<0cm,0cm>;<1.9cm,0cm>:<0cm,1.7cm>::} 
!{(-.7,1) }*+{X_{1}}="x1"
!{(.4,1.4) }*+{X_{2}}="x2"
!{(1.5,1)}*+{X_{3}}="x3"
 "x3":"x2"  "x3":"x1" }
   \caption{DIG of the system in \eqref{eq:ex2} and a corresponding tree approximation.}\label{fig:tr}
\end{figure}

A major approach to manage large high-dimensional problems is to keep a few edges of the causal network which together best approximate the dynamics of the system. 
In this case, instead of the true DIG, we obtain an approximation of it with less number of edges (sparser network). 
For example, Figure \ref{fig:tr} depicts the DIG of the system in \eqref{eq:ex2} and a tree\footnote{It is called tree because there is no cycle in the graph.} approximation.
\cite{quinn2013efficient} propose an efficient algorithm to identify the best directed tree approximation, where goodness of approximation is measured by Kullback-Leibler (KL) divergence from the full joint distribution to the distribution induced by the directed tree, i.e., 
\begin{align}
\min_{\PP^T_{\mathcal{R}}\in \mathcal{T}ree_{\mathcal{R}}}D\left(\PP_{\mathcal{R}}|| \PP^T_{\mathcal{R}}\right),
\end{align}
where $D(\cdot||\cdot)$ denote the KL divergence, $\PP_{\mathcal{R}}$ is joint distribution of $\mathcal{R}$, and $\mathcal{T}ree_{\mathcal{R}}$ is the set of all distributions on $\mathcal{R}=\{R_1,...,R_m\}$ that their DIGs are directed trees. They showed that the above optimization is equivalent to the following problem 
\begin{align}\label{eq:pair}
\max_{\PP^T_{\mathcal{R}}\in \mathcal{T}ree_{\mathcal{R}}}\sum_{i=1}^m I(R_{a_i}\rightarrow R_i),
\end{align}
where $a_i$ denotes the parent node of $R_i$ in the DIG of $\PP^T_{\mathcal{R}}\in \mathcal{T}ree_{\mathcal{R}}$. 
This result shows that inferring the best tree approximation of a DIG can be obtained by calculating the pair-wise DI, i.e., $I(R_j\rightarrow R_i)$ for all $i\neq j \in\{1,...,m\}$ instead of \eqref{cdif}. 
It is important to emphasize that the complexity of \eqref{cdif} increases with the network size. 
On the other hand, the complexity of the terms in \eqref{eq:pair} are independent of the network size and hence, in large networks, inferring the best directed tree approximation might be more suitable than the complete DIG.
A generalized version of this approximation is proposed by \cite{quinn2015bounded} in which they propose an algorithm to identify the optimal bounded in-degree approximations. 
In this algorithm, there is a trad off  between the complexity and the accuracy of the approximation. 

Using side information about the underlying dynamic (e.g., it is linear) or the structure (e.g., the true DIG is a directed tree) might help to develop alternative learning algorithms with lower complexities. 
As an example, \cite{acc2014} show that when the underlying dynamic is linear then the coefficient matrices of the model can determine the corresponding DIG of the system. 
Therefore, give the side information that the system is linear, one can infer its DIG by estimating the coefficient matrices of the model instead of estimating the DIs in \eqref{cdif}. 
Inferring the coefficients is a much simpler problem in terms of both computational and sample complexity compared to estimating the DIs.

Another approach to manage high-dimensional settings is substituting the conditioning set $\mathcal{R}_{-\{i,j\}}$ in \eqref{cdif} with a smaller set that contains the parents of node $R_j$. This will reduce both computational and sample complexity of the problem without introducing any approximation error. 
This is because, if $I(R_i\rightarrow R_j || \mathcal{R}_{-\{i,j\}})=0$, then for any subset $\mathcal{S}_j\subseteq \mathcal{R}_{-\{i,j\}}$ that contains all the parents of $R_j$, we have 
$
I(R_i\rightarrow R_j || \mathcal{S}_j)=0.     
$
However, forming such subsets may not be possible. In this case, one approach is to form a subset $\mathcal{S}'_j\subset\mathcal{R}_{-\{i,j\}}$ that contains the parents of $R_j$ with high probability. This will reduce the complexity but it may introduce approximation error.
We use this approach in our experiment to infer the causal network of major 124 financial institutions. More precisely, we formed $\mathcal{S}'_j$ by collecting 10 institutions that have the highest correlation with institute $R_j$.

\subsection{DIG of Econometric Models}\label{sec:models}
In the remaining of this section, we study well-known econometric models and introduce how their parameters are related to their corresponding DIGs.
These relationships can be used to infer the interconnections within these models without directly estimating the DIs. 
We also discuss the shortcomings of other approaches compared with DI facing the network identification problem in these models.

\subsubsection{VAR models:}
Consider a set of $m$ stationary time series, such that their relationships are captured by the following model:
\begin{equation}\label{ar}
\begin{aligned}
&\textbf{r}_t=\sum_{k=1}^p\textbf{A}_k\textbf{r}_{t-k}+\mathbf{\epsilon}_t,
\end{aligned}
\end{equation}
where $\textbf{r}_t=(R_{1,t},...,R_{m,t})^T$, $\mathbf{\epsilon}_t=(\epsilon_{1,t},...,\epsilon_{m,t})^T$, and $\{\textbf{A}_k\}$ is a set of $m\times m$ matrices. 
Moreover, we assume that the exogenous noises, i.e., $\{\epsilon_{i,t}\}$ are mutually independent and also independent from $\{R_{j,t}\}$. For simplicity, we also assume that the $\{\epsilon_{i,t}\}$ have mean zero. For this model, we have shown in \cite{acc2014} that
\begin{align}
    I(R_i\rightarrow R_j||\mathcal{R}_{-\{i,j\}})>0 \ \ \text{if and only if }\ \ \sum_{k=1}^p\big|[\textbf{A}_k]_{j,i}\big|>0,
\end{align}
where $[\textbf{A}_k]_{j,i}$ denotes the $(j,i)$-th entry. Thus, to learn the corresponding causal network (DIG) of this model, instead of estimating the DIs in (\ref{cdif}) which is a complex task, we can check whether the corresponding coefficients are zero or not.
To do so, one approach is to use information criterion for the model-selection and learn the parameter $p$ as described by \cite{schwarz1978estimating}, and then use F-tests as described by \cite{lomax2013statistical} to check the null hypotheses for the coefficients. 
 
\cite{materassi2012problem} use Wiener filtering as another alternative approach to estimate the coefficients and consequently learn the DIG. The idea of this approach is to find the coefficients by solving the following optimization problem,
 \begin{align}
\{\widehat{\textbf{A}}_1,...,\widehat{\textbf{A}}_p\}=\arg\min_{\textbf{A}_1,...,\textbf{A}_p}\mathbb{E}\left[\frac{1}{T}\sum_{t=1}^T||\textbf{r}_t-\sum_{k=1}^p\textbf{A}_k\textbf{r}_{t-k} ||^2\right].
 \end{align}
This leads to a set of Yule-Walker equations that can be solved efficiently by Levinson-Durbin algorithm introduced by \cite{musicus1988levinson}. Note that one can enforce sparsity by adding proper regularizer.

\subsubsection{GARCH models:}
The relationship between the coefficients of the linear model and the corresponding DIG can easily be extended to the financial data in which the variance of $\{\epsilon_{i,t}\}_{t=1}^T$ are no longer independent of $\{R_{i,t}\}$ but due to the heteroscedasticity, they are $\mathcal{R}_{i}^{t-1}$-measurable. More precisely, in financial data, the returns are modeled by GARCH that is given by 
\begin{align}\label{garch1}
\begin{aligned}
&R_{i,t}|\mathcal{R}^{t-1}\sim \mathcal{N}(\mu_{i,t},\sigma^2_{i,t}),\ \ i=1,...,m,\\
&\sigma^2_{i,t}=\alpha_0+\sum_{k=1}^q \alpha_k(R_{i,t-k}-\mu_{i,t})^2+\sum_{l=1}^s\beta_l \sigma^2_{i,t-l},
\end{aligned}
\end{align}
where $\{\alpha_k\}$ and $\{\beta_l\}$ are nonnegative constants. 

\begin{proposition}\label{prop1}
Consider a network of time series whose dynamic is given by \eqref{garch1}. In this case, $I(R_j\rightarrow R_i||\mathcal{R}_{-\{i,j\}})=0$ if and only if
\begin{align}\label{test1}
\mathbb{E}[R_{i,t}|\mathcal{R}^{t-1}]=\mathbb{E}[R_{i,t}|\mathcal{R}_{-\{j\}}^{t-1}], \ \forall t.
\end{align}
\end{proposition}
All proofs appear in \ref{sec:proofs}.
Multivariate GARCH models are a generalization of (\ref{garch1}) in which the variance of $\epsilon_{i,t}$ is $\mathcal{R}^{t-1}$-measurable. In this case, not only $\mu_{i,t}$ but also $\sigma^2_{i,t}$ encode the interactions between the returns. More precisely, in multivariate GARCH, we have
\begin{align}
\begin{aligned}\label{eq:garchi}
&\textbf{r}_t|\mathcal{R}^{t-1}\sim \mathcal{N}(\textbf{u}_t,\textbf{H}_t),\\
&vech[\textbf{H}_t]=\Omega_0+\sum_{k=1}^q \Omega_k vech[{\epsilon}_{t-k}{\epsilon}_{t-k}^T]+\sum_{l=1}^p\Gamma_l vech[\textbf{H}_{t-l}],
\end{aligned}
\end{align}
where ${\epsilon}_{t}=\textbf{r}_t-\textbf{u}_t$, $\textbf{u}_t\in\mathbb{R}^m$, $\textbf{H}_t\in\mathbb{R}^{m\times m}$ is  symmetric positive definite matrix and $\mathcal{R}^{t-1}$-measurable. Note that $vech$ denotes the vector-half operator, which stacks the lower triangular elements of an $m\times m$ matrix as an $\small{(m(m+1)/2)\times1}$ array.

\begin{proposition}\label{prop2}
Consider a network of time series whose dynamic is captured by a multivariate GARCH model in \eqref{eq:garchi}. In this case, $I(R_j\rightarrow R_i||\mathcal{R}_{-\{i,j\}})=0$ if and only if both the condition in Proposition \ref{prop1} and the following condition hold
\begin{align}\label{test2}
\mathbb{E}[(R_{i,t}-u_{i,t})^2|\mathcal{R}^{t-1}]=\mathbb{E}[(R_{i,t}-u_{i,t})^2|\mathcal{R}_{-\{j\}}^{t-1}], \ \forall t.
\end{align}
\end{proposition}
See \ref{sec:examples} for an example demonstrating the above result.
Recall that the pairwise Granger-causality calculation, in general, fails to identify the true causal network. 
It was proposed by \cite{billio2012econometric} that the returns of the $i$-th institution linearly depend on the past returns of the $j$-th institution, when
\begin{small}
 \begin{align}
\mathbb{E}[R_{i,t}|\mathcal{R}^{t-1}]=\mathbb{E}\big[R_{i,t}|R_{j,t-1},R_{i,t-1},\{R_{j,\tau}-u_{j,\tau}\}_{\tau=-\infty}^{t-2},\{R_{i,\tau}-u_{i,\tau}\}_{\tau=-\infty}^{t-2}\big].
\end{align}
\end{small}
This test is obtained based on pairwise Granger-causality calculation and does not consider non-linear causation through the variance of $\{\epsilon_i\}$. For instance, if both institutions $R_j$ and $R_k$ affect institution $R_i$, then the above equality does not hold.
This is because $R_k$ influences $R_i$ and it is included in the conditioning on the left but not on the right hand side.

\subsubsection{Moving-Average (MA) models:}\label{sec:ma}

\cite{pesaran1998generalized} show that the model in (\ref{ar}) can be represented as an infinite MA as long as $\textbf{r}_t$ is covariance-stationary, i.e., all the roots of $|\textbf{I}-\sum_{k=1}^p\textbf{A}_kz^k|$ fall outside the unit circle:
\begin{small}
\begin{align}\label{moving}
\textbf{r}_t=\sum_{k=0}^\infty\textbf{W}_k\mathbf{\epsilon}_{t-k},
\end{align}
\end{small}
where $\textbf{W}_k=0$ for $k<0$, $\textbf{W}_0=\textbf{I}$, and $\textbf{W}_k=\sum_{l=1}^p\textbf{A}_{l}\textbf{W}_{k-l}$. In this representation, $\{\epsilon_i\}$s are called shocks and when they are independent, they are also called orthogonal. 
Herein, we study the causal structure of a MA model of finite order $p$, i.e., the summation in \eqref{moving} is up to $p$.
In this case, Equation (\ref{moving}) can be written as $\textbf{W}_0^{-1}\textbf{r}_t=\Psi(L){\epsilon}_{t}$, where $\Psi(L):=\textbf{I}+\sum_{k=1}^p\textbf{W}_0^{-1}\textbf{W}_k L^{k}$ and $L$ is the lag operator, i.e., $L^{i}{\epsilon}_{t}={\epsilon}_{t-i}$. Subsequently, we have
\begin{small}
\begin{align}\label{inar}
&\Psi(L)^{-1}\textbf{W}_0^{-1}\textbf{r}_t=\textbf{W}_0^{-1}\textbf{r}_t-\sum_{k=0}^\infty\mathbf{J}_i L^i\textbf{W}_0^{-1}\textbf{r}_t=\mathbf{\epsilon}_t,\\
&\mathbf{J}_i:=\textbf{W}_0^{-1}\textbf{W}_i-\sum_{j=1}^{i-1}\mathbf{J}_{i-j}\textbf{W}_0^{-1}\textbf{W}_j.
\end{align}
\end{small}
This representation is equivalent to an infinite VAR model.
\begin{corollary}\label{cor1}
Consider a MA model described by (\ref{moving}) with orthogonal shocks such that $\textbf{W}_0$ is non-singular and diagonal. In this case, $I(R_j\rightarrow R_i||\mathcal{R}_{-\{i,j\}})=0$  if and only if the corresponding coefficients of $\{R_{j,t-k}\}_{k>0}$ in $R_i$'s equation in \eqref{inar} are zero. 
\end{corollary}
In the interest of simplicity and space, we do not present the explicit form of the coefficients, but show the importance of this result via an example in \ref{sec:examples}.

We studied the DIG of a MA model with orthogonal shocks. However, the shocks are rarely orthogonal in practice. To identify the causal structure of such systems, we can apply the whitening transformation to transform the shocks into a set of uncorrelated variables. More precisely, 
suppose $\mathbb{E}[\mathbf{\epsilon}_t\mathbf{\epsilon}_t^T]=\mathbf{\Sigma}$, where the Cholesky decomposition of $\mathbf{\Sigma}$ is $\textbf{V}\textbf{V}^T$. Hence, $\textbf{V}^{-1}\mathbf{\epsilon}_t$ is a vector of uncorrelated shocks. Using this fact, we can transform (\ref{moving}) with correlated shocks into 
$
\textbf{r}_t=\sum_{k=0}^p\widetilde{\textbf{W}}_k\tilde{\mathbf{\epsilon}}_{t-k},
$
 with uncorrelated shocks, where $\mathbf{\tilde{\epsilon}}_{t}:=\textbf{V}^{-1}\mathbf{\epsilon}_t$ and $\widetilde{\textbf{W}}_k:=\textbf{W}_k\textbf{V}$.

\begin{remark}\label{rem3}
\cite{diebold2014network} applied GVD method to identify the population connectedness or in another word the causal structure of a MA model with correlated shocks. Using this method, they monitor and characterize the network of major U.S. financial institutions during 2007-2008 financial crisis. 
In this method, the weight of $R_j$'s influence on $R_i$ in (\ref{moving}) was defined to be proportional to 
\begin{small}
\begin{align}
d_{i,j}=\sum_{k=0}^{p}\left(\left[\textbf{W}_k\mathbf{\Sigma}\right]_{i,j}\right)^2.
\end{align}
\end{small}
Recall that $\mathbb{E}[\mathbf{\epsilon}_t\mathbf{\epsilon}_t^T]=\mathbf{\Sigma}$. 
Applying the GVD method to Example \ref{example2}, where $\mathbf{\Sigma}=\textbf{I}$, we obtain that $d_{1,2}=d_{3,1}=0$. That is $R_2$ does not influence $R_1$ and $R_1$ does not influence $R_3$. This result is not consistent with the Granger-causality concept since the corresponding causal network (DIG) of this example is a complete graph, i.e., every node has influence on any other node.
Thus, the GVD analysis also seems to suffer from disregarding the entire network akin to pairwise analysis commonly used in traditional application of the Granger-causality. 
\end{remark}

\subsubsection{Switching models:}\label{sec:non}
Note that DIG as defined in  Definition \ref{DIG1} does not require any assumptions on the underlying model. 
However, side information about the model class can simplify computation of the DIs in (\ref{cdif}). 
For instance, assuming that $\mathcal{R}=\{R_1,...,R_m\}$ is a first-order Markov chain, then 
$I(R_i\rightarrow R_j||\mathcal{R}_{-\{i,j\}})=0$ if and only if $\PP(R_{j,t}|\mathcal{R}_{t-1})=\PP(R_{j,t}|\mathcal{R}_{-\{i\},t-1})$ for all $t$. 
Recall that $\mathcal{R}_{-\{i\},t-1}=\{R_{1,t-1},...,R_{m,t-1}\}\setminus\{R_{i,t-1}\}$.
Furthermore, suppose that the transition probabilities are represented through a logistic function similar to the work by \cite{billio2010measuring}. 
More specifically, for any subset of processes $\mathcal{H}:=\{R_{i_1},...,R_{i_h}\}\subseteq \mathcal{R}$, we have
\begin{small}
\begin{align}\label{eq:ssw}
\PP(R_{j,t}|\mathcal{H}_{t-1})=\PP(R_{j,t}|R_{i_1,t-1},...,R_{i_h,t-1}):=\frac{\exp(\mathbf{a}_{\mathcal{H}}^T \textbf{u}_\mathcal{H})}{1+\exp(\mathbf{a}_\mathcal{H}^T \textbf{u}_\mathcal{H})},
\end{align}
\end{small}
where $\textbf{u}^T_\mathcal{H}:=\bigotimes_{R_i\in\mathcal{H}}(1, R_{i,t-1})=(1, R_{i_1,t-1})\otimes(1, R_{i_2,t-1})\otimes\cdots\otimes(1, R_{i_h,t-1})$, $\otimes$ denotes the Kronecker product, and $\mathbf{a}_\mathcal{H}$ is a vector of dimension $2^h\times 1$.
Under these assumptions, the causal discovery in the network reduces to the following statement:
$R_i$ does not influence $R_j$ if and only if all the terms of $\textbf{u}_{R}$ depending on $R_i$ have zero coefficients. 

\begin{proposition}
Consider the model in \eqref{eq:ssw} and let
\begin{align}
\begin{aligned}
&\textbf{u}_{{R}}=\textbf{u}_{\mathcal{R}_{-\{i\}}}\otimes(1, R_{i,t-1})=(\textbf{u}_{\mathcal{R}_{-\{i\}}}, \textbf{u}_{\mathcal{R}_{-\{i\}}}R_{i,t-1}),
\end{aligned}
\end{align}
$\mathbf{a}^T_{\mathcal{R}}=(\mathbf{a}^T_1, \mathbf{a}^T_2)$, where $\mathbf{a}_1$ and $\mathbf{a}_2$ are the vectors of coefficients corresponding to $\textbf{u}_{\mathcal{R}_{-\{i\}}}$ and $\textbf{u}_{\mathcal{R}_{-\{i\}}}R_{i,t-1}$, respectively. Then, $I(R_i\rightarrow R_j||\mathcal{R}_{-\{i,j\}})=0$ if and only if $\mathbf{a}_2=\textbf{0}$.
\end{proposition}
    
Multiple chain Markov switching model (MCMS)-VAR of \cite{billio2015granger} is a family of non-linear models, in which the relationship among a set of time series is given by
\begin{small}
\begin{align}\label{switch}
Y_{i,t}=\mu_i(S_{i,t})+\sum_{k=1}^p \sum_{j=1}^m\ [\textbf{B}_{k}(S_{i,t})]_{i,j}Y_{j,t-k}+\epsilon_{i,t}, \  \ i\in\{1,...,m\},
\end{align}
\end{small}
and $\mathbf{\epsilon}_t:=(\epsilon_{1,t},...,\epsilon_{m,t})\sim\mathcal{N}(0,\mathbf{\Sigma}(\textbf{s}_t))$, 
where the mean $\mu_i(S_{i,t})$, the lag matrices $\textbf{B}_{k}(S_{i,t})$, and the covariance matrix of the error terms $\mathbf{\Sigma}(\textbf{s}_t)$ depend on a latent random vector $\textbf{s}_t$ known as the state of the system.
Random variable $S_{i,t}$ represents the state variable associated with $Y_{i,t}$ that can take values from a finite set $\mathcal{S}$. The random vector $\textbf{s}_t=(S_{1,t},...,S_{m,t})$ is assumed to be a time-homogeneous first-order Markov process with the transition probability 
$
\PP(\textbf{s}_{t}|\textbf{s}^{t-1},\mathcal{Y}^{t-1})=\PP(\textbf{s}_{t}|\textbf{s}_{t-1}).
$
Furthermore, given the past of the states, the presents are independent, i.e., 
\begin{small}
$\PP(\textbf{s}_{t}|\textbf{s}_{t-1})=\prod_{i=1}^m\PP(S_{i,t}|\textbf{s}_{t-1}).$ 
\end{small}
 Next result states a set of sufficient conditions under which by observing only the time series $Y_t$ and estimating the DIs, we are able to identify the causal relationships between the processes.

\begin{proposition}\label{prop3}
Consider a MCMS-VAR in which $\mathbf{\Sigma}(\textbf{s}_t)$ is diagonal for all $\textbf{s}_t$. In this case, $I(Y_j\rightarrow Y_i||\mathcal{Y}_{-\{i,j\}})=0$ if
\begin{itemize}
\item  $[\textbf{B}_k(s_{i,t})]_{i,j}=0$ for all $k$ and all realizations $s_{i,t}$,

\item $[\mathbf{\Sigma}(\textbf{s}_t)]_{i,i}=[\mathbf{\Sigma}(S_{i,t})]_{i,i}$,

\item $\PP(S_{k,t}|\textbf{s}_{t-1}, S_{1,t},...,S_{k-1,t},S_{k+1,t},...,S_{m,t})=\PP(S_{k,t}|S_{k,t-1})$, for all $k$.

\end{itemize}
\end{proposition}
Note that the conditions introduced in this proposition are different from the ones in \citep{billio2015granger}. 
More precisely, \cite{billio2015granger} study the causal relationships between the time series under an extra condition that the state variables are known. 
Such assumption is not realistic as they are often not observable in real applications.
Below, we present a simple example in which the above conditions do not hold and $Y_1$ does not functionally depend on $Y_2$.
Yet, observing the states leads to the deduction that $Y_2$ does not influence $Y_1$, but without observing the states the inference would be different.

\begin{example}
Consider a bivariate MCMS-VAR $\{Y_1,Y_2\}$ in which the states only take binary values and
\begin{small}
\begin{align}
& Y_{1,t}= b_{1,1}(S_{1,t})Y_{1,t-1}+\epsilon_{1,t},\\ \notag
& Y_{2,t}=\mu_2(S_{2,t})+ 0.5Y_{1,t-1}+\epsilon_{2,t},
\end{align}
\end{small}
where $(\epsilon_{1,t},\epsilon_{2,t})\sim\mathcal{N}(0,I)$, $\mu_2(0)=10$, $\mu_2(1)=-5$, $b_{1,1}(0)=0.5$, and $b_{1,1}(1)=-0.5$.
Moreover, the transition probabilities of the states are $\PP(S_{1,t}|S_{1,t-1},S_{2,t-1})=\PP(S_{1,t}|S_{1,t-1})=0.8$ whenever $S_{1,t}=S_{1,t-1}$, and $\PP(S_{2,t}=S_{1,t-1})=0.9$.
Based on \cite{billio2015granger}, in this setup, $Y_{2,t-1}$ does not Granger-cause $Y_{1,t}$ given $Y_{1,t-1}, S_{1,t-1}$, i.e., 
\begin{small}
\begin{align}
\PP(Y_{1,t}|Y_{2,t-1},Y_{1,t-1},S_{1,t-1})=\PP(Y_{1,t}|Y_{1,t-1},S_{1,t-1}).
\end{align}
\end{small}
However, without observing the states, we have $\PP(Y_{1,t}|Y_{2,t-1},Y_{1,t-1})\neq \PP(Y_{1,t}|Y_{1,t-1})$. This is because, $Y_{2,t-1}$ contains information about $S_{2,t-1}$ which in turn contains information about $S_{1,t-2}$. 
\end{example}

Another attempt to capture time-varying networks using Markov switching models is the work by \cite{bianchi2019modeling}.
In this work, the stock returns of the $i$-th firm in excess of the risk-free rate at time $t$, denoted by $Y_{i,t}$ is given by 
\begin{align}
Y_{i,t} = \textbf{Z}_{i,t}^T\textbf{b}_i(s_t) + \varepsilon_{i,t},  \ \ \ t=1, ..., T, \ \ i=1,...,n,
\end{align}
where $\textbf{Z}_{i,t}:=(1, \textbf{X}_{i,t}^T)^T$ and $\textbf{X}_{i,t}$ denotes the $m_i$-dimensional vector of systematic risk factors. $\textbf{b}_i(s_t)$ is a $(m_i + 1)$-vector of time-varying regression coefficients, and $\{\varepsilon_{i,t}\}$  are error terms that can be identified with a firm-specific idiosyncratic risk factor when $Cov(\textbf{Z}_{i,t}, \varepsilon_{i,t})=0$. 
Furthermore, it is assumed that the risk factors are common across stocks and the error terms have a full time-varying variance-covariance matrix and they are normally distributed conditionally on the latent state $s_t$, i.e., $(\varepsilon_{1,t},...,\varepsilon_{n,t})\sim\mathcal{N}(0, \mathbf{\Sigma}(s_t))$.
This model is a reduced-form approximation of a linear pricing kernel \citep{cochrane2009asset,vassalou2003news} and can be seen as a special variation of \eqref{switch}.
\cite{bianchi2019modeling} define the interconnection network of this model using the inverse covariance matrix of the error terms which is an undirected graph.  
More precisely, there is no interconnection between two firms $i$ and $j$ at time $t$, when $\varepsilon_{i,t}$ and $\varepsilon_{i,t}$ are independent given $\{\varepsilon_{l,t}: l\neq i,j\}$. Such network cannot capture the causal relationships among the firms. Next example illustrates a scenario in which the model by \cite{bianchi2019modeling} fails to capture the true influences. 

\begin{example}
Consider a setting with three firms $\{i,j,k\}$ in which both $i$ and $j$ influence $k$ but there is no influence between $i$ and $j$. The DIG of this system is $i\rightarrow k\leftarrow j$. This happens when $\varepsilon_{i,t}$ and $\varepsilon_{j,t}$ are independent but they become statistically dependent by conditioning on $\varepsilon_{k,t}$.
In this case, all the off-diagonal entries of the inverse covariance matrix are non-zero. Hence, the method by \cite{bianchi2019modeling} inaccurately infers the network of this system to be a complete graph.
\end{example}

\subsubsection{Quantiles:}\label{sec:quantile}
Quantile-on-quantile causality proposed by \cite{sim2015oil} and later discussed by \cite{bonaccolto2019estimation} aims to check whether the $\theta$-th quantile of variable $X$ causes the $\tau$-th quantile of $Y$, and vice versa. More precisely, in this work, they consider the following linear model on the conditional quantiles
\begin{align}
& Q_{Y_t}(\tau, \theta) =\beta_{0,1}(\tau, \theta) + \beta_{1,1}(\tau, \theta)Y_{t-1} + \beta_{2,1}(\tau, \theta)X_{t-1},\\
& Q_{X_t}(\tau, \theta) =\beta_{0,2}(\tau, \theta) + \beta_{1,2}(\tau, \theta)Y_{t-1} + \beta_{2,2}(\tau, \theta)X_{t-1},
\end{align}
where $Q_{Y_t}(\tau, \theta)$ and $Q_{X_t}(\tau, \theta)$ denote the conditional quantiles of $Y_t$ and $X_t$, respectively. The coefficients are obtained using quantile regression method from \citep{sim2015oil}. Note that this is a parametric approach that assumes a linear model for the conditional quantiles. 

\cite{jeong2012consistent} propose a non-parametric approach based on hypotheses testing for detecting causality in quantiles. This approach identifies $X$ as a cause of $Y$ in its $\tau$-th quantile when 
$
Q^Y_{\tau}(Y^{t-1}_{t-p}, X^{t-1}_{t-q})\neq Q^Y_\tau(Y^{t-1}_{t-p}),    
$ 
where $Q^Y_\tau(\mathcal{Z})$ denotes the $\tau$-th quantile of $Y_t$ conditional on set $\mathcal{Z}$. This definition of causality is analogous to the definition of DI in which $X$ was declared as a cause of $Y$ when there exists a time $t$ such that
$P(Y_t| Y^{t-1}, X^{t-1})\neq P(Y_t|Y^{t-1})$.
The non-parametric test developed by \cite{jeong2012consistent} is based on a measure of distance defined as:
\begin{align}
J_\tau:=\mathbb{E}\left[\left(F\left(Q^Y_{\tau}(Y^{t-1}_{t-p})\Big| Y^{t-1}_{t-p}, X^{t-1}_{t-q} \right)-\tau\right)^2f(Y^{t-1}_{t-p}, X^{t-1}_{t-q})\right],
\end{align}
where $F(u|v)$ denotes the conditional cumulative distribution function (CDF) of $u$ given $v$ and $f(v)$ is the marginal density function of $v$. \cite{jeong2012consistent} show that this measure can be estimated using the feasible kernel-based estimator.

The causal network recovered by this method may vary depending on the value of $\tau$ as it is possible that $X$ causes $Y$ in its quantile only for certain values of $\tau$. Hence, the quantile method detects causal relationship from $X$ to $Y$ only for such values of $\tau$. 
On the other hand, next result shows that DI robustly detect the causation. 
\begin{proposition}\label{pro4}
Consider a network of two time series $X$ and $Y$. If there exists $\tau\in(0,1)$ such that $X$ causes $Y$ in quantile $\tau$, i.e., $J_\tau>0$, then $I(X\rightarrow Y)>0$.
\end{proposition}

\subsection{Causal Effect Between Substes of Processes}
Often in practice, it is important to infer the causal effect between two disjoint subsets of processes. 
For instance, in market analysis, it is important to understand the effect of Cryptocurrency sector either on a specific market such as Apple company or on a different sector such as Real Estate. 
Herein, we show how DI can be used to quantify such effects. 

First, we quantify the causal effect of a subset of $l$ processes $\mathcal{R}_\mathcal{A}$ with index set $\mathcal{A}:=\{A_1,...,A_l\}\subseteq-\{i\}=\{1,...,m\}\setminus\{i\}$ on a time series $R_i$ denoted by $I(\mathcal{R}_{\mathcal{A}}\rightarrow R_i||\mathcal{R}_{-\mathcal{A}\cup\{i\}})$. 
To do so, we need to compare the effects of adding and removing the history of $\mathcal{R}_\mathcal{A}$ on predicting $R_i$ during a time horizon $T$ while the effect of the remaining processes $\mathcal{R}_{-\mathcal{A}\cup\{i\}}$ is omitted. This leads to 
\begin{small}
\begin{align}\label{eq:cluster_effect}
    & I(\mathcal{R}_{\mathcal{A}}\rightarrow R_i||\mathcal{R}_{-\mathcal{A}\cup\{i\}}):=\frac{1}{T}\sum_{t=1}^T\mathbb{E}\left[\log\frac{\mathbb{P}(R_{i,t}|R_i^{t-1},\mathcal{R}_{-\mathcal{A}\cup\{i\}}^{t-1},\mathcal{R}_\mathcal{A}^{t-1})}{\mathbb{P}(R_{i,t}|R_i^{t-1},\mathcal{R}_{-\mathcal{A}\cup\{i\}}^{t-1})}\right].
\end{align}
\end{small}
Equation \eqref{eq:cluster_effect} can be written in terms of DI as follows,
\begin{small}
\begin{align*}
    &I(R_{A_l}\rightarrow R_i||\mathcal{R}_{-\{i,A_l\}})+\frac{1}{T}\sum_{t=1}^T\mathbb{E}\left[\log\frac{\mathbb{P}(R_{i,t}|R_i^{t-1},\mathcal{R}_{\mathcal{A}\setminus\{A_l\}}^{t-1},\mathcal{R}_{-\mathcal{A}\cup\{i\}}^{t-1})}{\mathbb{P}(R_{i,t}|R_i^{t-1},\mathcal{R}_{-\mathcal{A}\cup\{i\}}^{t-1})}\right].
\end{align*}
\end{small}
By induction, the above expression will be  
\begin{small}
\begin{align}
    I(\mathcal{R}_{\mathcal{A}}\rightarrow R_i||\mathcal{R}_{-\mathcal{A}\cup\{i\}})=\sum_{j=0}^{l-1}I(R_{A_{l-j}}\rightarrow R_i||\mathcal{R}_{-\{i,A_l,...,A_{l-j}\}}).
\end{align}
\end{small}
It is noteworthy that the above equation resembles the chain rule. 
Given the above expression for the effect of a subset $\mathcal{R}_{\mathcal{A}}$ on a time series $R_i$, we can naturally define the causal effect of the subset $\mathcal{R}_{\mathcal{A}}$ on another disjoint subset $\mathcal{R}_{\mathcal{B}}$ in a network of $m$ processes. We denote this effect by $I(\mathcal{R}_{\mathcal{A}}\rightarrow \mathcal{R}_{\mathcal{B}}||\mathcal{R}_{-\mathcal{A}\cup\mathcal{B}})$, where $-\mathcal{A}\cup\mathcal{B}:=\{1,...,m\}\setminus(\mathcal{A}\cup\mathcal{B})$ and it is the average individual effects of $\mathcal{R}_{\mathcal{A}}$ on the processes in $\mathcal{R}_{\mathcal{B}}$, i.e.,
\begin{small}
\begin{align}\notag
    I(\mathcal{R}_{\mathcal{A}}\!\rightarrow\! \mathcal{R}_{\mathcal{B}}||\mathcal{R}_{-\mathcal{A}\cup\mathcal{B}})&:=\frac{1}{|\mathcal{B}|}\sum_{i\in\mathcal{B}}I(\mathcal{R}_{\mathcal{A}}\!\rightarrow\! R_i||\mathcal{R}_{-\mathcal{A}\cup\{i\}})
    \\
    &=\frac{1}{|\mathcal{B}|}\sum_{i\in\mathcal{B}}\sum_{j=0}^{l-1}I(R_{A_{l-j}}\!\rightarrow\! R_i||\mathcal{R}_{-\{i,A_l,...,A_{l-j}\}}).\label{eq:agg-di}
\end{align}
\end{small}
The last term in the above expression is the conditional DI between two processes and can be estimated similar to \eqref{estdi}.

The next important task is to quantify how much of the aggregated effect from $\mathcal{A}$ to $\mathcal{B}$ is by a specific time series $R_{A_j}$, $A_j\in\mathcal{A}$? 
Answering this question can clarify, for instance, within the influences of a financial sector (e.g., Cryptocurrency) on another sector (e.g., Real Estate), which asset contributes the most or the least. 
More precisely, to quantify the contribution of $R_{A_j}$, $A_j\in\mathcal{A}$, we compare the influences of the subsets $\mathcal{R}_{\mathcal{A}}$ and $\mathcal{R}_{\mathcal{A}\setminus\{A_j\}}$ on $\mathcal{R}_{\mathcal{B}}$, i.e., 
\begin{small}
\begin{align}\label{eq:contribution}
  C_{\mathcal{A}\rightarrow\mathcal{B}}(R_{A_j}):=I\big(\mathcal{R}_{\mathcal{A}}\!\rightarrow\! \mathcal{R}_{\mathcal{B}}||\mathcal{R}_{-\mathcal{A}\cup\mathcal{B}}\big)-I\big(\mathcal{R}_{\mathcal{A}\setminus\{A_j\}}\!\rightarrow\! \mathcal{R}_{\mathcal{B}}||\mathcal{R}_{-(\mathcal{A}\setminus\{A_j\})\cup\mathcal{B}}\big).
\end{align}
\end{small}

\subsection{Time-varying Causal Networks}\label{sec:time_v}
All aforementioned formulations of causal influences are based on an assumption that the underlying causal network is time invariant over the time horizon during which the causal effects are measured. 
However, it is quite possible that the causal relationships and information spillover among financial assets changes over time. Changes in causal relationships among financial assets usually indicate varying patterns, so it is important to detect and assess such dynamic relationships. Below, we introduce a framework using DI to capture such time-varying networks and compare it with a major approaches in literature for detecting time-varying causal relations. 


 \cite{lu2014time} propose a time-varying Granger causality test for the influence of $Y_2$ on $Y_1$ at time $t$ with rolling sample size $S$. 
 If the test statistic is larger than the critical value at a given level of significance, then there is significant Granger causality at time $t$. The test is defined as follows,
 \begin{small}
\begin{align}\label{eq:lu2014}
    &H_t(S):=\frac{S\sum_{j=1}^{S-1}k^2(j/M)r^2_{1,2,t}(j,S)- L_S(k)}{\sqrt{2D_S(k)}},
\end{align}
\end{small}
where $r_{1,2,t}(j,S)$ denotes lag $j$ sample cross correlation between standardized residuals of $Y_1$ and $Y_2$ in the sub-sample $[t-S+1, t]$. Furthermore,  $k(x)$ is the kernel function (e.g., Bartlett kernel), $M$ is a positive integer, and $L_S$ and $D_S$ are 
\begin{small}
\begin{align*}
    L_S(k):=\sum_{j=1}^{S-1}(1-j/S)k^2(j/M),\quad D_S(k):=\sum_{j=1}^{S-1}(1-j/S)(1-(j+1)/S)k^4(j/M).
\end{align*}
\end{small}
 \begin{figure}
    \centering
    \includegraphics[scale=.2]{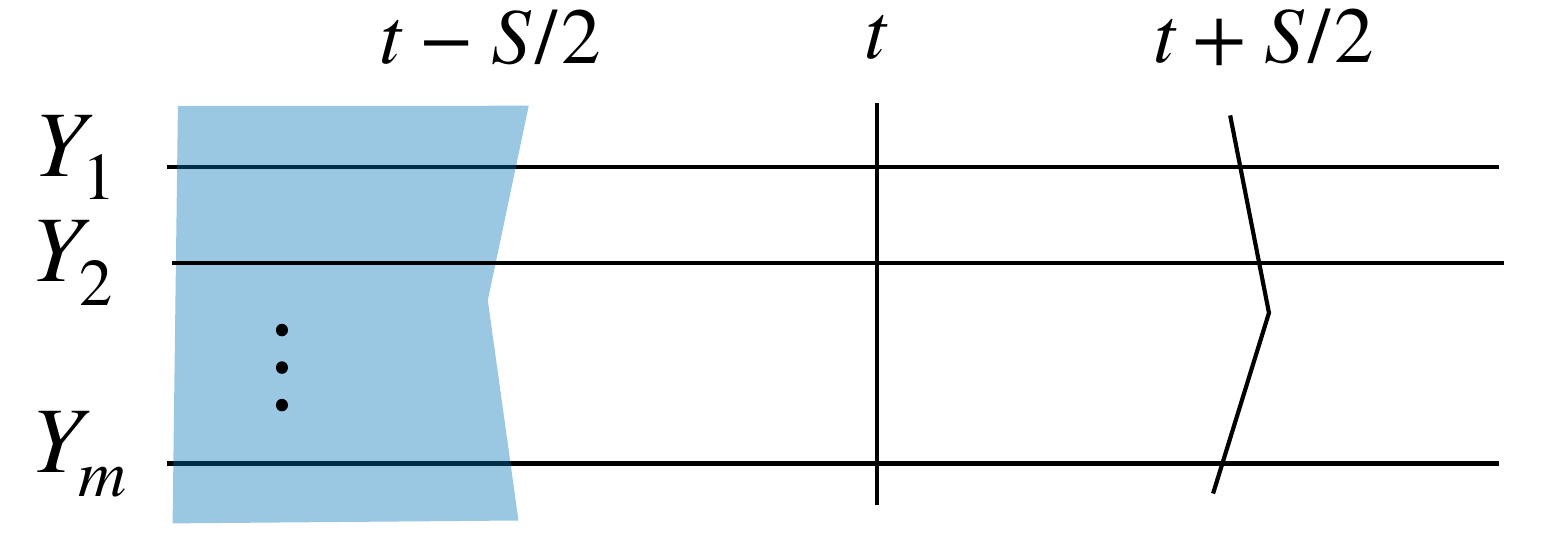}
    \caption{Measuring causal effect of $Y_2$ on $Y_1$ withing a time window of length $S$ around time $t$.}
    \label{fig:time_varying_rolling}
\end{figure}
It is important to emphasize that the above time-varying Granger test is limited to ARMA-GARCH, it cannot capture causalities in higher moments, and it is pairwise. 
On the other hand, we overcome these shortcomings by proposing the time-varying DI (TV-DI). 
More precisely, we propose the following quantity to detect the causal influence from $Y_2$ to $Y_1$ at time $t$ with rolling sample size $S$ (for simplicity, we assume $S$ is even) in a network of $m$ processes.
\begin{small}
\begin{align}\label{eq:time_DI}
    I_{t,S}(Y_2\rightarrow Y_1|| \mathcal{Y}_{-\{1,2\}})&:=\frac{1}{S}\sum_{j=-S/2}^{S/2}\mathbb{E}\left[\log\frac{\PP(Y_{1,t-j}|\mathcal{Y}^{t-S/2-1},\mathcal{Y}_{t-S/2}^{t-j-1})}{\PP(Y_{1,t-j}|\mathcal{Y}^{t-S/2-1},\mathcal{Y}^{t-j-1}_{-\{2\},t-S/2})}\right]. 
\end{align}
\end{small}
This quantity is developed by comparing the performance of two predictors for $Y_1$ at time $t-j$ ($j$ varies within $[t-S/2, t+S/2]$) both of which use the information of all processes up to time $t-S/2$.
Note that one of them uses additional information of all processes from $t-S/2$ to $t-j-1$ while the other one uses additional information of all processes excluding the information of $Y_2$ from $t-S/2$ to $t-j-1$. 
If the performance of these two predictors are the same, i.e., the expression in \eqref{eq:time_DI} is zero, it implies that $Y_2$ has no influence on $Y_1$ during the time window $[t-S/2, t+S/2]$.
It is noteworthy that by including the information of all variables up to time $t-S/2$, as it is shown in Figure \ref{fig:time_varying_rolling}, we ensure that the expression in \eqref{eq:time_DI} only detects impacts on $Y_1$ that are from after $t-S/2$.
When this quantity is greater than the critical value at given level of significance, then $Y_2$ influences $Y_1$ within this time window.

It is straightforward to see when time-varying Granger test proposed by \cite{lu2014time} or change-point detection methods detect a change in the causal network, our TV-DI test will pick that change as well, i.e., \eqref{eq:time_DI} will be positive. However, 
when the change occurs in higher moments, unlike the method of \cite{lu2014time}, our TV-DI test can detect it. We will use our TV-DI test in Section \ref{sec:exp} to detect time-varying causal relations in both synthetic and real-world experiments.

\subsection{Sample Path Causal Influences}
The definition of DI involves taking an expectation over the histories of the time series, and are thus well suited to address questions such as ``Does the past of time series $Y$ up to time $t-1$, i.e., $Y^{t-1}$ cause $X_t$?"
A natural next question that one might pose is ``Did the past realization of $Y$ up to time $t-1$, i.e., $y^{t-1}$ cause $X_t$?"
In other words, the goal of sample path causal influence is to identify the causal effect that particular values of $Y$ have on the distribution of the subsequent sample of $X$. Examples of this include ``When does the institute A have the greatest effect on another institute B?"

\cite{schamberg2019measuring} develop a framework using sequential prediction for estimating the sample path causal relationships.
In this framework, the influence from $Y$ to $X$ in the presence of other processes $Z$ at time $t$ is defined by
\begin{align}
&S_{Y\rightarrow X}(\mathcal{H}^{t-1}):=\mathbb{E}\left[\log\frac{\PP(x_t|\mathcal{H}^{t-1})}{\PP(x_t|\mathcal{H}^{t-1}_{-y})}\right],
\end{align}
where $\mathcal{H}^{t-1}$ is the $\sigma$-algebra generated by $\{x^{t-1},y^{t-1},z^{t-1}\}$ and  $\mathcal{H}^{t-1}_{-y}$ is the $\sigma$-algebra generated by $\{x^{t-1},z^{t-1}\}$. The relationship between the sample path causal measure and the DI is given by
$
\sum_{t}\mathbb{E}[S_{Y\rightarrow X}(\mathcal{H}^{t-1})]=I(Y\rightarrow X || Z).
$
Furthermore, \cite{schamberg2019measuring} discuss how sequential prediction theory may be leveraged to estimate the proposed causal measure and introduce a notion of regret for assessing the performance of such estimator.

\section{Simulation Studies}\label{sec:exp}
This section presents the results of our simulation experiments, which assess the efficacy of the TV-DIG model presented in Section \ref{sec:time_v} in capturing complex causal relationships within multivariate time series networks. These experiments focus on scenarios involving nonlinear dynamics and time-varying effects, highlighting the model's ability to infer network structures under diverse conditions. We particularly address scenarios where traditional models, constrained by linear, normal, and time-invariant assumptions, are inadequate. This underscores the importance of nonparametric approaches in deciphering the dynamic interconnections between nodes in complex networks.

\subsection{Setting 1: Time-Invariant Nonlinear Network}
In the first setting, our simulation targets a time-invariant graph with 10 nodes. We employed the Monte Carlo simulation to generate data from a nonlinear Data Generating Process (DGP) of 10 time series, aiming to demonstrate the generality of the DI method in capturing causal relationships within these series. Subsequently, we inferred the corresponding network using DI. This nonlinear DGP is uniquely designed, not aligning with any econometric model classes previously studied in related works, to robustly test our method's capability to handle diverse and complex scenarios. We simulate the following system
\begin{align}\label{eq:synth}
& X_{i,t} = cut_{\nu}(\textbf{x}^T_{t-1}\textbf{A}_i\textbf{x}_{t-1}) + \epsilon_{i,t}, \ \ 1\leq i\leq 10,
\end{align}
where $\textbf{x}_{t}=(X_{1,t},...,X_{10,t})^T\in\mathbb{R}^{10\times1}$, $\textbf{A}_i\in\mathbb{R}^{10\times10}$, $\epsilon_{i,t}$ is a standard Normal distribution, and the nonlinearity is introduced through the $cut_{\nu}$ function
\begin{equation}
\vcenter{\hbox{\begin{minipage}{5cm}
\centering
\includegraphics[width=4cm,height=3cm]{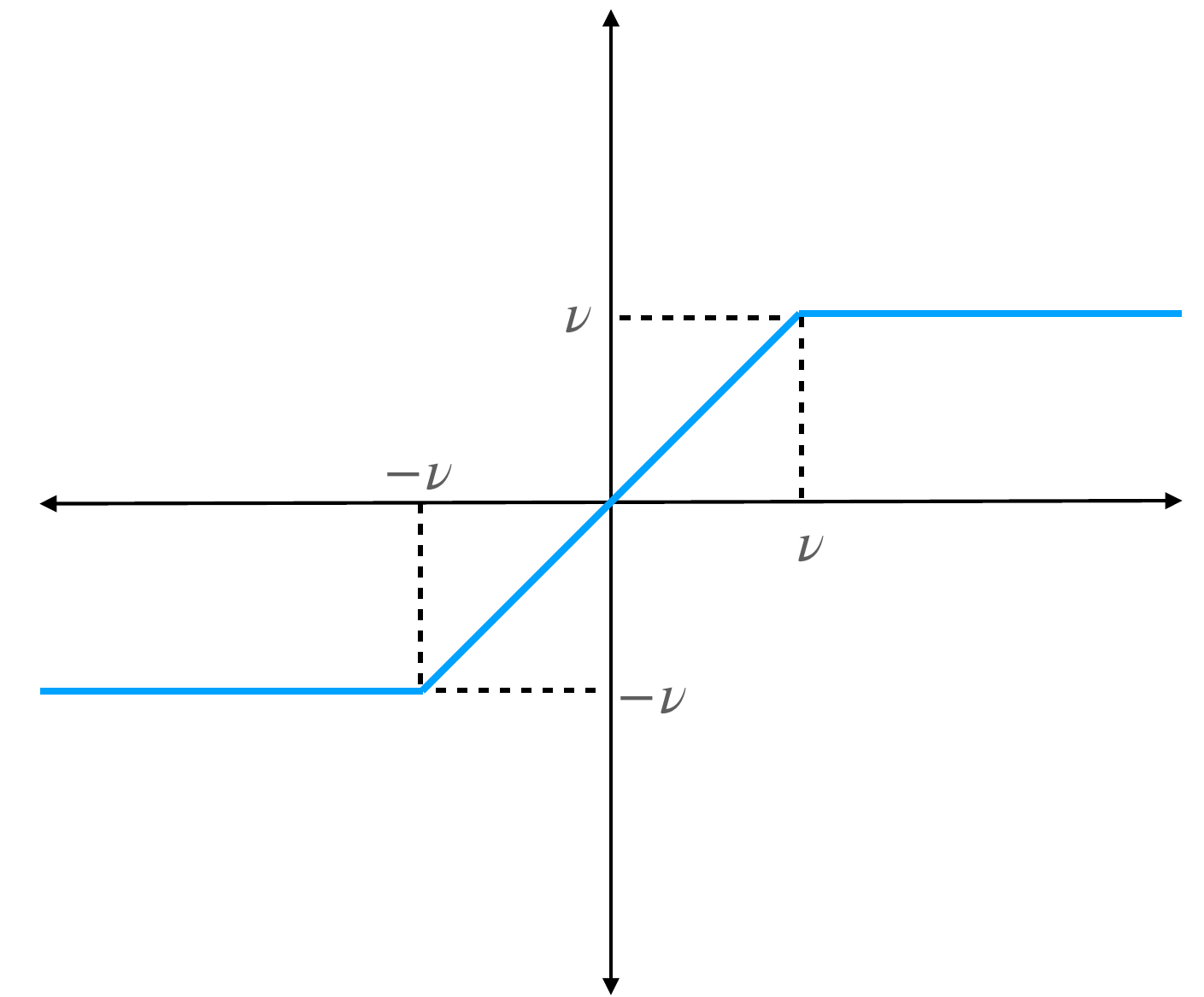}
\captionof{figure}{Function $cut_\nu(x)$.}
\label{fig:cut}
\end{minipage}}}
\qquad\qquad
\begin{aligned}
    cut_{\nu}(x):=\begin{cases} 
      -\nu & x\leq -\nu \\
      x & -\nu< x < \nu \\
      \nu & x\geq\nu 
   \end{cases}.
\end{aligned}
\end{equation}
This system is a composition of a quadratic system with a nonlinear function $cut_\nu(x)$ (depicted in Figure 5). 
The main reason for using this function is to prevent the system from diverging and having a stable system\footnote{In our experiment, we used $\nu=15$.}.
Clearly, the above system is neither VAR nor switching models and thus for example, the proposed methods in \citep{billio2012econometric,diebold2014network,billio2021networks,kalli2018bayesian,bianchi2019modeling} will fail to capture its corresponding interconnection network. 
On the other hand, as we discussed in Section \ref{sec:dig}, DI is capable of inferring the causal network of a dynamical system without any prior assumption on its underlying model. 
Hence, it can detect the interconnections within this system.
To demonstrate this capability of DI, we first generated $\{\textbf{A}_{1},...,\textbf{A}_{10}\}$ by selecting their entries from $\{-.5,0,.5\}$ uniformly at random. Afterwards, using Monte Carlo simulation, we generated $\{\textbf{x}_1,..., \textbf{x}_T\}$ from \eqref{eq:synth}. By applying the kernel estimator to the generated dataset, we estimated the DIs and consequently infer the DIG.

For a given threshold $\tau$, the corresponding DIG$_{\tau}$ is obtained by drawing an arrow from $X_j$ to $X_i$ when $I(X_j\rightarrow X_i||\mathcal{X}_{-\{i,j\}})>\tau$.
On the other hand, because the functional dependencies among the time series are given in (\ref{eq:synth}), it is possible to obtain the true causal network of this system. More precisely, $X_j$ is a parent of $X_i$ in the true network if and only if 
$X_{i,t+1}$ functionally depends on $X_{j,t}$ in \eqref{eq:synth}.
To measure the performance of our DIG method, we compared the inferred DIG with the true network and reported the precision and the recall in Figure \ref{fig:p-r}. This figure is obtained by averaging over 50 trails. 
\begin{align*}
    Precision:= \dfrac{TP}{TP+FP}, \ \ Recall:= \dfrac{TP}{TP+FN},
\end{align*}
where $TP, FP$, and $FN$ denote true positive, false positive, and false negative, respectively. $TP$ is the number of edges that are common between the estimated DIG and the true network. 
$FP$ denotes the number of edges that they are in the estimated DIG but do not exist in the true network. Finally, $FN$ represents the number of edges that do not exist in the estimated DIG but they are in the true network.
\begin{figure*}[t]
\centering
\subfigure[DIG$_\tau$]{\includegraphics[scale=.23]{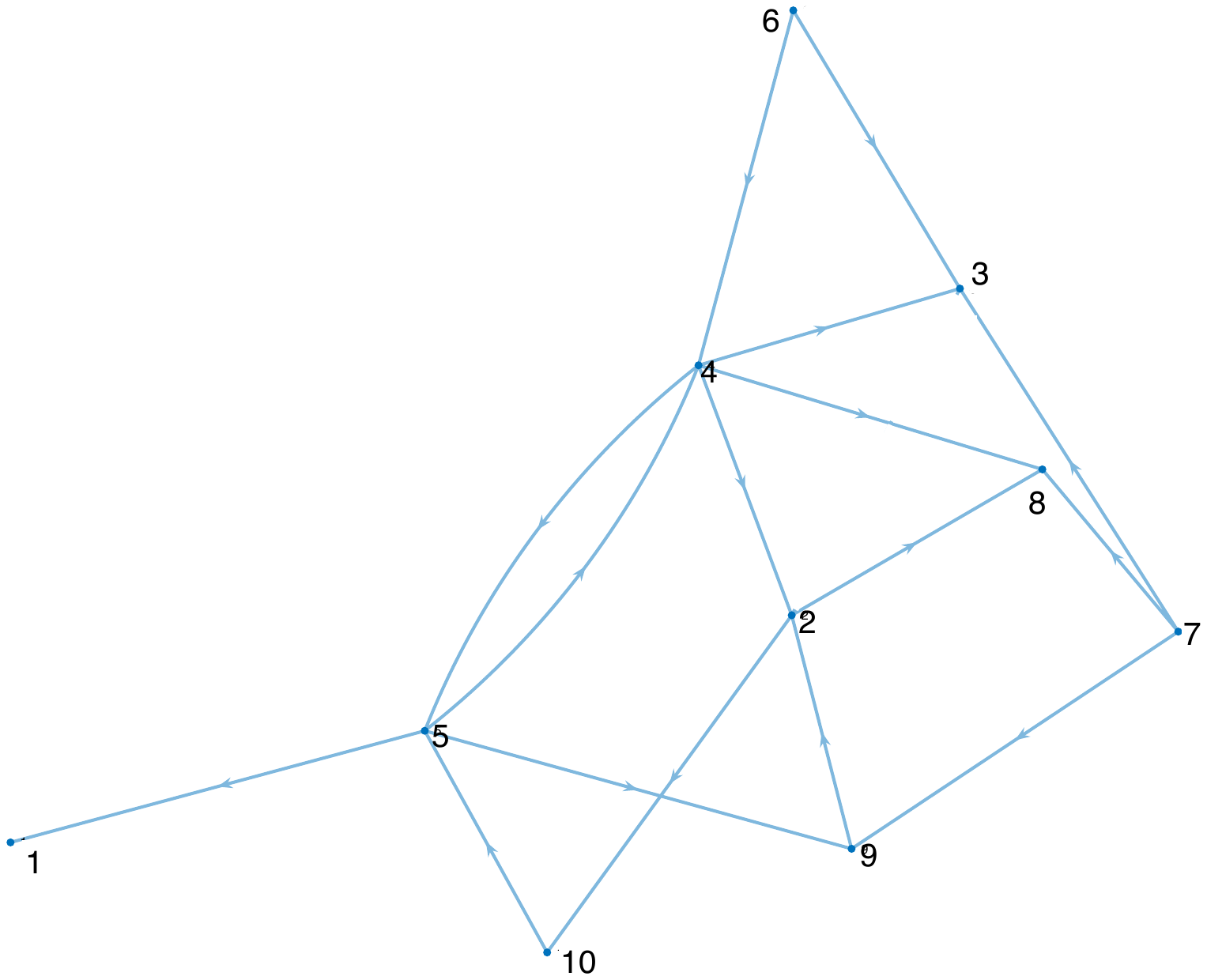}}
\subfigure[True Network]{\includegraphics[scale=.23]{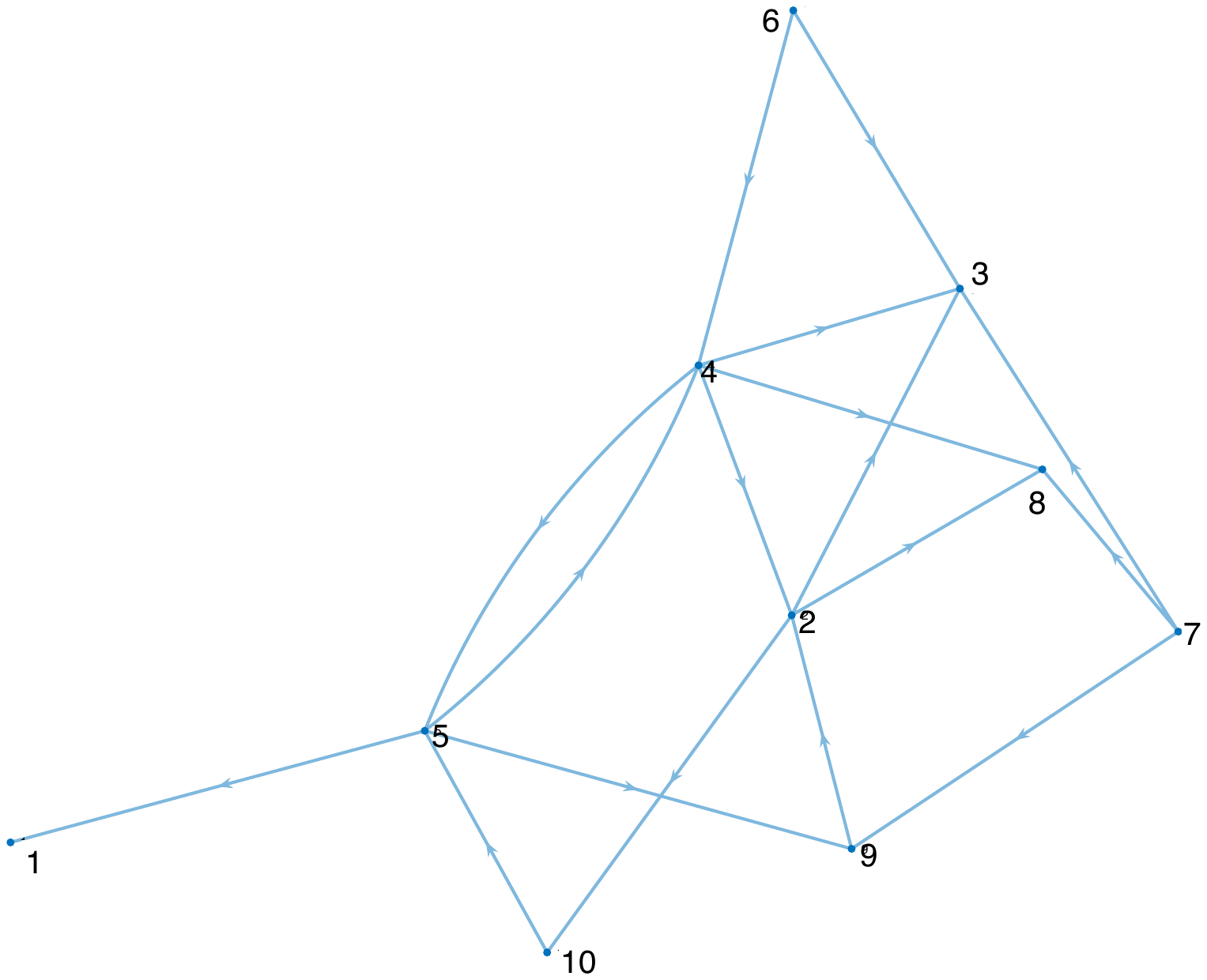}}
       \caption{The resulting estimated DIG of the synthetic data with threshold $\tau=0.05$ and its corresponding true network. Estimated DIG only misses one edge between nodes 2 and 3.}
\label{fig:synth}
 \end{figure*} 

\begin{figure}
    \centering
    \includegraphics[scale=.42]{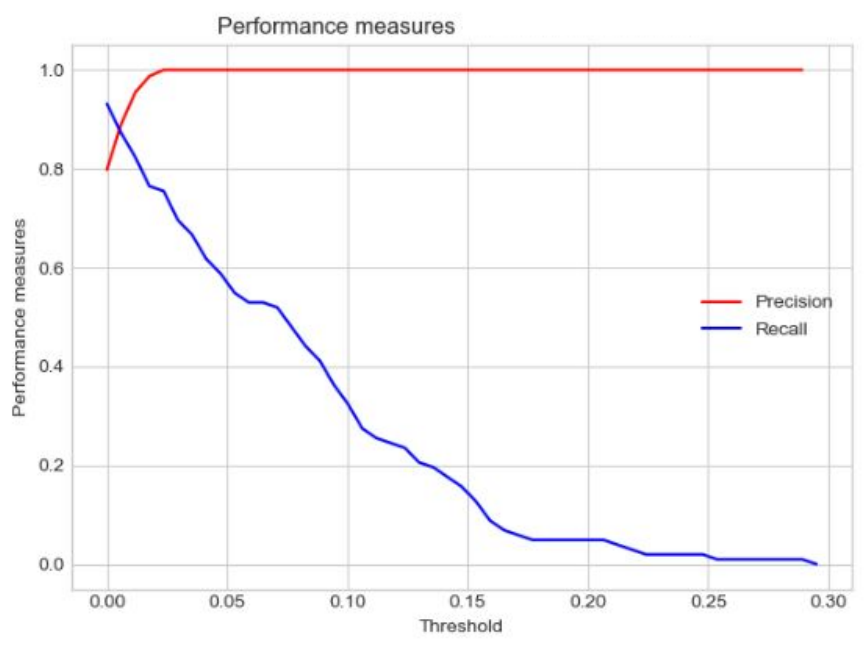}
    \caption{Precision and recall curves in nonlinear experiment for varying amount of thresholds.}
    \label{fig:p-r}
\end{figure}
The results, as depicted in Figures \ref{fig:synth} and \ref{fig:p-r},
demonstrate the robustness of the DI model in accurately inferring the network
structure. In Figure \ref{fig:synth}, the comparison between the inferred DIG and
the true network illustrates the model's precision in capturing most of the existing
connections, with only a single missed edge between nodes 2 and 3. This high level
of accuracy in network reconstruction is further substantiated by the precision and
recall curves shown in Figure \ref{fig:p-r}. The curves indicate an optimal balance
between precision and recall across different threshold values, underscoring the model's
effectiveness in identifying true relationships without overfitting. These results
validate the DI method's capability to detect complex causal interconnections in a
time-invariant nonlinear network, highlighting its potential application in similar
complex settings.

\subsection{Setting 2: Time-Varying Network}
In the second setting, we increase complexity by incorporating both nonlinearity and time-varying influences among nodes, simulating a dynamic scenario with evolving network relationships. The TV-DIG model's performance here is assessed based on its accuracy in tracking these time-varying causal influences, reflected in changes to the network structure. We compare these results with those from the time-varying Granger test method proposed by \cite{hong2001test}.  

We simulate a nonlinear, time-varying system using the following DGP:
\begin{align}\label{eq:sim:time}
    \textbf{x}_{t+1}=\sqrt{|\textbf{A}_{t}\textbf{x}_{t}|} + \textbf{n}_{t},
\end{align}
where $\textbf{x}_{t}=(X_{0,t},...,X_{4,t})^T\in\mathbb{R}^{5\times1}$, $\sqrt{|\cdot|}$ acts element-wise, $\textbf{n}_t\in\mathbb{R}^{5\times1}$ is a random vector distributed as Normal with mean zero and identity covariance matrix. The matrix $\textbf{A}_{t}$, a time-varying coefficient matrix, introduces dynamic interactions among the time series, as shown below:
\begin{small}
\begin{align*}
    \textbf{A}_{t}:=
    \begin{pmatrix}
    \frac{10}{1+t/200} & \frac{-20}{1+t/150} & -0.8 & 0 &0\\
    0 & 0 & 8\sin(1.4\frac{\pi t}{T} )& 0 & 8\sin(2.4\frac{\pi t}{T} +\pi)\\
    0 & 0 & 0 & 4u\big(\cos(   \frac{1.6\pi t}{T} )\big)& -5u\big(\sin(- \frac{1.6\pi t}{T} )\big)\\
    2u\big( 7.9 - \ln(t) \big) & 0& 0 & 0.7 & 0\\
    0&0.8&0 &0.9&-0.5
    \end{pmatrix},
\end{align*}
\end{small}
where $u(t)$ denotes the step function, i.e., $u(t)=1$ for $t\geq 0$ and zero otherwise. The matrix $\textbf{A}_t$, with its time-dependent elements, introduces a unique dynamic in the network's causal relationships. The function $u(t)$ selectively activates or deactivates these influences over time. 
 \begin{figure*}[h]
\centering
\subfigure[$|\textbf{A}_{t}|_{4,1}$ over time. ]{\includegraphics[width=4.3cm, height=2.4cm]{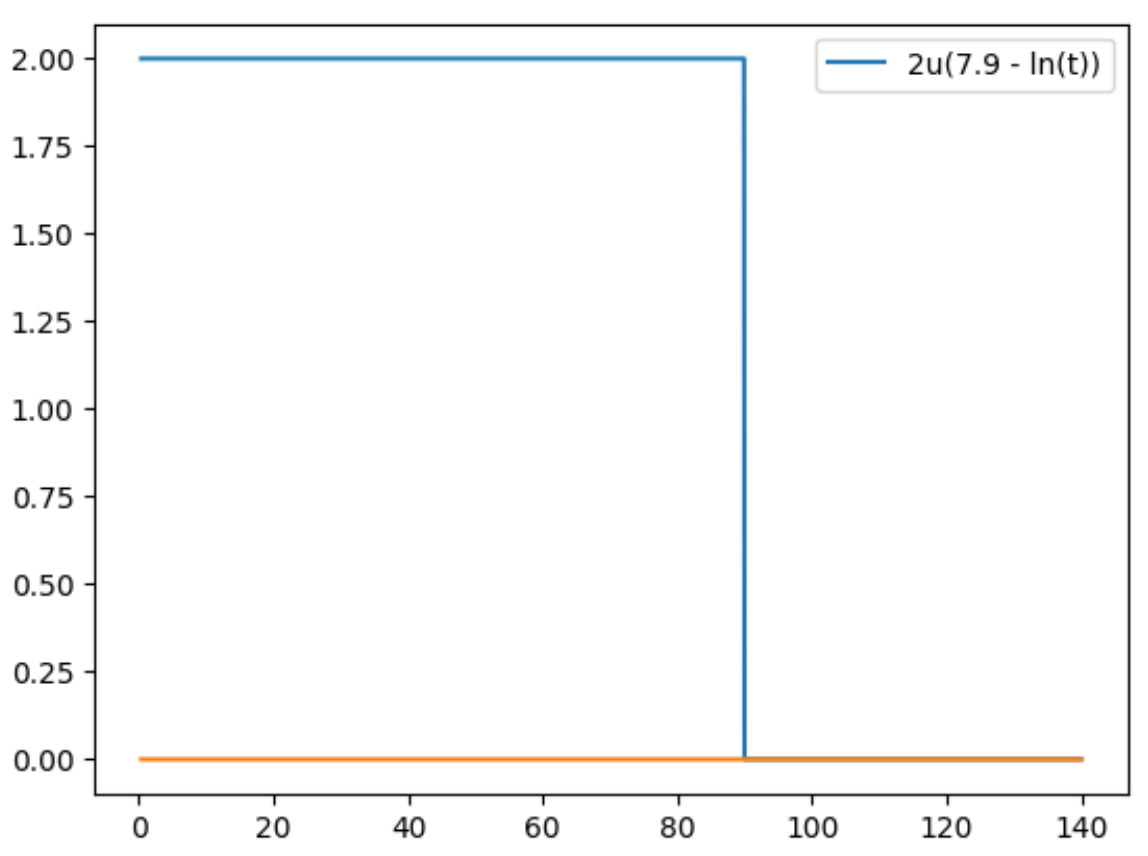}}
\subfigure[$|\textbf{A}_{t}|_{1,2}$ over time. ]{\includegraphics[width=4.3cm, height=2.4cm]{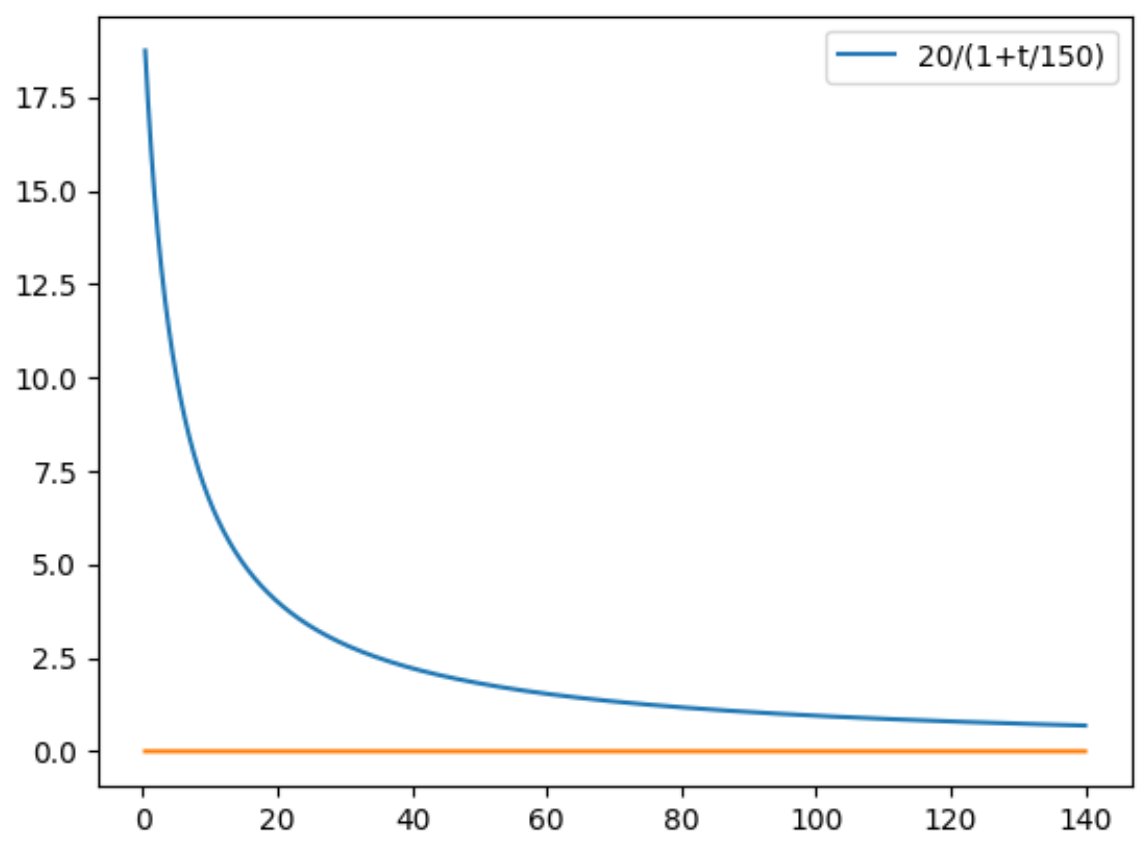}}
\subfigure[$|\textbf{A}_{t}|_{2,3}$ over time. ]{\includegraphics[width=4.3cm, height=2.4cm]{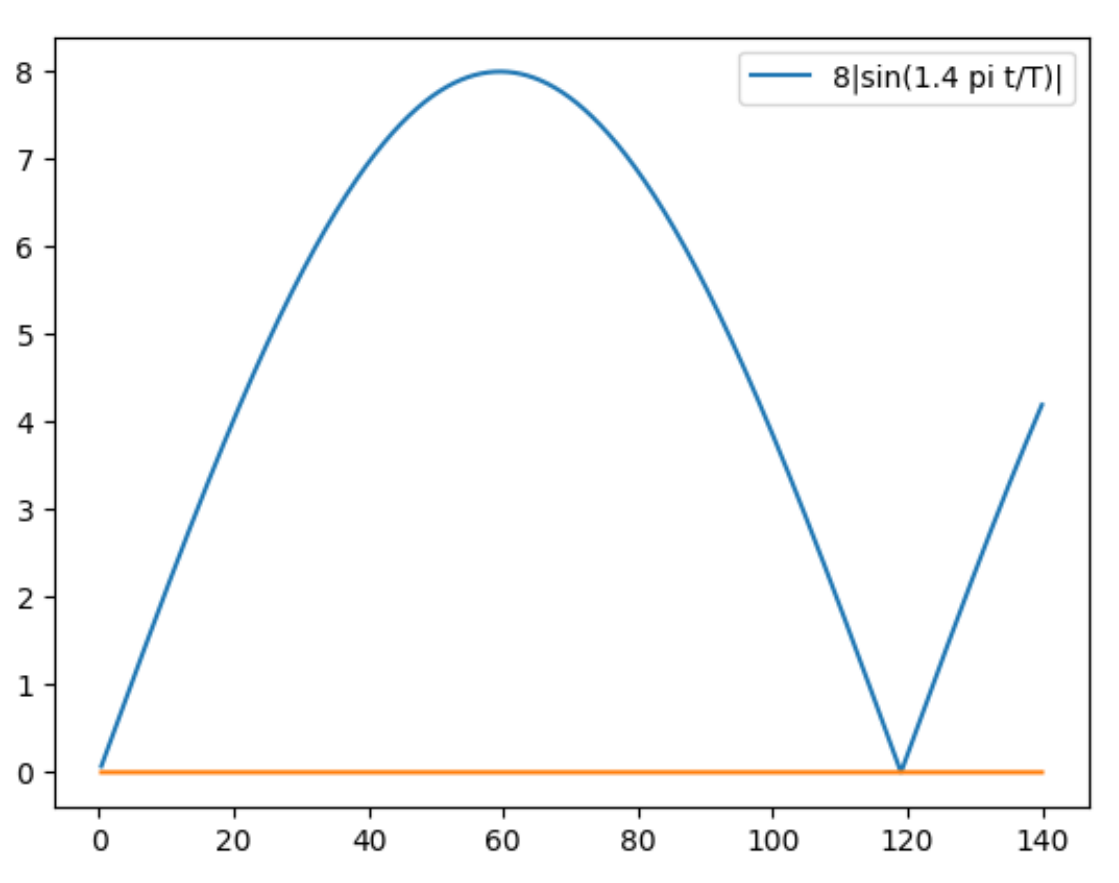}}
\subfigure[$X_0\rightarrow X_3$ over time. ]{\includegraphics[width=4.3cm, height=2.4cm]{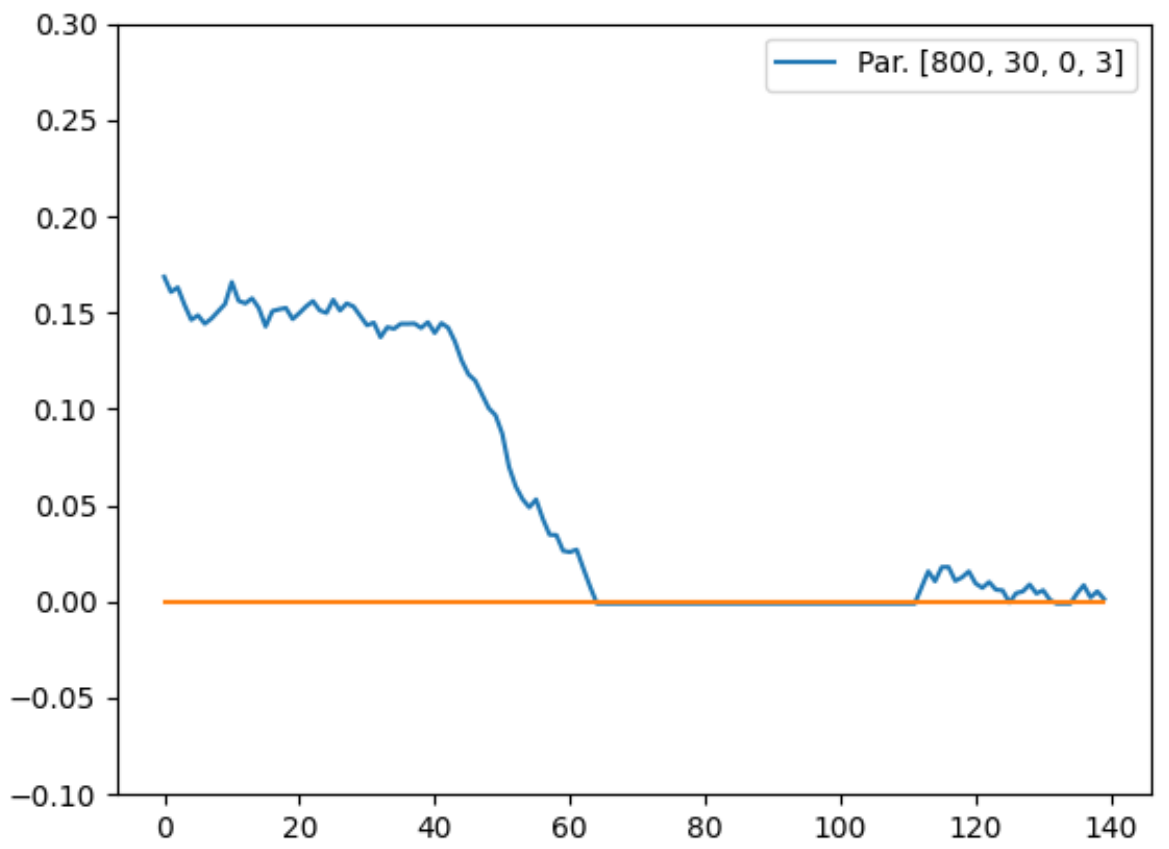}}
\subfigure[$X_1\rightarrow X_0$ over time.]{\includegraphics[width=4.3cm, height=2.4cm]{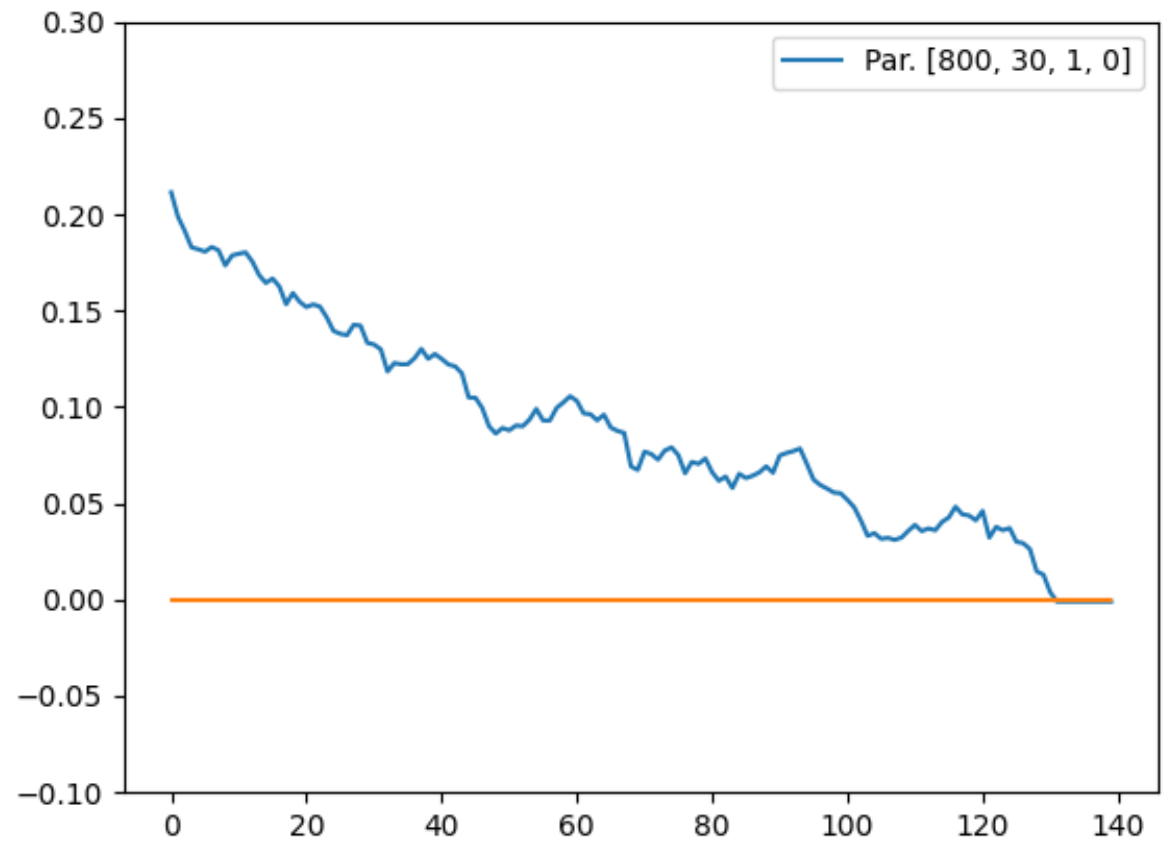}}
\subfigure[$X_2\rightarrow X_1$ over time.]{\includegraphics[width=4.3cm, height=2.4cm]{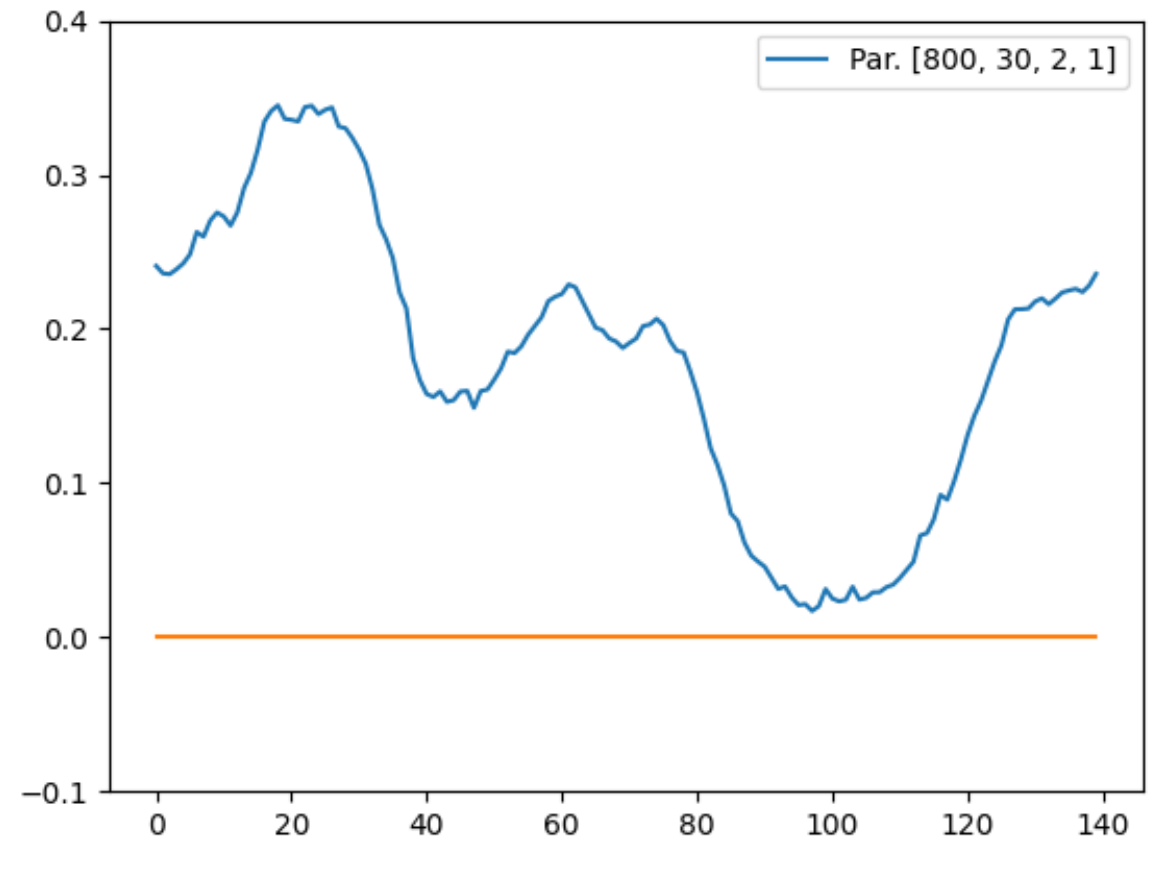}}
\subfigure[$H_t$ from $X_0$ to $X_3$. ]{\includegraphics[width=4.3cm, height=2.4cm]{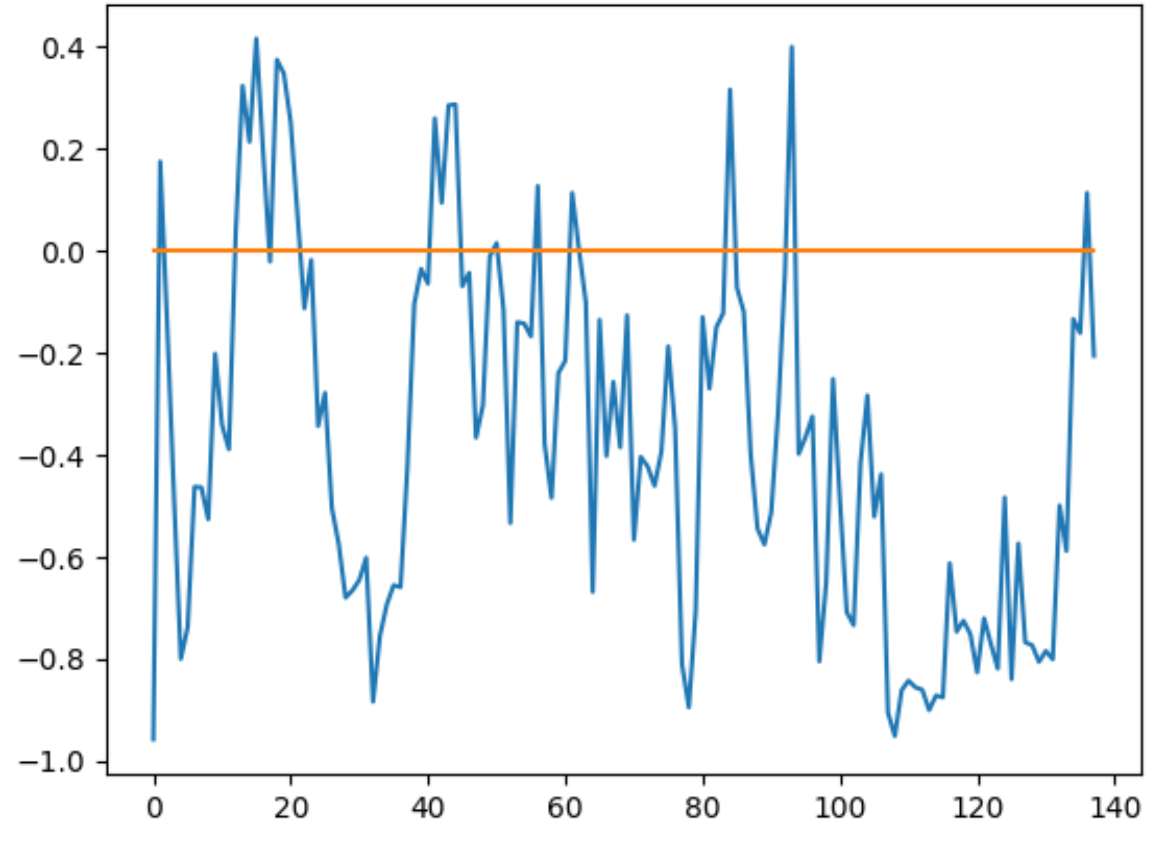}}
\subfigure[$H_t$ from $X_1$ to $X_0$.]{\includegraphics[width=4.3cm, height=2.4cm]{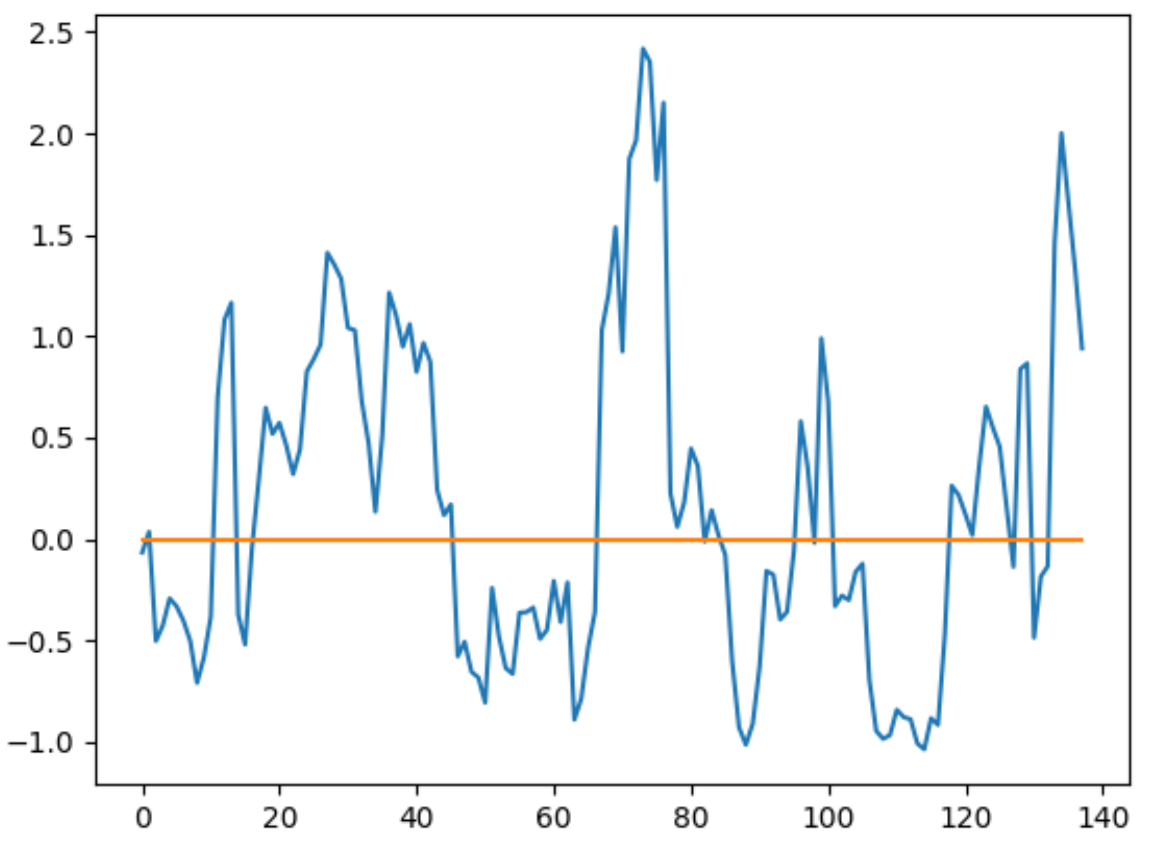}}
\subfigure[$H_t$ from $X_2$ to $X_1$.]{\includegraphics[width=4.3cm, height=2.4cm]{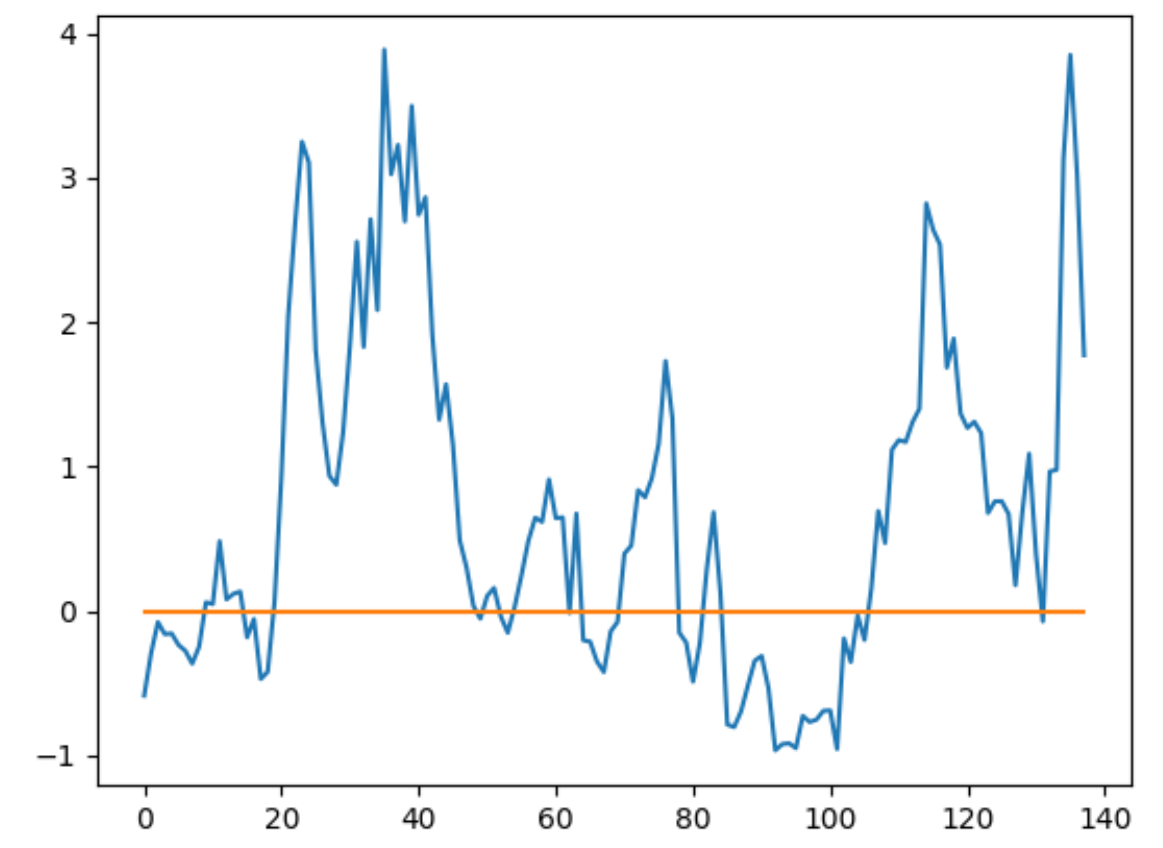}}
       \caption{A selection of the resulting plots of the time-varying network experiment. First row illustrates the functional dependencies between the selected time series, the second and the third rows present the resulting TV-DI in \eqref{eq:time_DI} and the time-varying Granger test in \eqref{eq:lu2014}, respectively. }
\label{fig:time_varying}
 \end{figure*} 

This setup enables us to observe and analyze how causal influences evolve over time, thus providing insights into the adaptability and effectiveness of our method in adjusting to changing network dynamics and capturing evolving causal relationships. In order to capture the time-varying causal effects among these five time series, we used the Monte Carlo simulation and generated data using the nonlinear DGP in \eqref{eq:sim:time} for a period of $T$ time steps. Afterwards, using the method introduced in Section \ref{sec:estimation}, we estimated \eqref{eq:time_DI} between all pairs of time series for $t\in\{30\tau: \tau\geq0\}$ and $S=800$. Some of the resulting plots\footnote{The x-axis present time with the scaling $30\tau$, i.e., $60$ represent $1800$.} are depicted in Figure \ref{fig:time_varying}. The remaining plots are presented in \ref{sec:app-emp0}. 

For instance, the influences $X_0\rightarrow X_3$ and $X_1\rightarrow X_0$ are encoded in $[\textbf{A}_{t}]_{4,1}=2u\big( 7.9 - \ln(t) \big)$ and $[\textbf{A}_{t}]_{1,2}=\frac{-20}{1+t/150}$, respectively. Therefore, the influence $X_1\rightarrow X_0$ decays over time and the influence $X_0\rightarrow X_3$ vanishes after certain time when $\ln(t)>7.9$. These changes of influences can be seen in Figures \ref{fig:time_varying}(d) and (e). 
Similarly, the influence $X_2\rightarrow X_1$, captured by $[\textbf{A}_{t}]_{2,3}=8\sin\left(1.4\frac{\pi t}{T}\right)$, showcases a different type of dynamic interaction. This term, reliant on a sinusoidal function of time, implies a periodic influence between these nodes, with the strength of the influence oscillating over time. 
This influence is captured by our TV-DI in Figure \ref{fig:time_varying}(f).
The model's ability to track these periodic changes, closely mirroring the sinusoidal pattern defined in the DGP, further demonstrates the TV-DI's precision in capturing time-variant and nonlinear relationships. The consistency between the estimated influences and the theoretical expectations from the sinusoidal term in $\textbf{A}_t$ emphasizes the model's capability to adapt to and accurately represent cyclic or oscillating causal effects in complex networks.
 It is noteworthy that the time-varying Granger test of \cite{hong2001test} defined by $H_t$ in \eqref{eq:lu2014} fails to accurately capture the time-varying causal influences as shown in figures \ref{fig:time_varying}(g)-(i).

\section{Empirical Studies}\label{sec:emp}

In this section, we apply the TV-DIG framework to identify and monitor the evolution of interconnectedness and systemic risk among major assets and industrial sectors within the financial network. Our primary objective is to support policymakers and regulators in managing systemic risk, thus maintaining financial market stability and integrity through macroprudential approaches. While there is a rich and growing body of literature on static or time-invariant network estimation and systemic risk, research on estimating dynamic networks from time series data remains limited. This research focuses on understanding how cryptocurrencies might affect financial stability. We explore the changing interactions between different sectors in the financial industry, especially how cryptocurrencies influence other sectors. The study also delves into the impact of the COVID-19 crisis and the Federal Reserve's emergency actions during the coronavirus outbreak in March 2020 on these interactions.

To verify any possible non-linear relations in the data, we applied a non-linearity test based on principal component analysis, as introduced by \cite{kruger2008developments}, detailed in \ref{sec:app-nonlinea}. This test is particularly effective in identifying non-linear dependencies in multivariate time series. \cite{kruger2008developments}'s method involves using principal component analysis to transform the data, followed by applying specific statistical tests to these transformed components to detect non-linearity. The test’s strength lies in its ability to uncover complex, non-linear relationships that traditional linear tests might miss. Our application of this test to our data set revealed significant non-linear interactions, thereby rejecting the null hypothesis of linear dependency between the series. This finding is crucial as it indicates that the underlying structure of the relationships between the series is inherently non-linear and any model attempting to estimate the influences within the network must be capable of capturing this non-linearity.

\subsection{Data}
In support of this research question, we obtained the daily logarithmic returns of 124 assets, 113 stocks from Global Industry Classification Standard (GICS) sectors: 17 stocks from Financials/Banks (GICS 4010), 25 stocks from Financials/Diversified Financials (GICS 4020), 22 stocks from Financials/Insurance (GICS 4030), 31 stocks from Real Estate/Equity Real Estate Investment Trusts REITs (GICS 6010), and 18 stocks from Fintech. Notably, Fintech is not a GICS-defined sector, and the selected companies are based on the KBW Nasdaq Financial Technology Index (KFTX). We also included a representative sample of the cryptocurrency market to determine the impact this new sector has on systemic risk within the entire financial industry. Our sample has 11 major cryptocurrencies with the largest market capitalization and trading volume which together account for approximately 75\% of total crypto market capitalization. The complete list of assets, including Banks, Diversified Financial Services, Insurance, REITs, FinTech, and Crypto, is presented in \ref{sec:app:ex}.

The period of study spans five years, from January 1, 2018, to January 1, 2023, which covers the recent global pandemic and the Fed’s response to pandemic. The Fed’s aggressive interest rate cuts, aimed at stimulating the economy, increased investors’ propensity to seek higher returns through riskier investments. This phenomenon, known as “reaching for yield,” has been well-documented in existing literature, with significant findings both in institutional (e.g., \cite{di2017unintended, andonov2017pension}) and individual investor behaviors \citep{lian2019low}.

\subsection{Related Literature}
Over the past decade, the cryptocurrency sector has evolved from an obscure asset to a wildly popular investment. The market capitalization of cryptocurrencies experienced a meteoric rise from approximately US\$7 billion at the end of 2015 to nearly US\$3 trillion in November 2021, as per data from CoinMarketCap.com. This was followed by a significant contraction, where the market capitalization plummeted to around US\$1 trillion during the recent cryptocurrency winter. The market value of cryptocurrencies has been characterized by substantial volatility.
In early 2023, the cumulative market capitalization witnessed a resurgence after the decline in 2022, which was precipitated by the Terra/Luna collapse and the FTX crisis. The TerraUSD collapse eradicated over \$50 billion in value as shown in \cite{uhlig2022luna}, and the subsequent failure of FTX, the second-largest cryptocurrency exchange, intensified skepticism towards the industry, highlighting  issues related to regulatory oversight.

The dynamics of market capitalization in the cryptocurrency sector are indicative of the capital flow within this market. Although, the cryptocurrency market's size is relatively small in comparison to the nearly \$300 trillion global financial system, its impact cannot be underestimated. Historical precedents, such as the financial crisis of 2008, demonstrate that even a small segment of the financial sector can catalyze significant stability concerns. For instance, the subprime mortgage market, va    lued at approximately \$1.2 trillion in 2008, was a key factor in the crisis (\cite{pinto2010sizing}). This underscores the importance of understanding the implications of the cryptocurrency market's evolution, not just in terms of its size but also in the context of its potential impact on broader financial stability.
In response to the growing prominence of cryptocurrencies, governments worldwide, including the United States, are examining how to regulate them. A notable example is the executive order by the U.S. President on March 9, 2022, which mandates a comprehensive review of digital assets, including cryptocurrencies. This review, conducted by federal agencies, focuses on evaluating the impact of digital currencies on overall financial stability and other relevant factors.

Building on the governmental efforts to regulate the burgeoning cryptocurrency sector, it's important to recognize that modeling and analyzing the cryptocurrency market, along with its interactions with other financial markets, presents a significant challenge. This complexity arises from the market's inherent volatility, its multifaceted nature, and the rapid pace of its evolution. 
Many cryptocurrency assets are characterized by a lack of clear fundamental value or cash flows, as noted by \cite{makarov2020trading}. These assets are prone to fragmentation, offering opportunities for arbitrage and being vulnerable to market manipulation, as indicated by studies like \cite{griffin2020bitcoin} and \cite{gandal2018price}.

The crypto market's periods of rapid growth have drawn the attention of both retail and investors, as well as politicians and regulators, as \cite{auer2022distrust} points out. Additionally, fluctuations in the cryptocurrency market may increasingly align with other asset classes.
Research by \cite{chuen2017cryptocurrency, borri2019conditional, petukhina2021investing} highlights a generally low correlation between cryptocurrencies and other asset classes. However, \cite{iyer2022cryptic} and \cite{dong2023tracing} present evidence of growing links between U.S. equity markets and the prices of Bitcoin and Ethereum, suggesting a deeper integration of crypto markets with the equity cycle.
Specifically, \cite{iyer2022cryptic} and \cite{dong2023tracing} demonstrate that the correlation between Bitcoin and the S\&P 500 was initially low but has significantly increased since 2020. Most of these studies compare market indices, typically representing the stock market, with Bitcoin as the crypto market proxy.

The methodologies in the aforementioned work are often basic, relying on visual inspection and correlation-based analyses. Such methods can be misleading due to their sensitivity to the chosen time frame and the distribution of the sample data. They generally assume linearity and a time-invariant distribution of model parameters, which may limit their accuracy.
Our understanding of the primary factors driving crypto asset prices in this immature industry, as well as the elements influencing the correlation between cryptocurrency and other markets, remains limited. The existing literature offers a variety of potential and compatible explanations for these dynamics. 

To elucidate why crypto assets may have become more correlated with other asset classes post-COVID, studies indicate an increase in retail trading during the pandemic lockdowns, encompassing both crypto and stock markets. This is highlighted in the research by \cite{divakaruni2023uncovering} regarding crypto trading and \cite{ozik2021flattening} in the context of stock trading. Furthermore, \cite{toczynski2022fiscal} estimates that approximately US\$15 billion of the federal stimulus checks were invested in trading crypto assets.
The landscape for crypto asset trading evolved to accommodate increasing demand. Notably, well-known mobile payment applications like Revolut and PayPal, along with trading platforms such as Robinhood, began offering crypto trading services to their clients. April 2021 saw Coinbase, a prominent centralized crypto exchange, make its debut on the Nasdaq. Additionally, new investment vehicles like the Grayscale Bitcoin Trust emerged, providing investors with exposure to crypto assets without the need to hold the tokens directly.

Amid these developments, institutional investor engagement in crypto assets also saw a rise. Research by \cite{liu2022common} indicates a notable decrease in the volatility of crypto returns from 2011 to 2018. This trend suggests a diminishing risk factor, potentially making the crypto markets more appealing to institutional investors over time. Utilizing a supervisory database, \cite{auer2023banking} highlights the increasing role of traditional financial intermediaries in the crypto sphere. They observe that banks' involvement in crypto assets, although still a small fraction of their total balance sheets, has grown and holds significant implications for a market that was once dominated largely by retail investors.

However, the dynamics between crypto and stock markets may not be predominantly influenced by retail investors, as their trading patterns for these assets differ (\cite{kogan2023cryptos}). It is posited that the post-COVID monetary policies of the Federal Reserve and the resulting response from institutional investors are more crucial factors. These investors, now more active in the crypto markets, tend to trade cryptocurrencies and other high-risk assets in a correlated manner. In a low-interest environment, it's logical for institutions to pursue higher returns by opting for riskier assets (\cite{di2017unintended, andonov2017pension}). This trend is underscored by findings from \cite{devault2021embracing}, who note that investors in crypto assets often outperform their peers.
Thus, the increasing institutional participation in crypto markets raises the possibility of heightened risk for spillovers into the broader economy. This is particularly relevant as these institutions' portfolios often include a mix of crypto and traditional assets, and they are less likely to rebalance their portfolios as frequently as retail investors.

 \subsection{DIG of the Industry groups} 
 \begin{figure}[t]
\centering
\subfigure[Network of (3/6/2018 - 6/5/2018).]{\includegraphics[scale=.18]{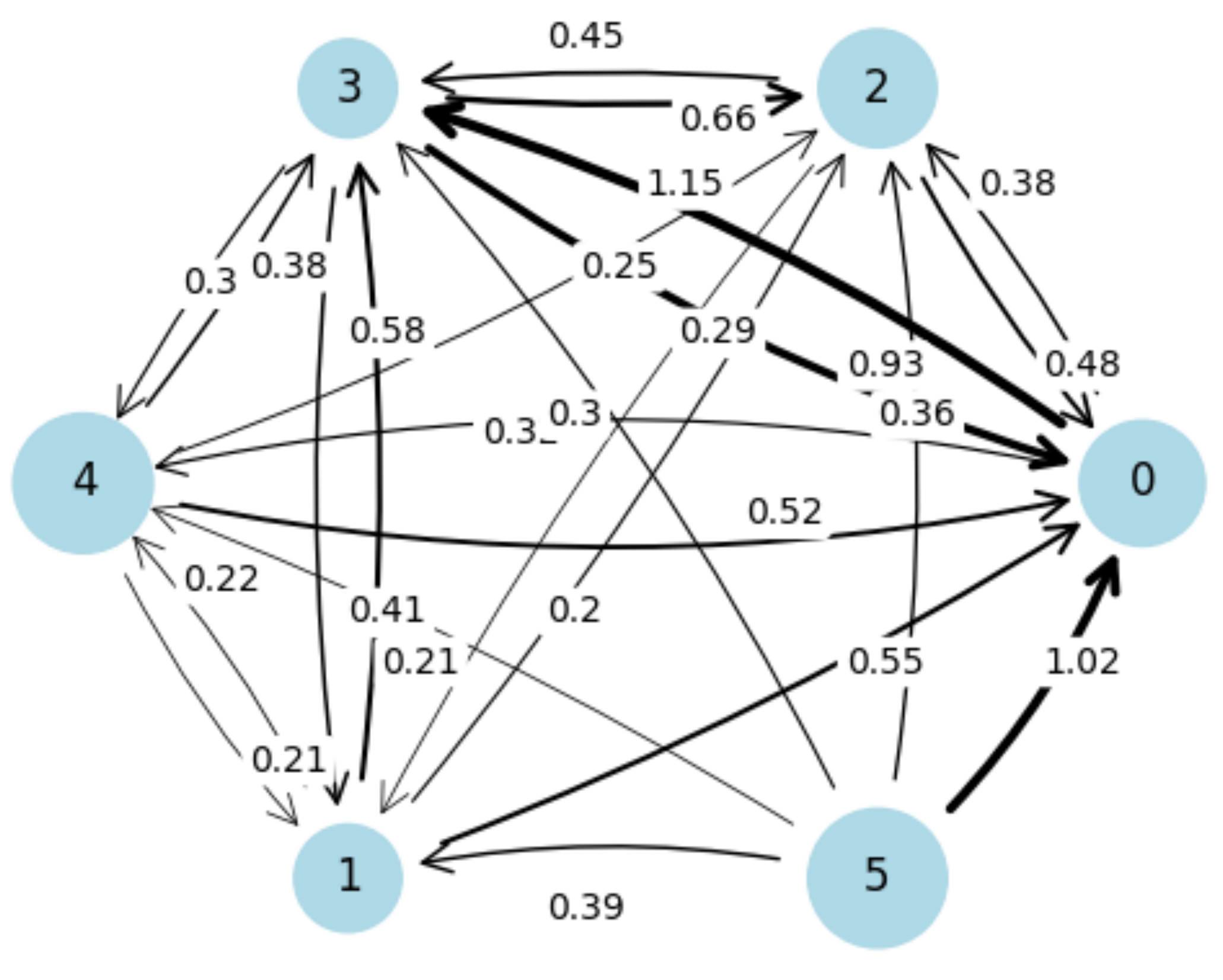}}
\hspace{.1cm}
\subfigure[Network of (11/4/2019 - 2/5/2020).]{\includegraphics[scale=.18]{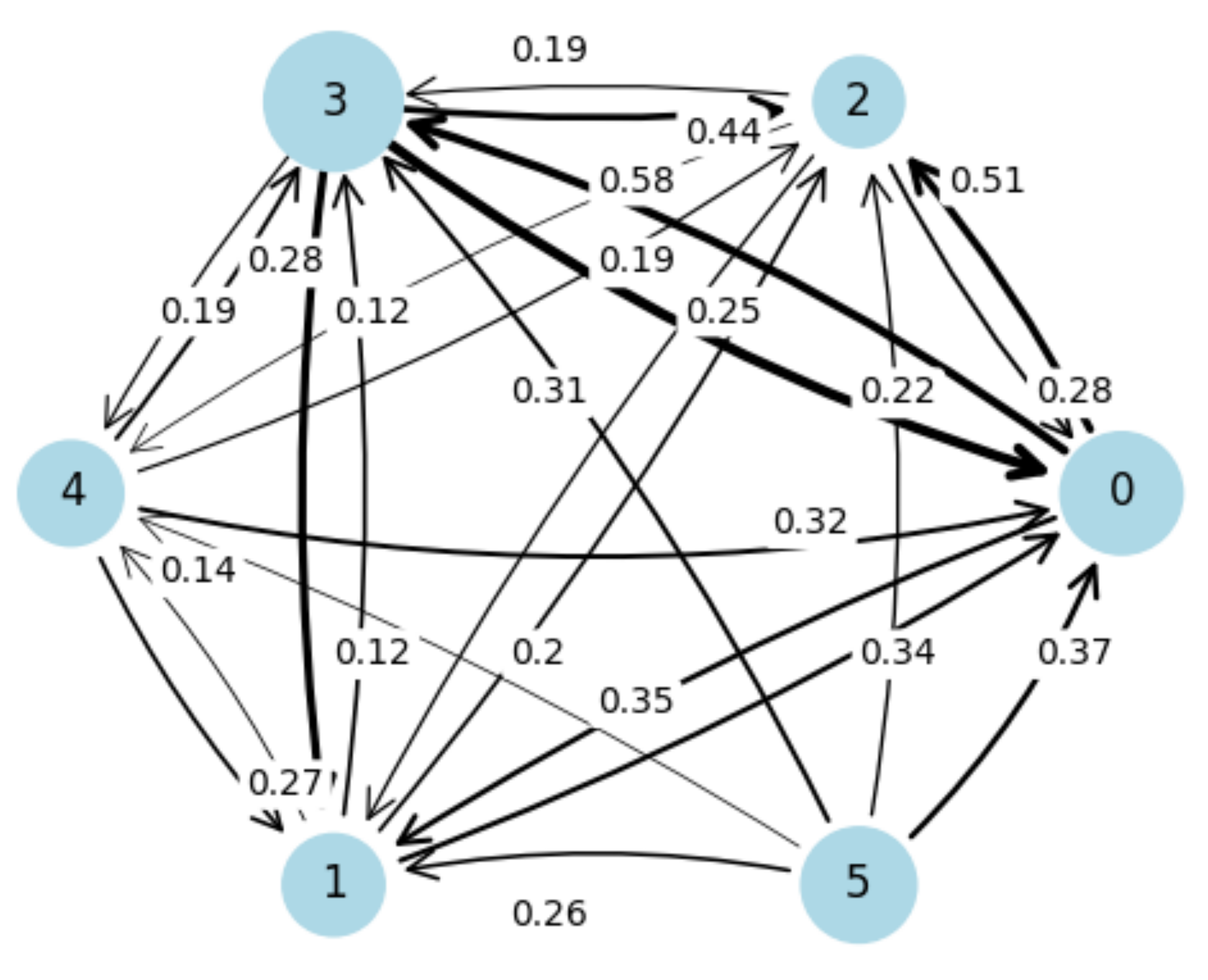}}
\subfigure[Network of (9/3/2020 - 12/3/2020).]{\includegraphics[scale=.18]{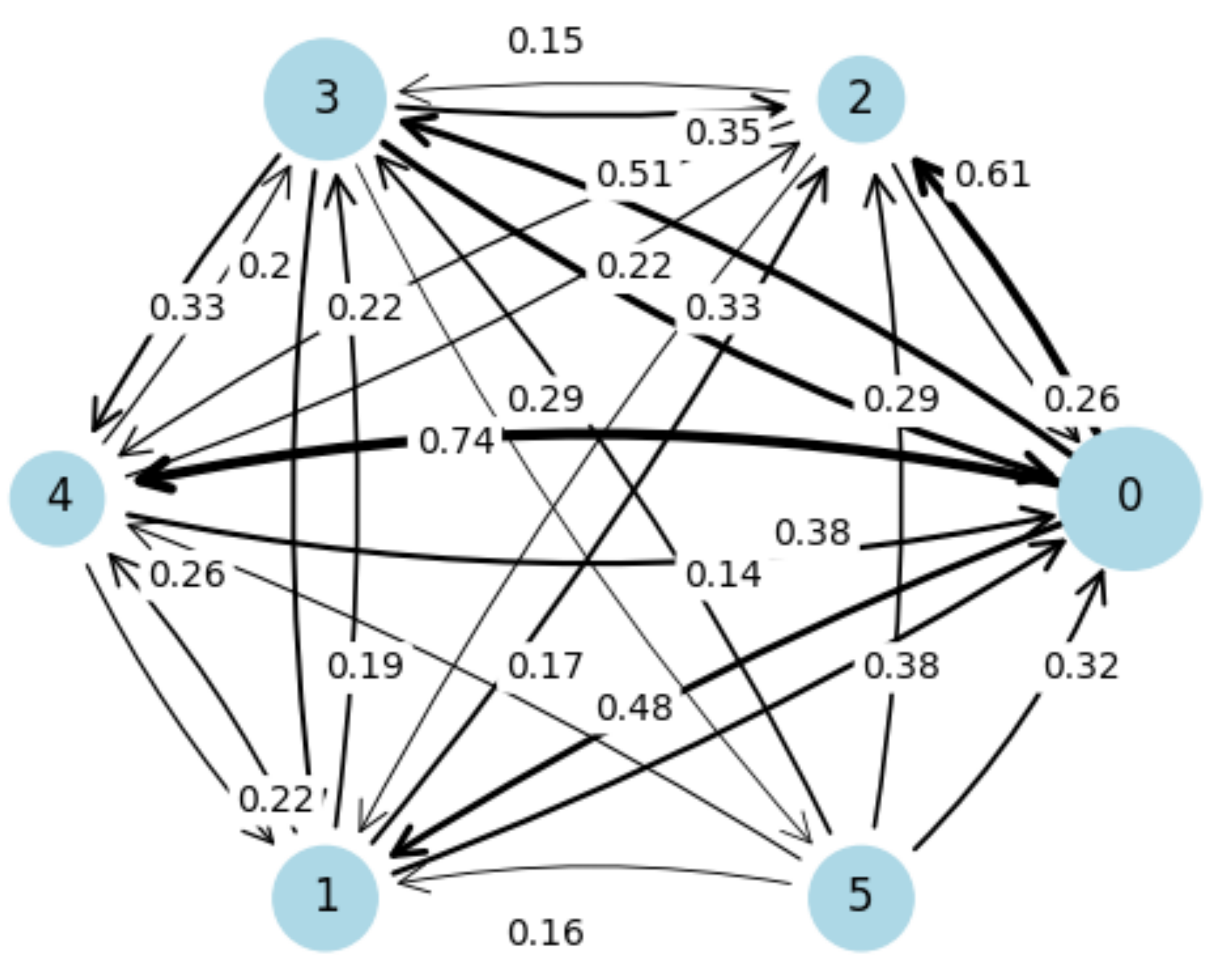}}
\hspace{.1cm}
\subfigure[Network of (5/5/2022 - 8/5/2022).]{\includegraphics[scale=.18]{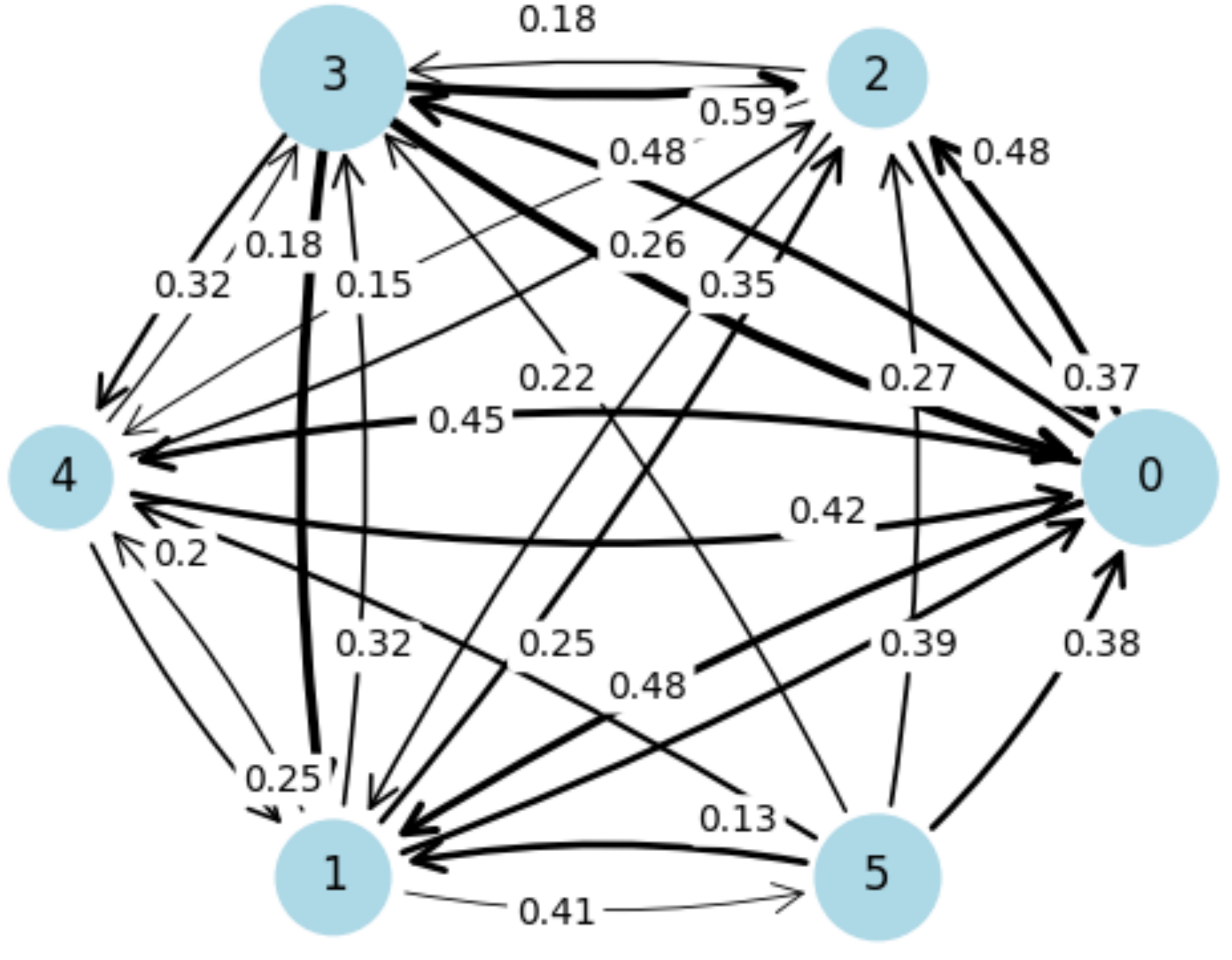}}
\caption{The causal networks among the industries in different time periods. }
\label{fig:time_varying_ind}
\end{figure}

We use this literature to explain the time-varying and dynamic nature of the interconnection between crypto and other assets in the financial network, and to investigate potential cross-sector spillovers. This approach enables us to indirectly track the flow of funds across different investment avenues, such as real estate, capital markets, the cryptocurrency market, and bank accounts. The insights gained from this model are valuable for policymakers and regulators. They can use this information to devise effective strategies and take actions that reduce systemic risk within financial networks.  

To detect the cross-sector causal relationships, we obtained the DIG between them by estimating the DIs in Equation  \eqref{eq:agg-di} in a rolling window moving block bootstrapping scheme (See Section \ref{sec:estimation}) between 01/2018 and 12/2022. 
More precisely, we used rolling windows of size $T_{rw}$=63 that is almost three months, $L_B$=3, and constructed batches of size 20. The windows are shifted $r=21$ (about one month) steps into the future which leads to $57$ point estimates of DIs.

\begin{figure}[t]
\centering
\subfigure{\includegraphics[scale=.15]{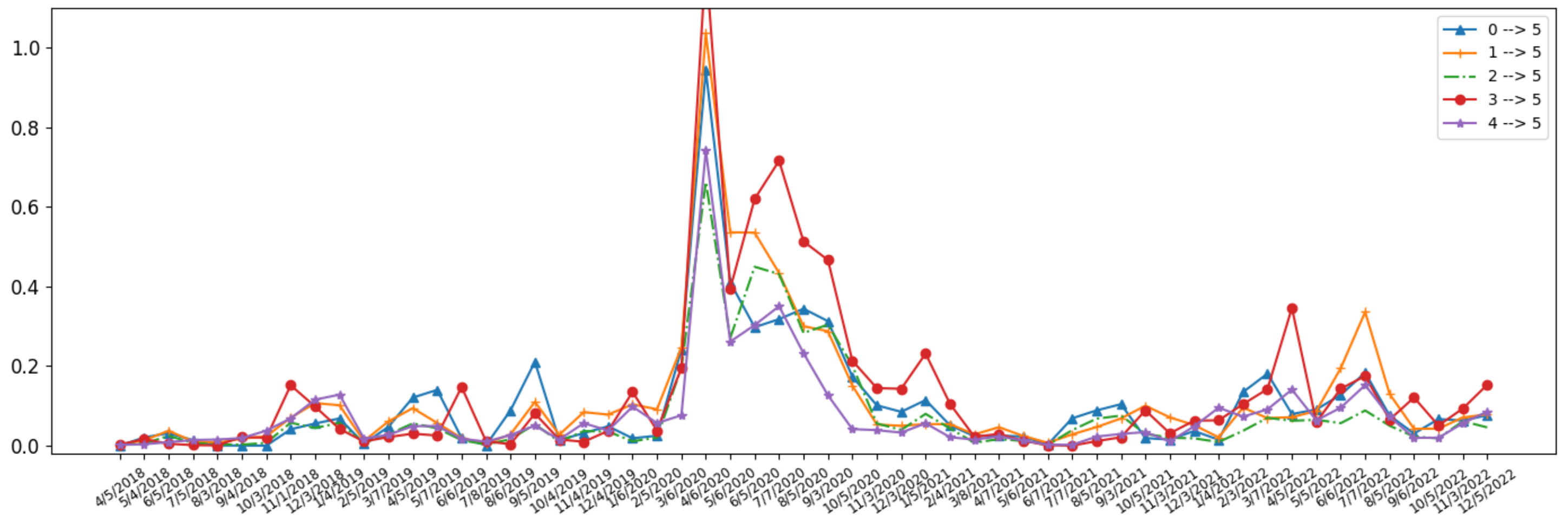}}
\subfigure{\includegraphics[scale=.15]{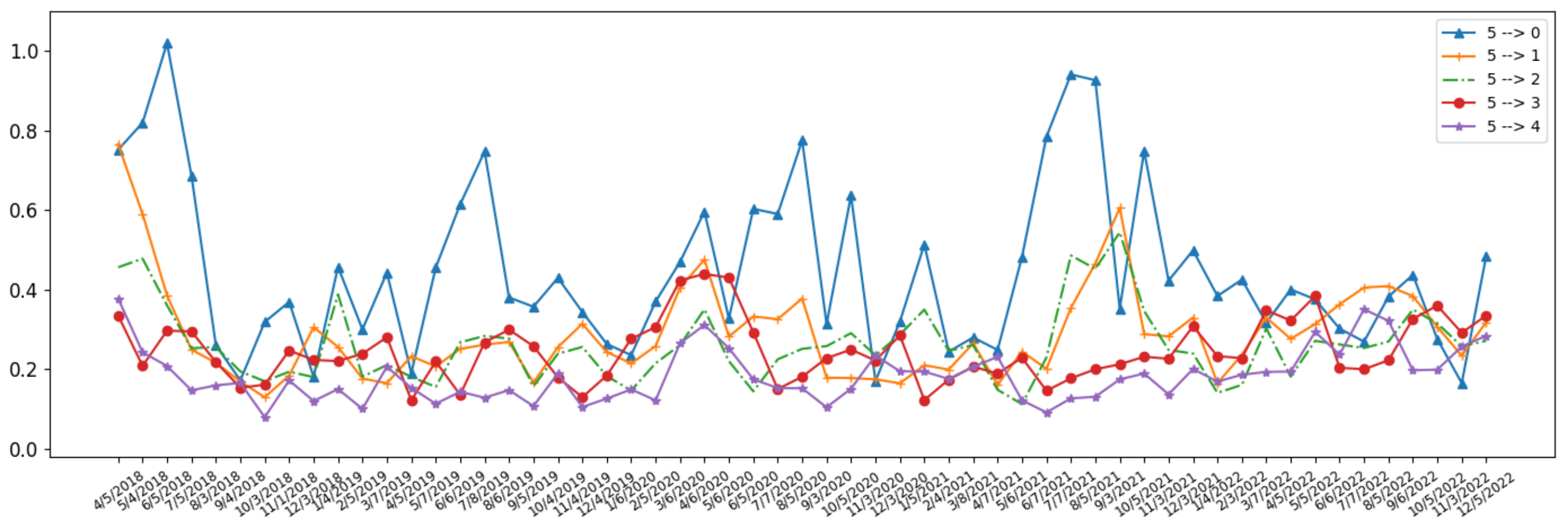}}
\subfigure{\includegraphics[scale=.15]{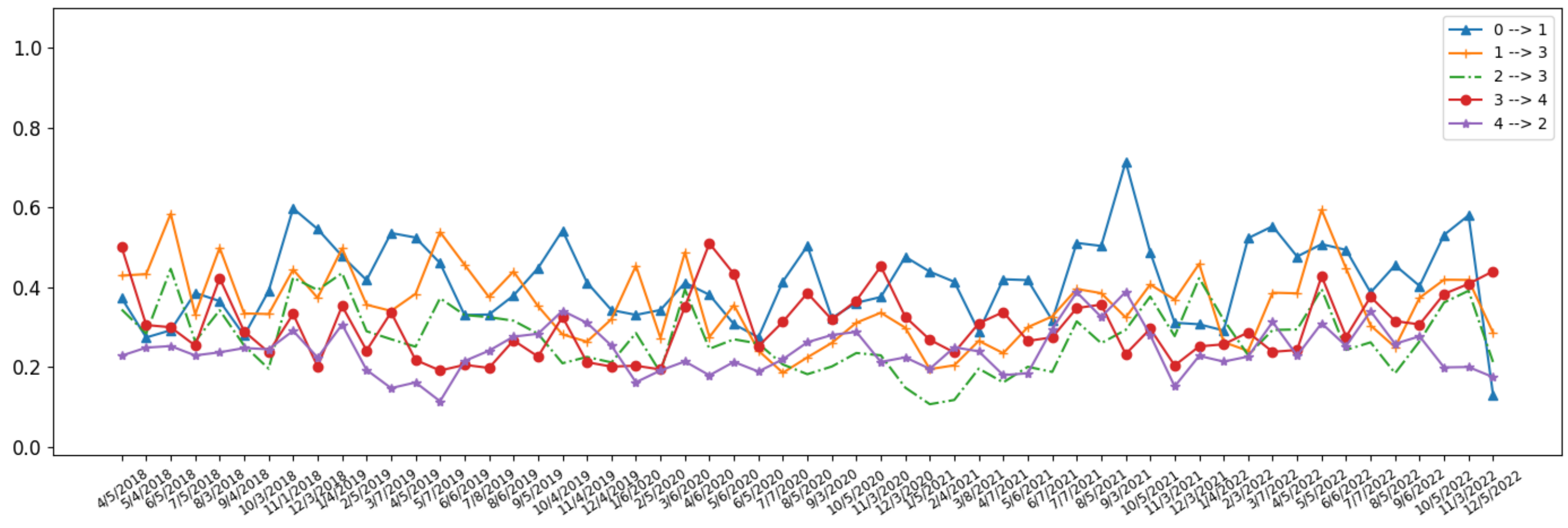}}
\subfigure{\includegraphics[scale=.15]{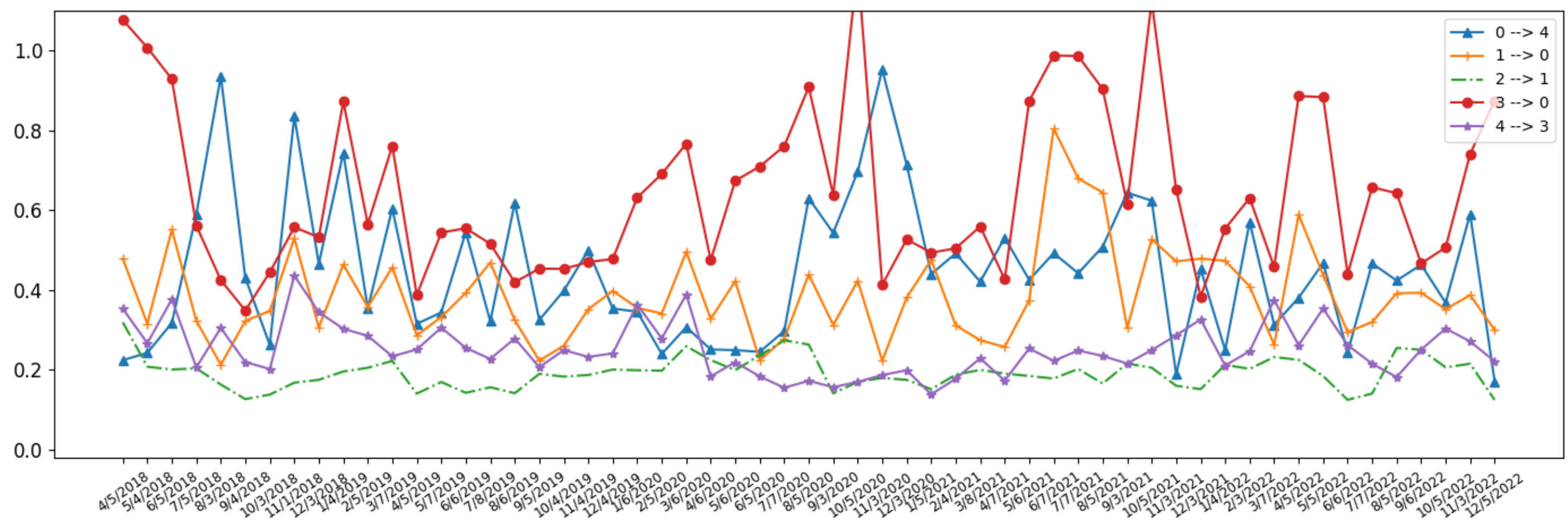}}
\caption{A selection of the resulting plots of the time-varying influences between the industries.}
\label{fig:time_varying_empi}
\end{figure} 
 Figure \ref{fig:time_varying_ind} illustrate four of the learned networks. In these networks, nodes with label $[0,1,2,3,4,5]$ denote industries [Banks, Diversified Financial, Insurance, Real Estate Investment, Financial Technology, Crypto], respectively. 
 More precisely, the causal influence from Banks to Insurance is $I(0\rightarrow 3 || 1, 2, 4, 5)$ which is computed at $57$ different time periods between 2018 and 2022. For instance, as depicted in Figure \ref{fig:time_varying_ind}(a), this influence between 3/6/2018 to 6/5/2018 is 1.15 which is shown by a weighted edges from node $0$ to $3$.
 In particular, node $5$ which represents the Crypto has several outgoing edges that is an indication that the cryptocurrencies have non-negligible impacts on other industries. 
 Please note that the thickness of the edges are proportional to the estimated DIs, i.e., the larger the causal influence from one node to another node is, the thicker the edge between the two nodes will be. 
 Furthermore, the size of the nodes are proportional to their centrality, that is the ratio of their outgoing influences to the overall influences. 

 Table \ref{table:contribution} reports the top three influential assets in each industry groups for four time windows that is $\small{\arg\max_{A_j\in\mathcal{A}}C_{\mathcal{A}\rightarrow\mathcal{B}}(R_{A_j})}$, where $\small{C_{\mathcal{A}\rightarrow\mathcal{B}}(R_{A_j})}$ is defined in \eqref{eq:contribution}.

 \begin{tiny}
 \begin{table}[!htb]
\centering
  \begin{tabular}{|p{1.9cm} |p{2.6cm}|p{2.7cm}|p{2.7cm}|p{2.7cm}|}  \hline
 & \footnotesize{3/6/2018- 6/5/2018}  & \footnotesize{11/4/2019- 2/5/2020}  & \footnotesize{9/3/2020- 12/3/2020}   &  \footnotesize{5/5/2022- 8/5/2022}   \\ \hline\hline
\textit{Banks}   &  
    \small{JPMorgan, \newline PNC, \newline M\&T Bank}
  & \small{M\&T Bank, US Bancorp, Comerica Inc.}  &  \small{PNC,\newline Zions,\newline JPMorgan}  & \small{PNC,\newline JPMorgan,\newline M\&T Bank}  \\ \hline
\textit{Div. Fin.}    &   \small{Morgan St.,\newline American Exp.,\newline BlackRock}  &  \small{BlackRock, Morgan St., Ameriprise Fin.}  &  \small{MSCI, American Exp., Discover Fin.}              & \small{Goldman Sachs, Nasdaq, Morgan St.}  \\ \hline
\textit{Insurance}  &  \small{Allstate,\newline Chubb Lim.,\newline Marsh\& McLennan}  &  \small{Travelers Co.,\newline Allstate,\newline Marsh\& McLennan}  &  \small{Allstate,\newline American Int.,\newline Progressive Co.}             & \small{Prudential Fin.,\newline Employers Holdings, \newline Chubb Lim.}  \\ \hline
\textit{Real Est.}  &  \small{Prologis Inc.,\newline Equity Residential, Federal Realty Inv.}  &  \small{Prologis Inc.,\newline Equinix Inc,\newline Simon Property}  &  \small{Equity Residential, Extra Space Storage, SL Green}      & \small{American Tower,\newline Welltower,\newline AvalonBay}  \\ \hline
\textit{Fin. Tech.}  &  \small{Mastercard,\newline Verisk Analytics, Visa}  & \small{Fair Isaac,\newline Global Pay., Mastercard}   &    \small{Fair Isaac,\newline CoStar, Verisk Analytics}      & \small{Thomson Reuters,\newline Visa, Euronet}  \\ \hline
\textit{Crypto}  &  \small{BTC-USD, ETH-USD, DOGE-USD}  &  \small{ETH-USD, BTC-USD, BNB-USD}  &  \small{ETH-USD, LTC-USD, DOGE-USD}             & \small{BTC-USD, ETH-USD, BNB-USD}  \\ \hline
  \end{tabular}   \caption{Top three influential assets in each Industry group for four time windows.}\label{table:contribution}
\end{table}
 \end{tiny} 
For a better visualisation of the time-varying causal influences, we present a selection of estimated DIs between 01/2018 and 12/2022 in Figure \ref{fig:time_varying_empi}. 
The x-axis of this figure presents the beginning of the $57$ time periods in which the DIs defined in  \eqref{eq:agg-di} are estimated. Note that each time period is about three months and the y-axis presents the estimated values of the DIs.

In the top right of Fig. \ref{fig:time_varying_empi}, we observe that our method detects a heightened level of interaction between cryptocurrencies and assets in other industry groups during the first half of 2020. This increase coincides with the COVID-19 crisis and the Federal Reserve's response to it. Additionally, notable periods in 2022 are aligned with significant events, such as the collapse of Terra in May 2022, which erased \$50 billion in valuation, and the FTX crisis in November 2020. The plot illustrates the influence of assets from other industry groups on cryptocurrencies, where all groups exhibit almost similar patterns. To quantify this influence, we applied a threshold to include only significant influences. Generally, as observed in the plot, the influence from assets in other industry groups on cryptocurrencies was comparatively lower before 2020. For complete list of the results, please see \ref{sec:ap-emp1}.

In the top left of Fig. \ref{fig:time_varying_empi}, we observe a surprising deviation from existing studies, which predominantly suggest that the influence of cryptocurrencies on the stock market was insignificant before 2020. These studies often focus on Bitcoin and Ethereum, representing the crypto market, and their interactions with stock market indices using relatively simplistic models. However, our results reveal a different scenario when we concentrate on a specific segment of the market, including only assets in financial sectors, and incorporate a broader range of crypto assets. We found that, even before 2020, there were significant influences from cryptocurrencies to these industry groups, particularly Banks and Diversified Finance, indicative of the capital market. In 2020, an increase in the influence of cryptocurrencies on Banks and other sectors was observed. Although our study did not focus on other sectors and industries, similar analyses could be applied to them. Given these findings, we can assert that cryptocurrencies, as a new sector, have the potential to pose systemic risks within financial networks. These changes in influence can be tracked using our time-varying approach.

\subsection{DIG of the Assets}\label{sec:ddi}

An alternative measure of the aggregated influences between different sectors and industry groups is the number of interconnections between the assets of those industries.
For instance, the number of edges from assets of Financial/Banks to the assets of Financial/Insurances.  Such measure has been used in various work such as \citep{billio2012econometric}. 
This requires the complete causal network between all 124 assets.
Hence, we inferred the DIGs of the assets by estimating the time-varying causal network among all 124 assets by estimating the DIs in (\ref{eq:time_DI}) in a rolling window moving bock bootstrapping scheme for $57$ different time instances within the time horizon between 01/2018 and 12/2022. 

\begin{figure*}[h]
\centering
\subfigure{\includegraphics[scale=.3]{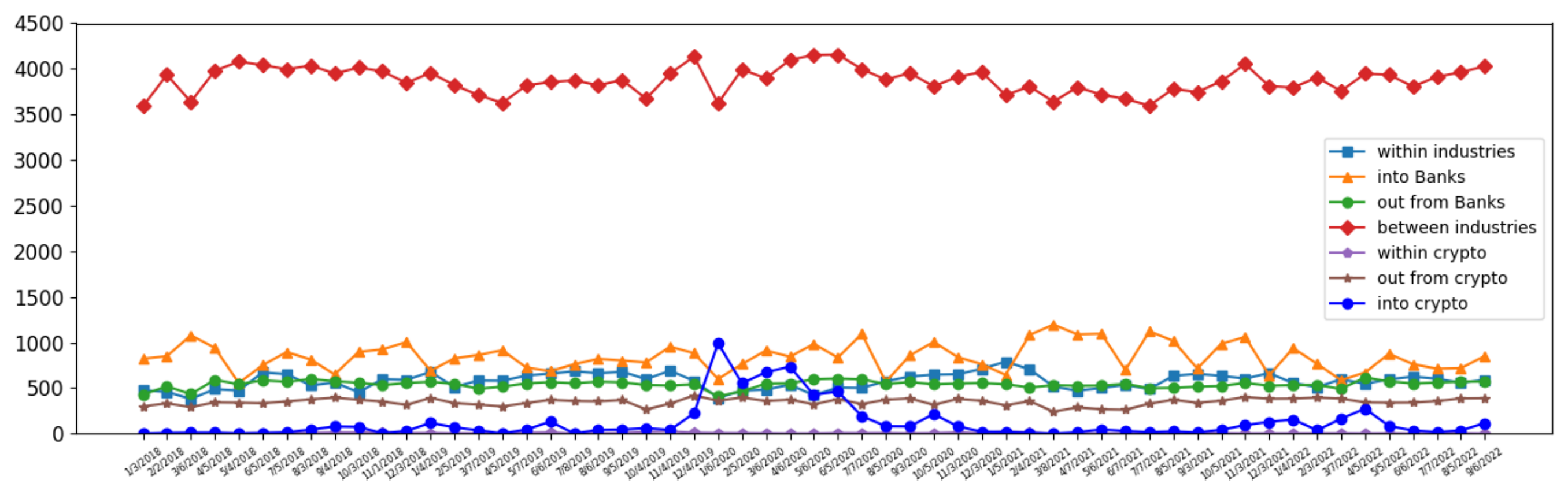}}
\caption{Number of connections over time between or within industry groups.  }
\label{fig:time_varying_edges}
\end{figure*} 
Estimating the complete network presents the challenge of high-dimensionality. As discussed in Section \ref{sec:dig}, to determine the influence of node \( i \) on node \( j \) at time \( t \), it's necessary to estimate \( I_{t,S}(R_i \to R_j \,||\, R_{-\{i,j\}}) \). In this study, this required estimating a joint distribution with a dimension of 124. Generally, such non-parametric estimation, without any knowledge and assumption about the underlying distribution, demands a substantial volume of independent samples, which were limited in our empirical study. To address this, we approximated the DIs by dimension reduction. Instead of conditioning on the entire set \( R_{-\{i,j\}} \) with 122 elements, we conditioned on a smaller subset \( S'_j \subset R_{-\{i,j\}} \) with only 10 elements, comprising assets most correlated with \( R_j \). Specifically, we ranked institutions in \( R_{-\{i,j\}} \) by their correlation with \( R_j \) and selected the top 10 for the conditioning set. This approximation is based on the fact that the conditioning set \( R_{-\{i,j\}} \) can be reduced to only the set of all direct causes of \( R_j \) within the set \( R_{-\{i,j\}} \). By postulating that these causes are among the highly correlated assets with \( R_j \), we obtained the approximation. 

Figure \ref{fig:time_varying_edges} presents the resulting number of connections within and between different industry groups and crypto. For instance, the curve labeled with `into Banks' demonstrates the number of edges from assets that are not categorized as Financial/Banks to assets that are categorized as Financial/Banks. 
The curve labeled with 'within industries' shows the number of edges that both of its end nodes are assets of the same industries, e.g., edges between assets of Insurances or edges between assets of Cryptos.
The curve `between industries' is the total number of edges from one industry to an another industry. 

Comparing the results in Figure \ref{fig:time_varying_edges} and Figure \ref{fig:time_varying_empi}, we observe similarities. For instance, as it is illustrated in Figure \ref{fig:time_varying_empi} (top right), there is higher amount of influences from other industries on Crypto during 11/4/2019 to 11/3/2020 and similarly in this time window, the number of edges coming into the Crypto's assets from other assets are higher according to the `into crypto' curve in Figure \ref{fig:time_varying_edges}.

\section{Conclusion}
As cryptocurrencies increasingly cement their position in the global financial landscape, understanding their dynamic interactions with other parts of the financial network has become crucial for policymakers and regulators. These assets, once peripheral, are now integral to financial markets, necessitating a deeper comprehension of their influence and interconnectedness. Addressing this need, our paper introduces a novel analytical framework that captures the evolving nature of these interactions through time series data.

Methodologically, our study introduces groundbreaking approaches in network analysis. The Time-Varying Directed Information Graph (TV-DIG) measure, a first-of-its-kind information-theoretic framework, allows for capturing time-varying causal relationships in a general network of time series. We complement this with a robust, non-parametric method for estimating TV-DI, utilizing rolling window moving block bootstrapping combined with advanced machine learning estimators for mutual information. This methodological innovation not only enhances the estimation process of Directed Information (DI) but also provides a powerful tool for inferring aggregated causal effects. Such tools are crucial for understanding complex influences within financial networks, like identifying the most or least influential sectors in the global financial network.

Empirically, this study marks a pioneering effort in analyzing the systemic risk posed by crypto assets to financial networks. We reveal the intricate and dynamic nature of interactions between cryptocurrencies and other financial assets, shedding light on potential cross-sector spillovers. The insights gleaned are invaluable for policymakers and regulators, providing a framework to track fund flows across various investment channels and to detect significant influences of cryptocurrencies on other sectors. Our findings challenge existing notions by illustrating significant pre-2020 influences of cryptocurrencies on certain financial sectors, underscoring the potential systemic risks posed by this new asset class. The time-varying approach we developed offers a novel perspective in tracking these evolving influences, thereby contributing to the understanding and management of systemic risks in financial networks.


\bibliography{ref}

\newpage
\appendix

\section{Examples}\label{sec:examples}

\begin{example}\label{r1}
Consider a network of three times series $\{X,Y,Z\}$ with the following VAR model also known as structural equation model (SEM):
\begin{small}
\begin{align}\label{model1}
& \begin{pmatrix}
  Z_{t} \\
  X_{t} \\
  Y_t
 \end{pmatrix}= 
  \begin{pmatrix}
  -0.5 & 0 & 0 \\
0.4 & 0.5 & 0\\
-\frac{1}{\sqrt{2}} & 0 & -0.2
 \end{pmatrix}  \begin{pmatrix}
  Z_{t-1} \\
  X_{t-1} \\
  Y_{t-1}
 \end{pmatrix}+  \begin{pmatrix}
 \epsilon_{z,t} \\
  \epsilon_{x,t} \\
  \epsilon_{y,t}
 \end{pmatrix},
\end{align}
\end{small}
where $\epsilon_x$, $\epsilon_y$, and $\epsilon_z$ are three independent stationary Gaussian processes with mean zero and covariance matrix $\text{diag}(0.8, 0.9, 1)$ that is a diagonal matrix. 
Due to the functional relationships between these time series, the causal network of this model is $X\leftarrow Z\rightarrow Y$, i.e., there are influences from $Z$ to $X$ and from $Z$ to $Y$ because $X_t$ and $Y_t$ both depend on $Z_{t-1}$.
This can also be inferred using the DIs in (\ref{cdif}); it is straightforward to show that 
\begin{align*}
&I(X\rightarrow Y||Z)=0,\ \ I(X\rightarrow Z||Y)=0,\ \ I(Y\rightarrow X||Z)=0,\ \ I(Y\rightarrow Z||X)=0,\\
&I(Z\rightarrow Y||X)\approx 0.18,\ \ \ \ I(Z\rightarrow X||Y)\approx 0.07.
\end{align*}
In this example, since the model is linear and the exogenous noises are Gaussian, the expression in \eqref{cdif} can be computed in terms of the covariance matrices, namely
\begin{equation}\label{simueq}
I(Z\rightarrow Y||X)=\frac{1}{2T}\sum_{t=1}^{T}\log\frac{|\mathbf{\Sigma}_{Y_{t-1} Y_{t}X_{t-1}}||\mathbf{\Sigma}_{Z_{t-1}Y_{t-1}X_{t-1}}|}{|\mathbf{\Sigma}_{Y_{t-1}X_{t-1}}||\mathbf{\Sigma}_{Z_{t-1}Y_{t-1}Y_{t}X_{t-1}}|},
\end{equation}
where $|\mathbf{\Sigma}_{Y_{t-1}Y_{t}X_{t-1}}|$ denotes the determinant of the covariance matrix of $(Y_{t-1},Y_{t},X_{t-1})$.
Note that none of the above DIs are pairwise as they have conditioned on the remaining time series. 
If we were to consider the pairwise causal relationships, for instance between $X$ and $Y$, we would have
 \begin{align}
 I(X\rightarrow Y)=\frac{1}{T}\sum_{t=1}^T\mathbb{E}\left[\log\frac{\PP(Y_t|Y^{t-1},X^{t-1})}{\PP(Y_t|Y^{t-1})}\right]\approx 0.0013>0.
 \end{align}
This wrongly implies that $X$ directly causes $Y$. 
\end{example}

\begin{example}\label{example2}
Consider the following multivariate GARCH model
\begin{small}
\begin{align}
& \begin{pmatrix}
  R_{1,t} \\
  R_{2,t} 
 \end{pmatrix}= 
  \begin{pmatrix}
  0.2 & 0.3 \\
0 & 0.2 
 \end{pmatrix}  \begin{pmatrix}
  R_{1,t-1} \\
  R_{2,t-1} 
 \end{pmatrix}+  \begin{pmatrix}
 \epsilon_{1,t} \\
  \epsilon_{2,t} 
 \end{pmatrix},\\ \notag
&\begin{pmatrix}
  \sigma^2_{1,t} \\
  \rho_t\\
  \sigma^2_{2,t}  
 \end{pmatrix}= 
 \begin{pmatrix}
0 \\
   0.3 \\
0.1
 \end{pmatrix}+
  \begin{pmatrix}
  0.2 & 0& 0.3 \\
0 & 0.2 & 0.7\\
0.1 & 0.4&0
 \end{pmatrix}  \begin{pmatrix}
  \epsilon^2_{1,t-1} \\
   \epsilon_{1,t-1} \epsilon_{2,t-1} \\
  \epsilon^2_{2,t-1} 
 \end{pmatrix} + \begin{pmatrix}
  0.3 & 0.5& 0 \\
0.1 & 0.2 & 0\\
0 & 0&0.4
 \end{pmatrix}  \begin{pmatrix}
  \sigma^2_{1,t-1} \\
  \rho_{t-1}\\
  \sigma^2_{2,t-1}
 \end{pmatrix},
\end{align}
\end{small}
where $\rho_t=\mathbb{E}[\epsilon_{1,t} \epsilon_{2,t}]$.
The corresponding DIG of this model is $R_1\leftrightarrow R_2$. This is because $R_2$ influences $R_1$ through the mean and variance and $R_1$ influences $R_2$ only through the variance.
\end{example}

\begin{example}\label{example20}
Consider a MA with order one and dimension three such that $\textbf{W}_0=\textbf{I}$, and
\begin{small}
\begin{align}\notag
\textbf{W}_1=
  \begin{pmatrix}
  0.3 & 0& 0.5 \\
0.1 & 0.2 & 0.5\\
0 & 0.4 & 0.1
 \end{pmatrix},\ \ 
 \textbf{W}^2_1=
  \begin{pmatrix}
  0.09 & 0.2& 0.2 \\
0.05 & 0.24 & 0.2\\
0.04 & 0.12 & 0.21
 \end{pmatrix},
\end{align}
\end{small}
Using the expression in (\ref{inar}), we have $\textbf{r}_t=\mathbf{\epsilon}_t+\sum_{k=1}^\infty(-1)^{k-1}\textbf{W}_1^k \textbf{r}_{t-k}$. Because, $\textbf{W}_1^2$ has no nonzero entry, the causal network (DIG) of this model is a complete graph.
\end{example}

\section{$k$-nearest Method}\label{sec:k-nearest}
We used the $k$-nearest method of \cite{sricharan2011k} to estimate $I(X;Y|Z)$.

Suppose that $N+M$ i.i.d. realizations $\{(x_1,y_1,z_1),..., (x_{N+M},y_{N+M},z_{N+M})\}$ are available from $\PP_{X,Y,Z}$.
The data sample is randomly divided into two subsets $\mathcal{S}_1$ and $\mathcal{S}_2$ of $N$ and $M$ points, respectively. 
In the first stage, an k-nearest density estimator $\widehat{\PP}_{X,Y,Z}$ at the $N$ points of $\mathcal{S}_1$ is estimated using the $M$ realizations of $\mathcal{S}_2$ as follows: 
Let $d(\textbf{x},\textbf{y})$ denote the Euclidean distance between two vectors $\textbf{x}$ and $\textbf{y}$ and $d_k(\textbf{x})$ denotes the Euclidean distance between a point $\textbf{x}$ and its k-th nearest neighbor among $\mathcal{S}_2$. 
The k-nearest region is $\mathcal{S}_k(\textbf{x}) := \{\textbf{y} :d(\textbf{x},\textbf{y})\leq d_k(\textbf{x}) \}$ and the volume of this region is $V_k(\textbf{x}):=\int_{\mathcal{S}_k(\textbf{x})}1dn$. 

The standard k-nearest density estimator of \cite{sricharan2011k} is defined as 
\begin{align}
&\widehat{\PP}_{X,Y,Z}(x,y,z):=\frac{k-1}{M}\frac{1}{V_k(x,y,z)}.
\end{align}
Similarly, we obtain k-nearest density estimators $\widehat{\PP}_{X,Z}, \widehat{\PP}_{Y,Z}$, and $\widehat{\PP}_{Z}$.
Subsequently, the $N$ samples of $\mathcal{S}_1$ are used to approximate the conditional mutual information:
\begin{small}
\begin{align}
&\widehat{I}(X;Y|Z):=\frac{1}{N}\sum_{i\in \mathcal{S}_1} \log \widehat{\PP}_{X,Y,Z}(x_i,y_i,z_i) + \log \widehat{\PP}_{Z}(z_i) -\log \widehat{\PP}_{X,Z}(x_i, z_i) -\log \widehat{\PP}_{Y,Z}(y_i, z_i).
\end{align}
\end{small}
For more details corresponding this estimator including its bias, variance, and confidence, please see the works by \cite{sricharan2011k} and \cite{loftsgaarden1965nonparametric}.

\section{Technical Proofs}\label{sec:proofs}

\subsection*{Proof of Proposition \ref{prop1}}\label{p:prop1}
Note that in this model, since the variance of each $e_{i,t}$ is $\mathcal{R}_{i}^{t-1}$-measurable, the only term that contains the effect of the other returns on the $i$-th return is $\mu_{i,t}$.
Hence, if (\ref{test1}) holds, then $\mu_{i,t}$ is independent of $R_j$. This implies the result. Moreover, when $\mu_{i,t}=\sum_{k=1}^p\sum_{l=1}^m a^{(k)}_{i,l}R_{l,t-k}$, using the result in \cite{acc2014}, we declare $R_j$ affects $R_i$ if and only if $\sum_{k=1}^p\sum_{l=1}^m |a^{(k)}_{i,l}|>0$, where $a^{(k)}_{i,l}$ denotes the $(j,l)$-th entry of matrix $\textbf{A}_k$ in (\ref{ar}).

\subsection*{Proof of Proposition \ref{prop2}}\label{p:prop2}

First, we need to show that if there is no arrow from $R_j$ to $R_i$ in the corresponding DIG, then (\ref{test1}) and (\ref{test2}) hold. This case is straight forward, since when $I(R_j\rightarrow R_i||\mathcal{R}_{-\{i,j\}})=0$, then for all $t$, $R_{i,t}$ is independent of $R_j^{t-1}$ given $\mathcal{R}^{t-1}_{-\{j\}}$. This concludes both (\ref{test1}) and (\ref{test2}).

To show the converse, we use the fact that in multivariate GARCH model, $\textbf{r}_t|\mathcal{R}^{t-1}$ is a multivariate Gaussian random process. Thus, if the corresponding mean and variance of $R_{i,t}$ do not contain any influence of $R_j^{t-1}$ given the rest of the network, then $R_{i,t}$ is independent of $R_j^{t-1}$ given $\mathcal{R}^{t-1}_{-\{j\}}$. This holds if both conditions in (\ref{test1}) and (\ref{test2}) that are corresponding to the mean and the variance, respectively, are satisfied.

\subsection*{Proof of Proposition \ref{prop3}}\label{p:prop3}
Suppose the conditions in Proposition \ref{prop3} hold. We show that $I(Y_j\rightarrow Y_i|| \mathcal{Y}_{-\{i,j\}})$=0.
\begin{small}
\begin{align}
&\PP(Y_{i,t}|\mathcal{Y}^{t-1})=\sum_{S_{i,t}} \PP(Y_{i,t}|\mathcal{Y}^{t-1}, S_{i,t})\PP(S_{i,t}|\mathcal{Y}^{t-1})\\
&=\sum_{S_{i,t}} \PP(Y_{i,t}|\mathcal{Y}_{-\{j\}}^{t-1}, S_{i,t})\PP(S_{i,t}|\mathcal{Y}^{t-1}).
\end{align}
\end{small}
The second equality holds because given $S_{i,t}$, $Y_{i,t}$ is a linear function of $(\mu_i(S_{i,t}), \textbf{y}_{t-p},..., \textbf{y}_{t-1})$ plus the error term $\epsilon_{i,t}$. From the first and second conditions in Proposition \ref{prop3}, we have the coefficients corresponding to $Y_{j}$ are zero and also the error term is independent of $Y_{j}$. Thus, $Y_{i,t}$ is independent of $Y_{j}^{t-1}$ given $\{\mathcal{Y}_{-\{j\}}^{t-1}, S_{i,t}\}$.
\\
If we show $\PP(S_{i,t}|\mathcal{Y}^{t-1})=\PP(S_{i,t}|\mathcal{Y}_{-\{j\}}^{t-1})$, using the above equality, we obtain that $\PP(Y_{i,t}|\mathcal{Y}^{t-1})=\PP(Y_{i,t}|\mathcal{Y}_{-\{j\}}^{t-1})$ for all $t$ which implies $I(Y_j\rightarrow Y_i|| \mathcal{Y}_{-\{i,j\}})=0$. To do so, we have
\begin{small}
\begin{align}
&\PP(S_{i,t}|\mathcal{Y}^{t-1})=\sum_{S_{i,t-1}}\PP(S_{i,t}|\mathcal{Y}^{t-1},S_{i,t-1})\PP(S_{i,t-1}|\mathcal{Y}^{t-1})\\
&=\sum_{S_{i,t-1}}\PP(S_{i,t}|\mathcal{Y}_{-\{j\}}^{t-1},S_{i,t-1})\PP(S_{i,t-1}|\mathcal{Y}^{t-1})\\
&=\sum_{S_{i,t-1}}\PP(S_{i,t}|\mathcal{Y}_{-\{j\}}^{t-1},S_{i,t-1})\PP(S_{i,t-1}|\mathcal{Y}_{-\{j\}}^{t-1})=\PP(S_{i,t}|\mathcal{Y}_{-\{j\}}^{t-1}).
\end{align}
\end{small}
The second equality is due to condition three and the fact that $\textbf{s}_t$ is conditionally independent of $\mathcal{Y}_{t-1}$ given $\textbf{s}_{t-1}$. The third equality is true because 
\begin{small}
\begin{align}
&\PP(S_{i,t-1}|\mathcal{Y}^{t-1})=\PP\left(S_{i,t-1}|\mathcal{Y}^{t-2},\mathcal{Y}_{-\{j\},t-1},Y_{j,t-1}\right)\\
&=\PP\left(S_{i,t-1}|\mathcal{Y}^{t-2},\mathcal{Y}_{-\{j\},t-1},F_j(\mathcal{Y}^{t-2},S_{j,t-1})\right)\\
&=\PP\left(S_{i,t-1}|\mathcal{Y}^{t-2},\mathcal{Y}_{-\{j\},t-1}\right),
\end{align}
\end{small}
where $F_j$s represent the functional dependency between time series given in (\ref{switch}), i.e., $Y_{m,t-1}:=F_m(\mathcal{Y}^{t-2},S_{m,t-1})$. 
The above equality holds due to the third condition. 
Note that the third condition implies that states $S_{k,t}$ and $S_{k',t'}$ are independent for any pairs $(k,k')$ and $(t,t')$. Same line of reasoning implies
\begin{small}
 \begin{align}
 &\PP\left(S_{i,t-1}|\mathcal{Y}^{t-2},\mathcal{Y}_{-\{j\},t-1}\right)=\PP\left(S_{i,t-1}|\mathcal{Y}^{t-3},\mathcal{Y}_{-\{j\},t-2}^{t-1}, F_j(\mathcal{Y}^{t-3},S_{j,t-2})\right)\\
  &=\PP\left(S_{i,t-1}|\mathcal{Y}^{t-3},\mathcal{Y}_{-\{j\},t-2}^{t-1}\right)=\cdots=\PP\left(S_{i,t-1}|\mathcal{Y}_{-\{j\}}^{t-1}\right).
 \end{align}
 \end{small}

Recall that $\mathcal{Y}_{\mathcal{K}}^t$ denotes the time series with index set $\mathcal{K}$ up to time $t$.

\subsection*{Proof of Proposition \ref{pro4}}\label{p:pro4}
Assume that there exists a $\tau$, such that $J_\tau>0$, then the $\tau$-th quantiles of $Y_t$ conditional on $Y^{t-1}_{t-p}, X^{t-1}_{t-q}$ and $Y^{t-1}_{t-p}$ are not equal, i.e., 
\begin{align}
Q^Y_{\tau}(Y^{t-1}_{t-p}, X^{t-1}_{t-q})\neq Q^Y_\tau(Y^{t-1}_{t-p}),    
\end{align}
that is 
\begin{align}
\inf\{y : F(Y_t=y|Y^{t-1}_{t-p}, X^{t-1}_{t-q})\geq \tau\}:= y_1 \neq y_2:=\inf\{y' : F(Y_t=y'|Y^{t-1}_{t-p})\geq \tau\}.
\end{align}
Without loss of generality, we assume $y_1<y_2$.
Since $F$ is continuous then there exists a non-zero measure set around $y_1$ such that 
\begin{align}
\PP(Y_t\leq y |Y^{t-1}_{t-p}, X^{t-1}_{t-q})\geq \tau  > \PP(Y_t\leq y|Y^{t-1}_{t-p}).
\end{align}
Therefore, the ratio in (\ref{cdif}) will differ from one which leads to $I(X\rightarrow Y)>0$.

\section{Additional Experimental Results}\label{sec:app:ex}

\subsection{List of stocks in our experiment} Next table represents the list of 124 stocks from 6 Global Industry Classification Standard (GICS) sectors:
Banks, DivFin, Insurance, REITs, FinTech, and Crypto.

\begin{scriptsize} 
\begin{longtable}{@{}p{0.3\textwidth}p{0.3\textwidth}p{0.3\textwidth}@{}}

\caption{ List of Studied Assets Across Sub-sectors} \\ 
\toprule
\textbf{Banks - 4010} & \textbf{Diversified Financials - 4020} & \textbf{Insurance - 4030}  \\
\midrule
\endfirsthead
\multicolumn{3}{c}%
{{\bfseries Table \thetable\ Continued from previous page}} \\
\toprule
\textbf{Banks - 4010} & \textbf{Diversified Financials - 4020} & \textbf{Insurance - 4030}\\

\midrule
\endhead
\bottomrule
\multicolumn{3}{r}{{Continued on next page}} \\
\endfoot
\bottomrule
\endlastfoot
Bank of America Corporation & Ameriprise Financial, Inc. & Arch Capital Group Ltd. \\
Citigroup Inc. & American Express Company & Aflac Incorporated \\
Citizens Financial Group, Inc. & Franklin Resources, Inc. & American International Group, Inc. \\
Comerica Incorporated & BNY Mellon Corporation & Assurant, Inc. \\
Fifth Third Bancorp & BlackRock, Inc. & Arthur J. Gallagher \& Co. \\
First Republic Bank & Berkshire Hathaway Inc. & Allstate Corporation \\
Huntington Bancshares Incorporated & Cboe Global Markets, Inc. & Aon plc \\
JPMorgan Chase \& Co. & CME Group Inc. & Chubb Limited \\
KeyCorp & Capital One Financial Corporation & Cincinnati Financial Corporation \\
M\&T Bank Corporation & Discover Financial Services & Globe Life Inc. \\
The PNC Financial Services Group & The Goldman Sachs Group, Inc. & Hartford Financial Services Group \\
Regions Financial Corporation & Intercontinental Exchange, Inc. & Loews Corporation \\
SVB Financial Group & Invesco Ltd. & Lincoln National Corporation \\
Truist Financial Corporation & Moody's Corporation & MetLife, Inc. \\
U.S. Bancorp & MarketAxess Holdings Inc. & Marsh \& McLennan Companies, Inc. \\
Wells Fargo \& Company & Morgan Stanley & Principal Financial Group, Inc. \\
Zions Bancorporation & MSCI Inc. & Progressive Corporation \\
 & Nasdaq, Inc. & Prudential Financial, Inc. \\
 & Northern Trust Corporation & Employers Holdings Inc. \\
 & Raymond James Financial, Inc. & Travelers Companies, Inc. \\
 & The Charles Schwab Corporation & Unum Group \\
 & S\&P Global Inc. & W. R. Berkley Corporation \\
 & State Street Corporation & \\
 & Synchrony Financial & \\
 & T. Rowe Price Group Inc. & \\
\midrule
\textbf{REITs - 6010} & \textbf{Fintech} & \textbf{Cryptocurrencies} \\
\midrule
Aimco & ACI Worldwide, Inc. & Cardano to US Dollar \\
American Tower Corporation & Broadridge Financial Solutions, Inc. & Bitcoin Cash to US Dollar \\
Alexandria Real Estate Equities, Inc. & CoStar Group, Inc. & Binance Coin to US Dollar \\
AvalonBay Communities, Inc. & Euronet Worldwide, Inc. & Bitcoin to US Dollar \\
Boston Properties, Inc. & Envestnet, Inc. & Dogecoin to US Dollar \\
Crown Castle International Corp. & EVERTEC, Inc. & Ethereum to US Dollar \\
CareTrust REIT, Inc. & Fair Isaac Corporation & Filecoin to US Dollar \\
Digital Realty Trust, Inc. & Fidelity National Information Services & Litecoin to US Dollar \\
Equinix, Inc. & FleetCor Technologies, Inc. & TRON to US Dollar \\
Equity Residential & Global Payments Inc. & Monero to US Dollar \\
Essex Property Trust, Inc. & Jack Henry \& Associates, Inc. & Ripple to US Dollar \\
Extra Space Storage Inc. & Mastercard Incorporated & \\
Federal Realty Investment Trust & SS\&C Technologies Holdings, Inc. & \\
Host Hotels \& Resorts, Inc. & Thomson Reuters Corporation & \\
Iron Mountain Incorporated & Visa Inc. & \\
Kimco Realty Corporation & Verisk Analytics, Inc. & \\
Mid-America Apt. Communities & WEX Inc. & \\
Realty Income Corporation & The Western Union Company & \\
Healthpeak Properties, Inc. & & \\
Prologis, Inc. & & \\
Public Storage & & \\
Regency Centers Corporation & & \\
SBA Communications Corporation & & \\
Sabra Health Care REIT, Inc. & & \\
SL Green Realty Corp. & & \\
Simon Property Group, Inc. & & \\
UDR, Inc. & & \\
Vornado Realty Trust & & \\
Ventas, Inc. & & \\
Welltower Inc. & & \\
Weyerhaeuser Company & & \\
\end{longtable}

\end{scriptsize}

\subsection{Non-linearity Test}\label{sec:app-nonlinea}
We applied a non-linearity test on the data to determine whether the underlying structure within the recorded data is linear or nonlinear. 
The non-linearity test applied is based on non-linear principle component analysis (PCA) of \cite{kruger2008developments}. This test  works as follows, the range of recorded data is divided into smaller disjunct regions; and accuracy bounds are determined for the sum of the discarded eigenvalues of each region. If this sum is within the accuracy bounds for each region, the process is assumed to be linear. Conversely, if at least one of these sums is outside, the process is assumed to be nonlinear.

More precisely, the second step in this test requires computation of the correlation matrix for each of the disjunct regions. Since the elements of this matrix are obtained using a finite dataset, applying $t$-distribution and $\chi^2$-distribution establish confidence bounds for both estimated mean and variance, respectively. 
Subsequently, these confidence bounds can be utilized to determine thresholds for each element in the correlation matrix. 
Using these thresholds, the test calculates maximum and minimum eigenvalues relating to the discarded score variables, which in turn allows the determination of both a minimum and a maximum accuracy bound for the variance of the prediction error of the PCA model. This is because the variance of the prediction error is equal to the sum of the discarded eigenvalues.
 If this sum lies inside the accuracy bounds for each disjunct region, a linear PCA model is then appropriate over the entire region. Alternatively, if at least one of these sums is outside the accuracy bounds, the error variance of the PCA model residuals then differs significantly for this disjunct region and hence, a nonlinear model is required. 

 \begin{figure}[H]
\centering
\subfigure[Region 1. ]{\includegraphics[width=4cm,height=3cm]{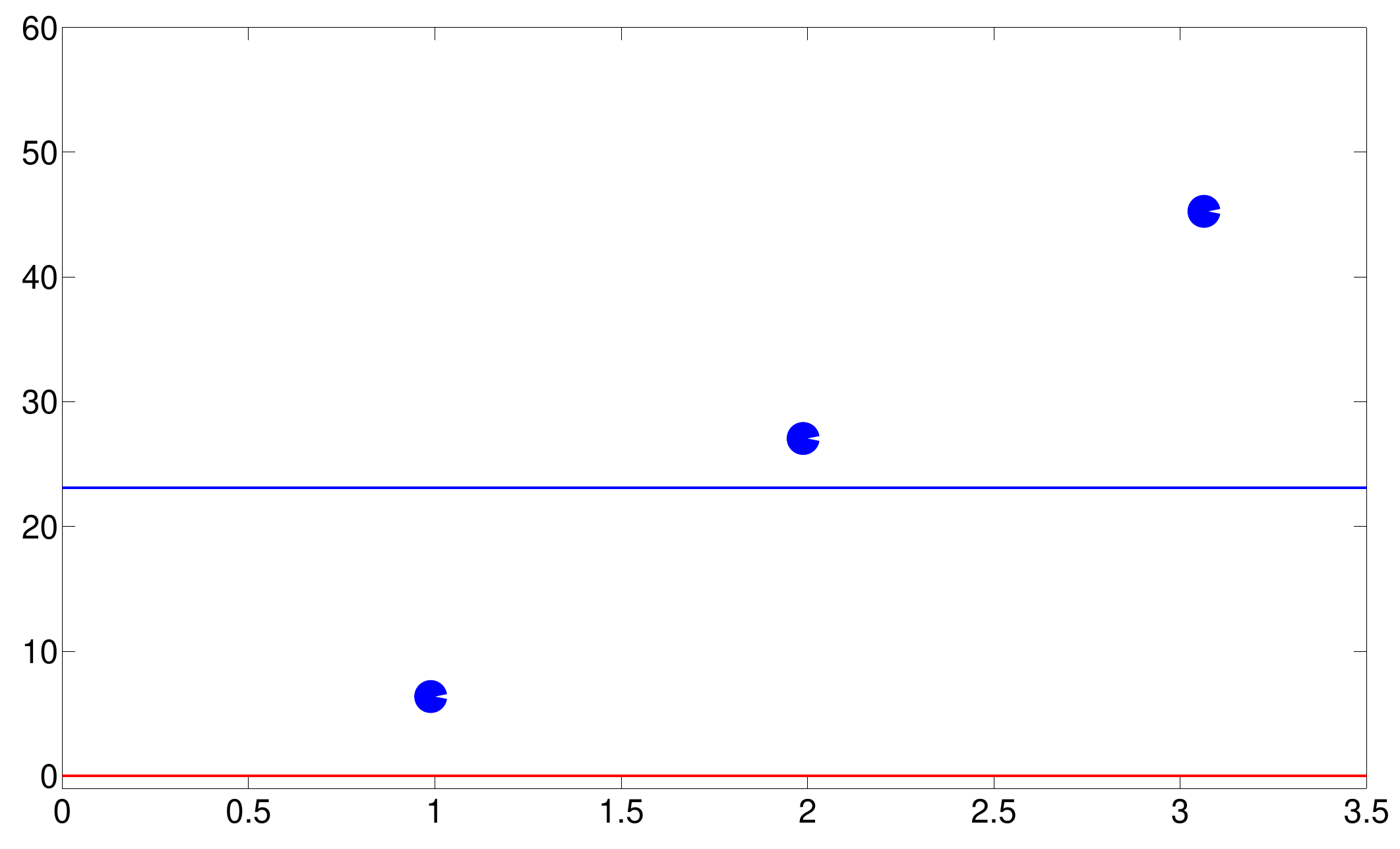}}
\subfigure[Region 2.]{\includegraphics[width=4cm,height=3cm]{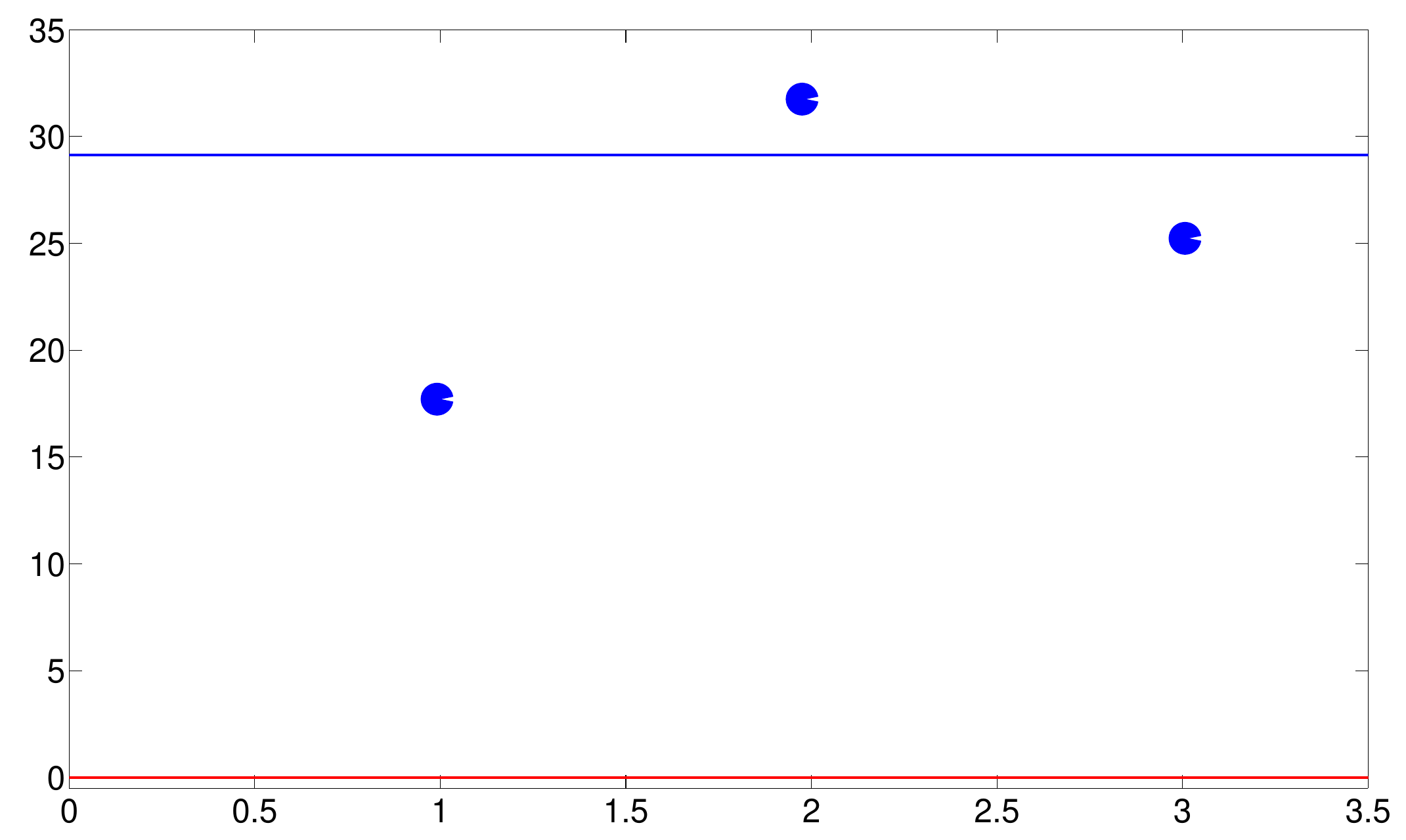}}
\subfigure[Region 3.]{\includegraphics[width=4cm,height=3cm]{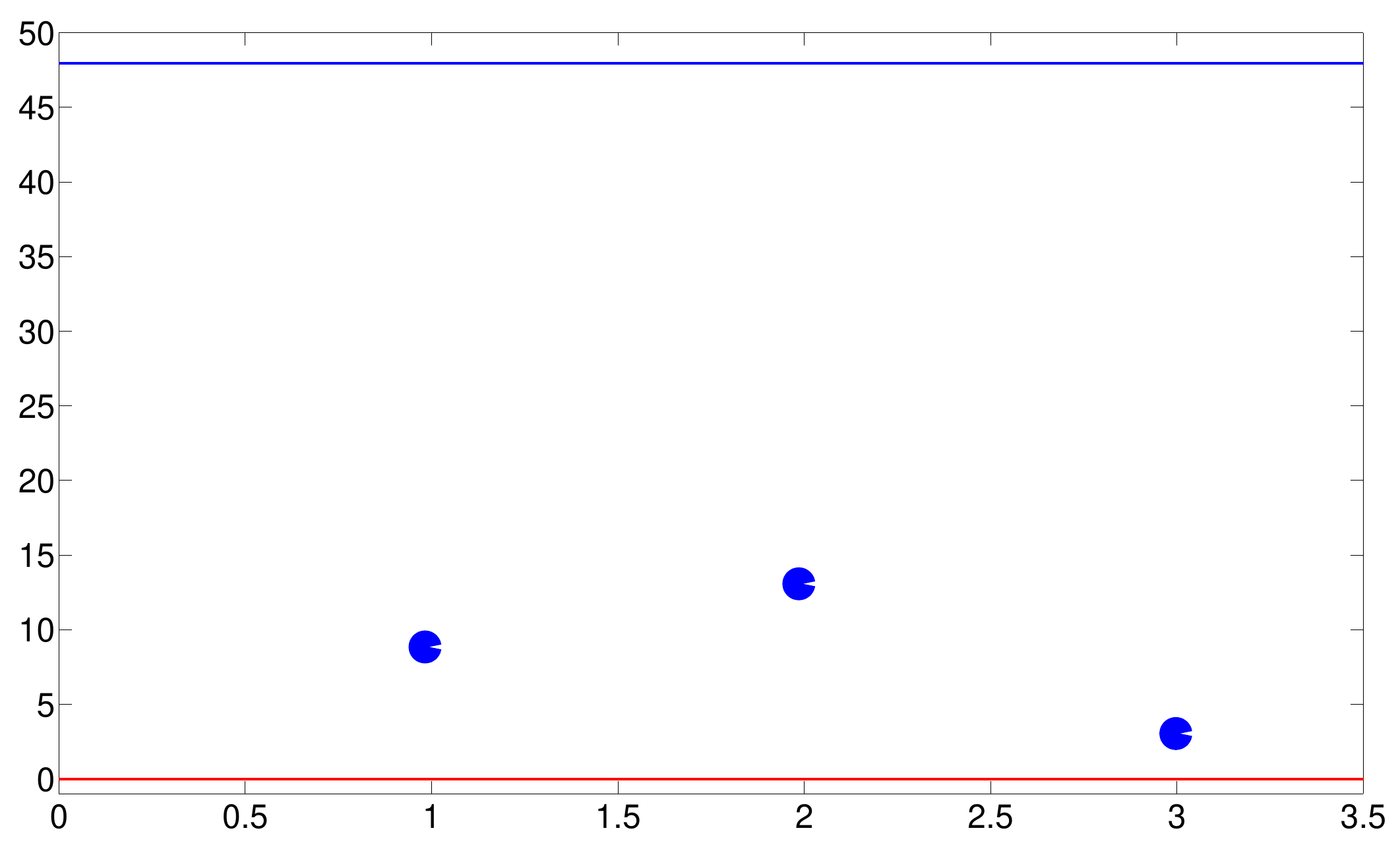}}
       \caption{Benchmarking of the residual variances against accuracy bounds of each disjunct region.}
\label{nontest}
 \end{figure}  
We divided the operating region into 3 disjunct regions. The accuracy bounds for each disjuct region and also sum of the discarded eigenvalues were computed. These bounds were based on thresholds for each element of the correlation matrix corresponding to confidence level of 95\%.
The processes were normalized with respect to the mean and variance of the regions for which the accuracy bounds were computed.
Figure \ref{nontest} shows the accuracy bounds and the sum of the discarded eigenvalues. As figures \ref{nontest}-(a) and \ref{nontest}-(b) illustrate, the recorded data shows nonlinearity between series.


 \subsection{Complete Results of Time-varying Experiment}\label{sec:app-emp0}

 \begin{figure}[H]
\centering
\subfigure[$X_0\rightarrow X_1$]{\includegraphics[width=4cm,height=3.6cm]{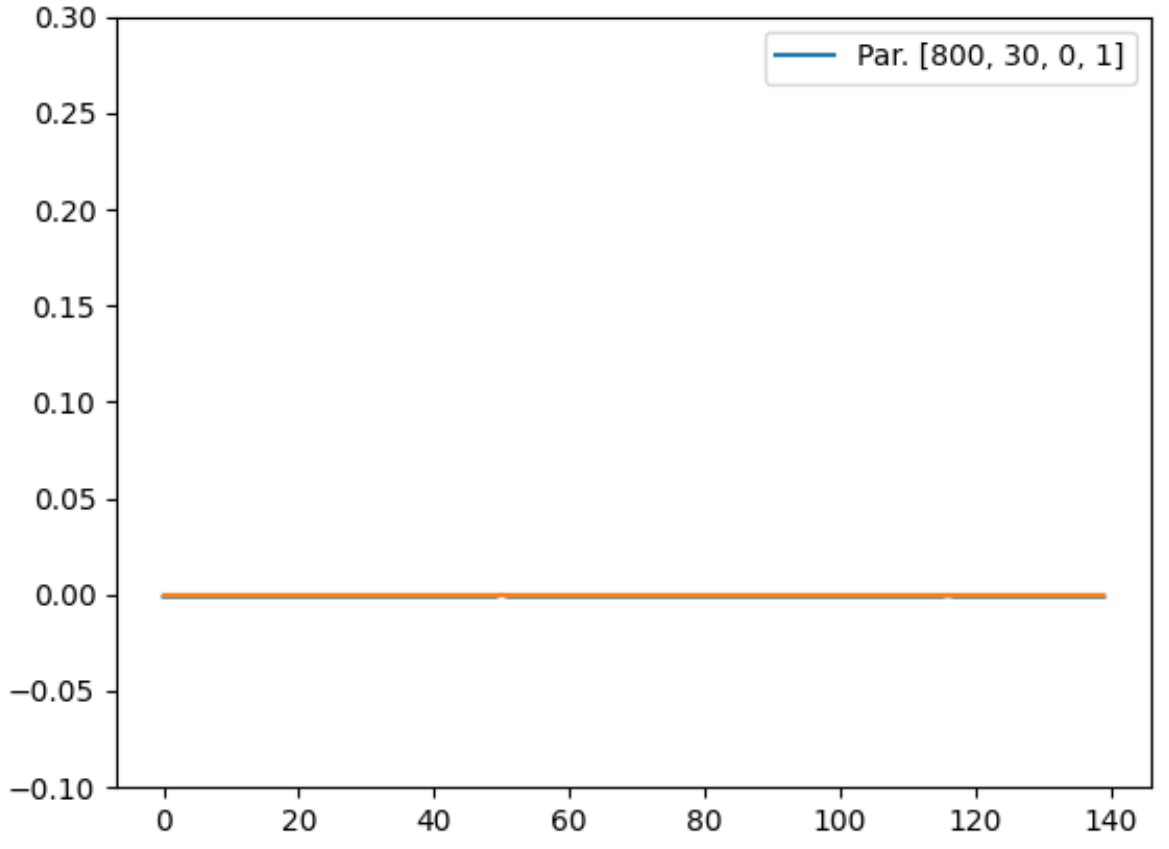}}
\subfigure[$X_0\rightarrow X_2$]{\includegraphics[width=4cm,height=3.6cm]{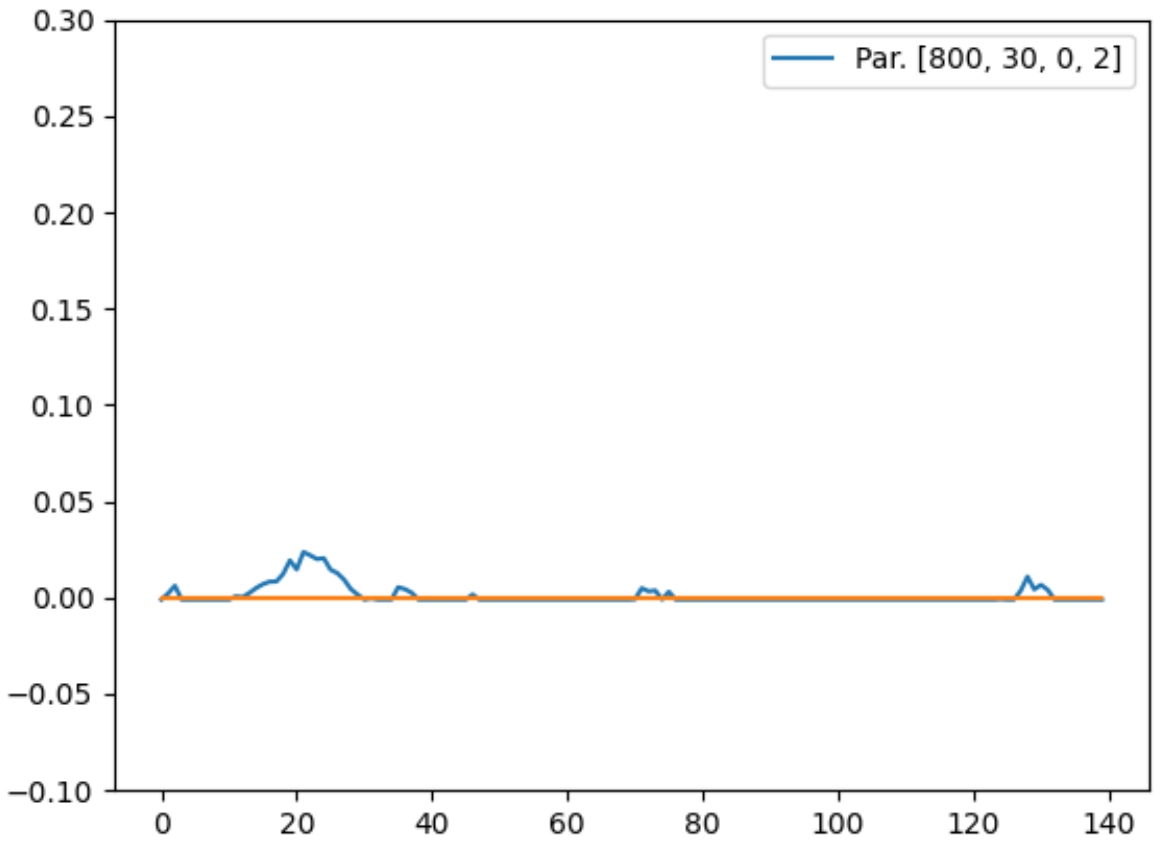}}
\subfigure[$X_0\rightarrow X_3$]{\includegraphics[width=4cm,height=3.6cm]{0-3.pdf}}
\subfigure[$X_0\rightarrow X_4$]{\includegraphics[width=4cm,height=3.6cm]{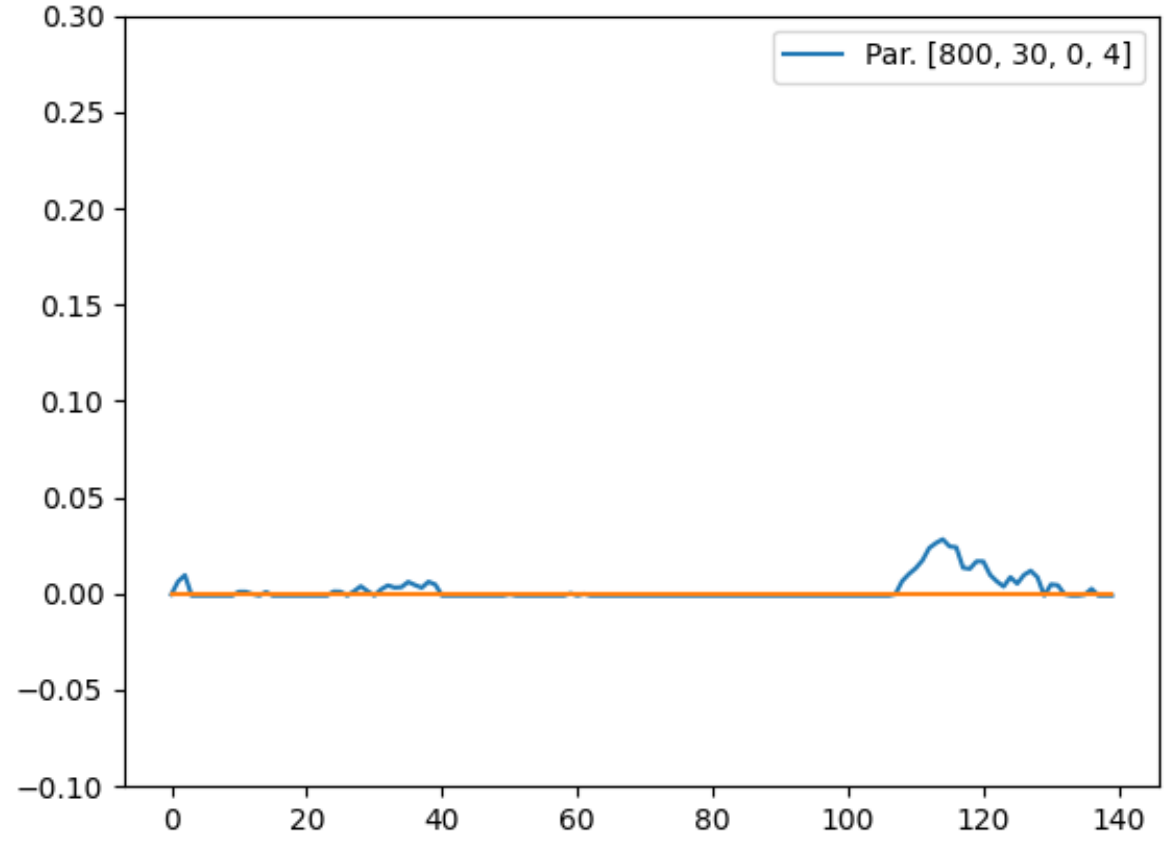}}
\subfigure[$X_1\rightarrow X_0$]{\includegraphics[width=4cm,height=3.6cm]{1-0.pdf}}
\subfigure[$X_1\rightarrow X_2$]{\includegraphics[width=4cm,height=3.6cm]{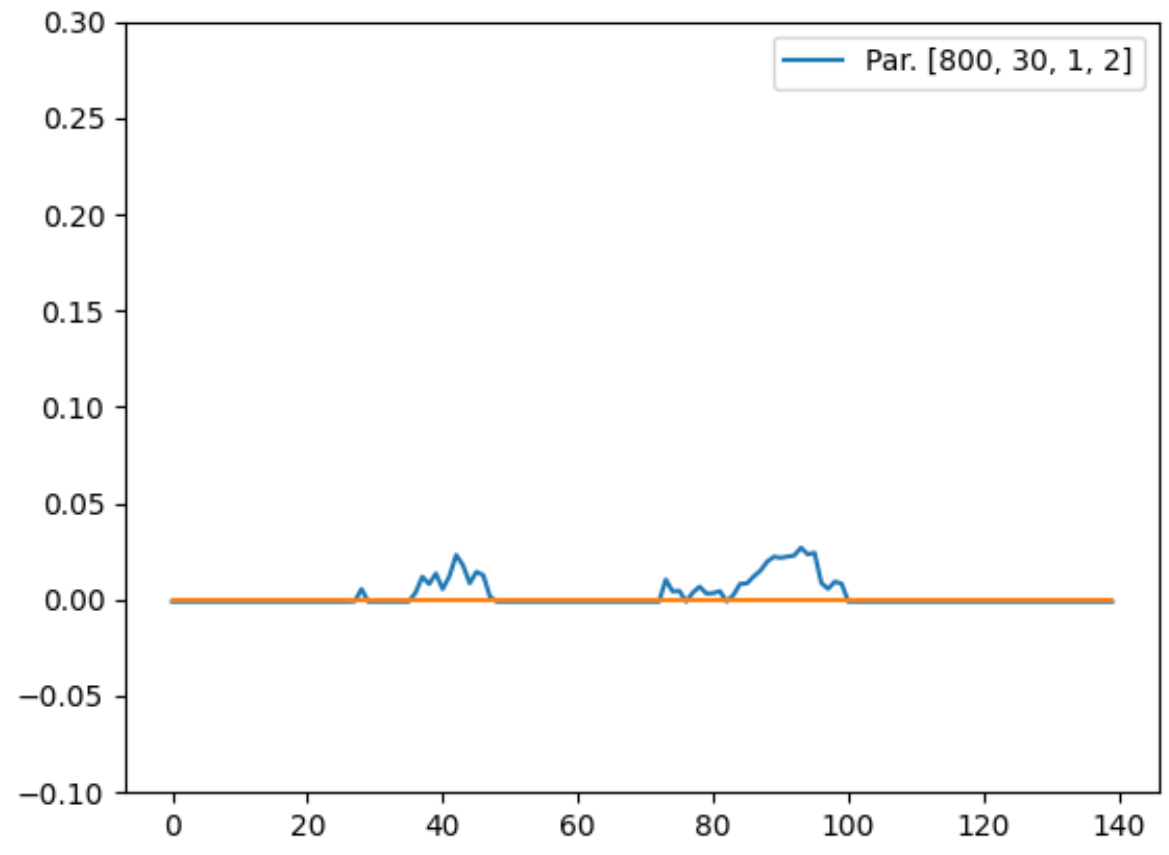}}
\subfigure[$X_1\rightarrow X_3$]{\includegraphics[width=4cm,height=3.6cm]{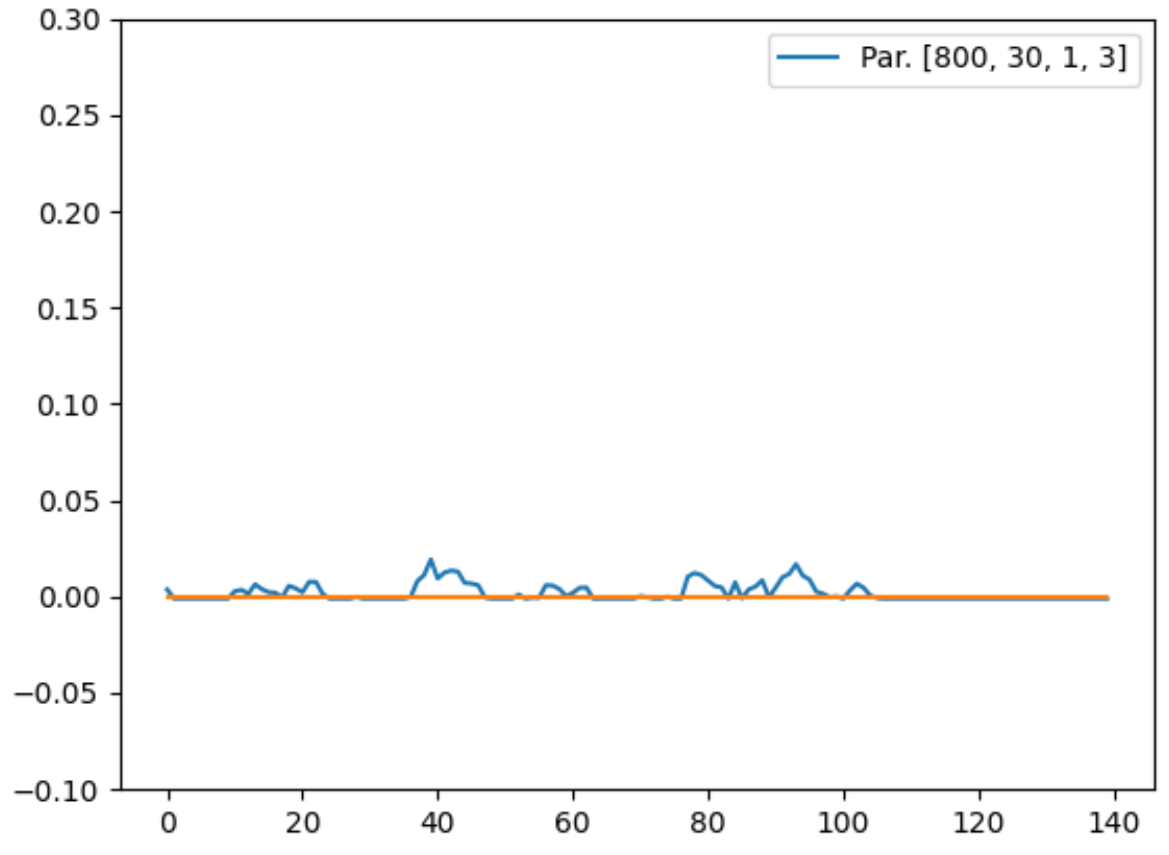}}
\subfigure[$X_1\rightarrow X_4$]{\includegraphics[width=4cm,height=3.6cm]{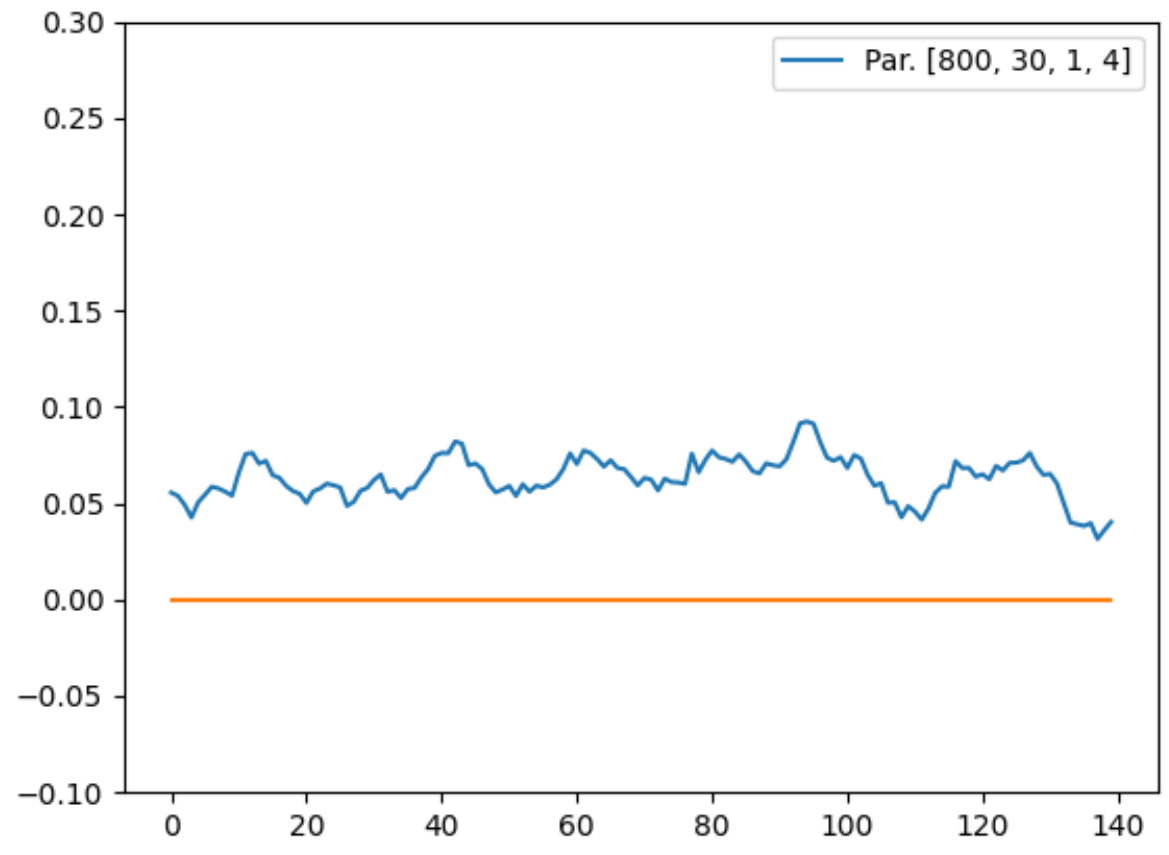}}
\subfigure[$X_2\rightarrow X_0$]{\includegraphics[width=4cm,height=3.6cm]{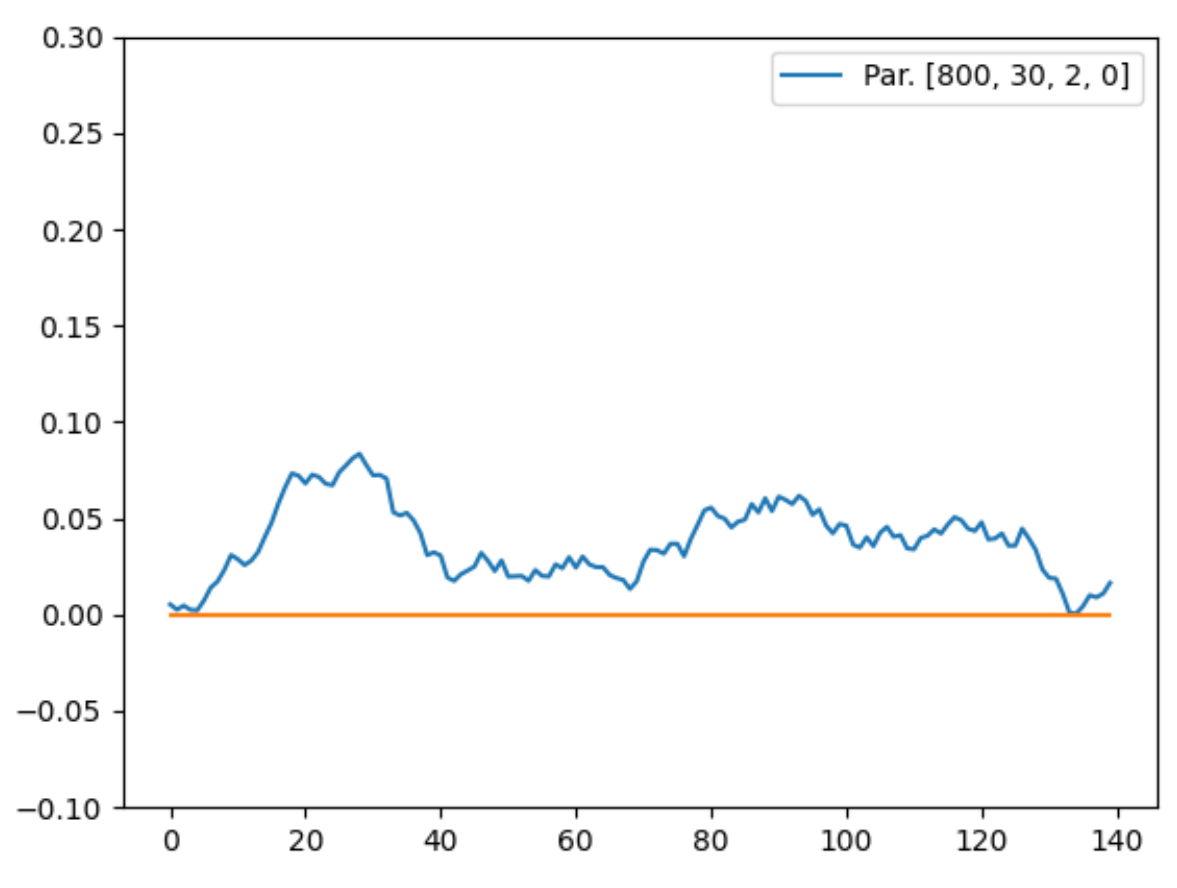}}
\caption{Estimated DIs of the time-varying experiment.}
\end{figure}

 \begin{figure}[H]
\centering
\subfigure[$X_2\rightarrow X_1$]{\includegraphics[width=4cm,height=3.6cm]{2-1.pdf}}
\subfigure[$X_2\rightarrow X_3$]{\includegraphics[width=4cm,height=3.6cm]{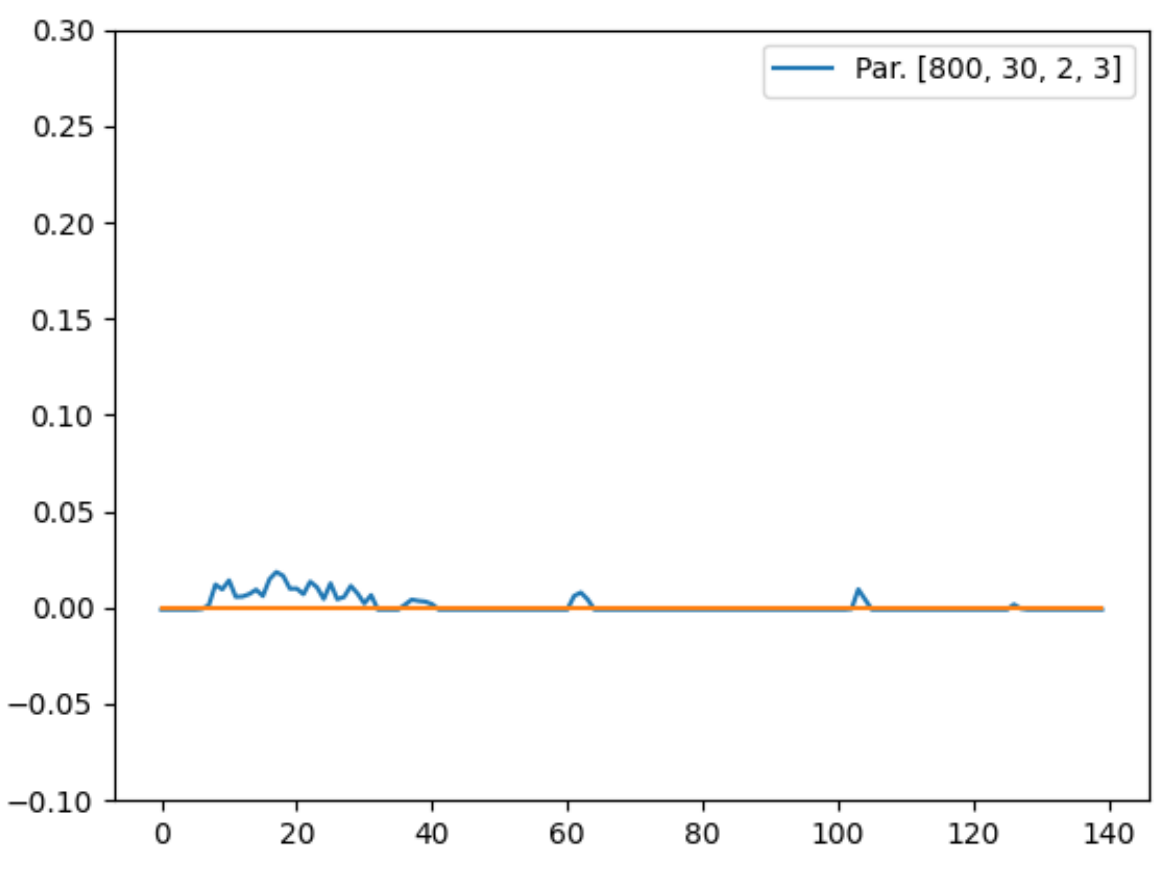}}
\subfigure[$X_2\rightarrow X_4$]{\includegraphics[width=4cm,height=3.6cm]{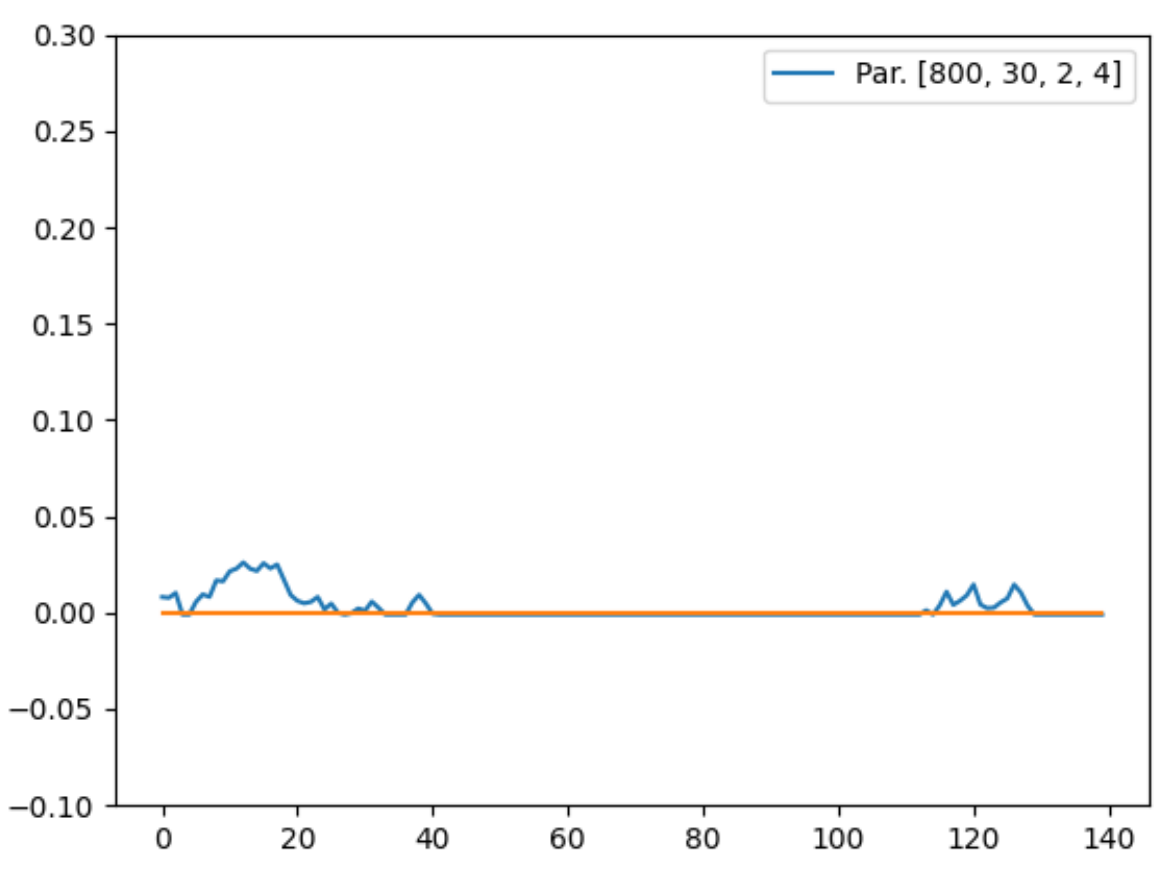}}
\subfigure[$X_3\rightarrow X_0$]{\includegraphics[width=4cm,height=3.6cm]{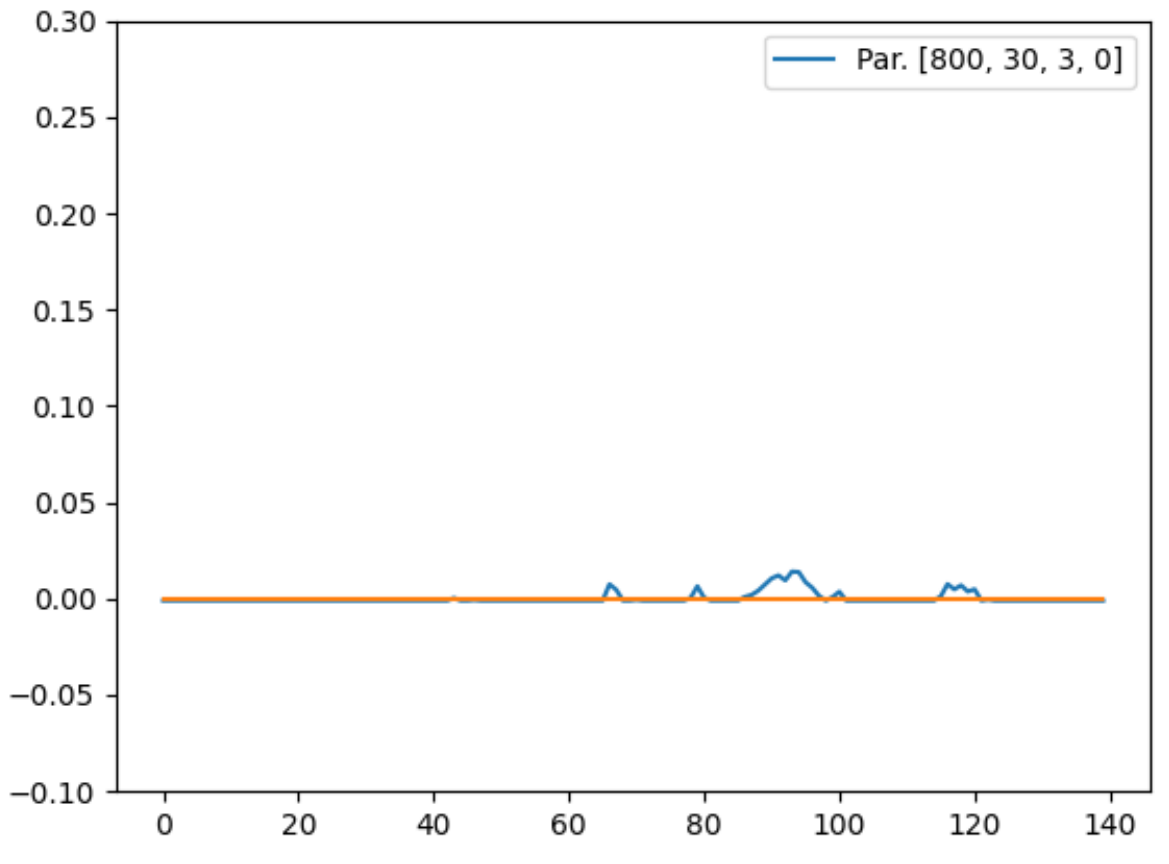}}
\subfigure[$X_3\rightarrow X_1$]{\includegraphics[width=4cm,height=3.6cm]{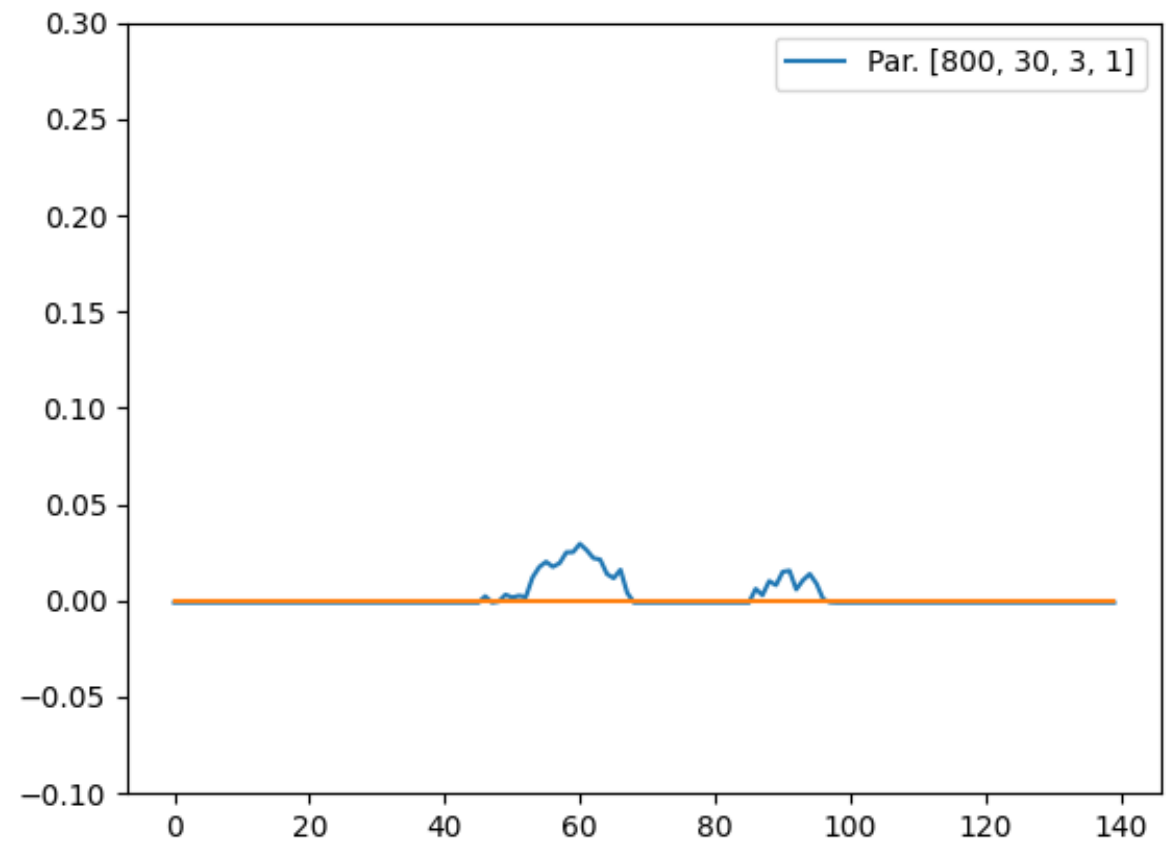}}
\subfigure[$X_3\rightarrow X_2$]{\includegraphics[width=4cm,height=3.6cm]{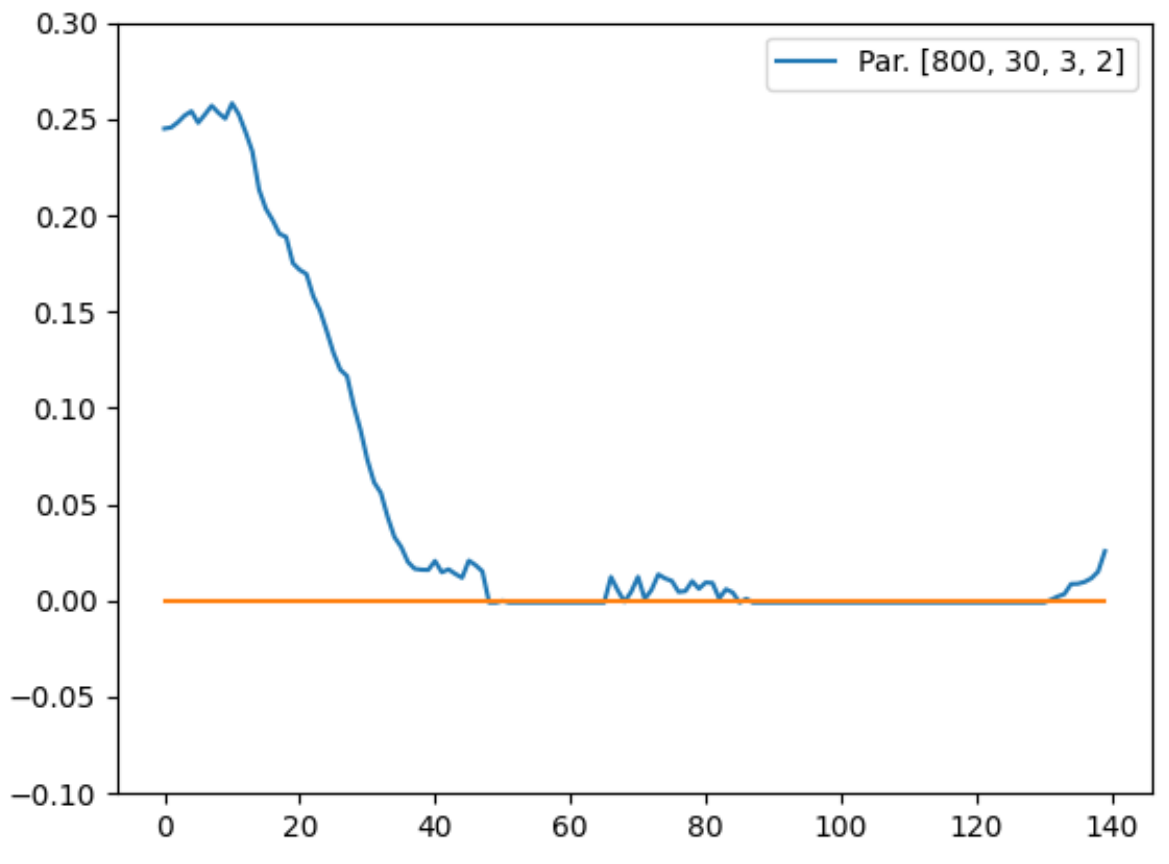}}
\subfigure[$X_3\rightarrow X_4$]{\includegraphics[width=4cm,height=3.6cm]{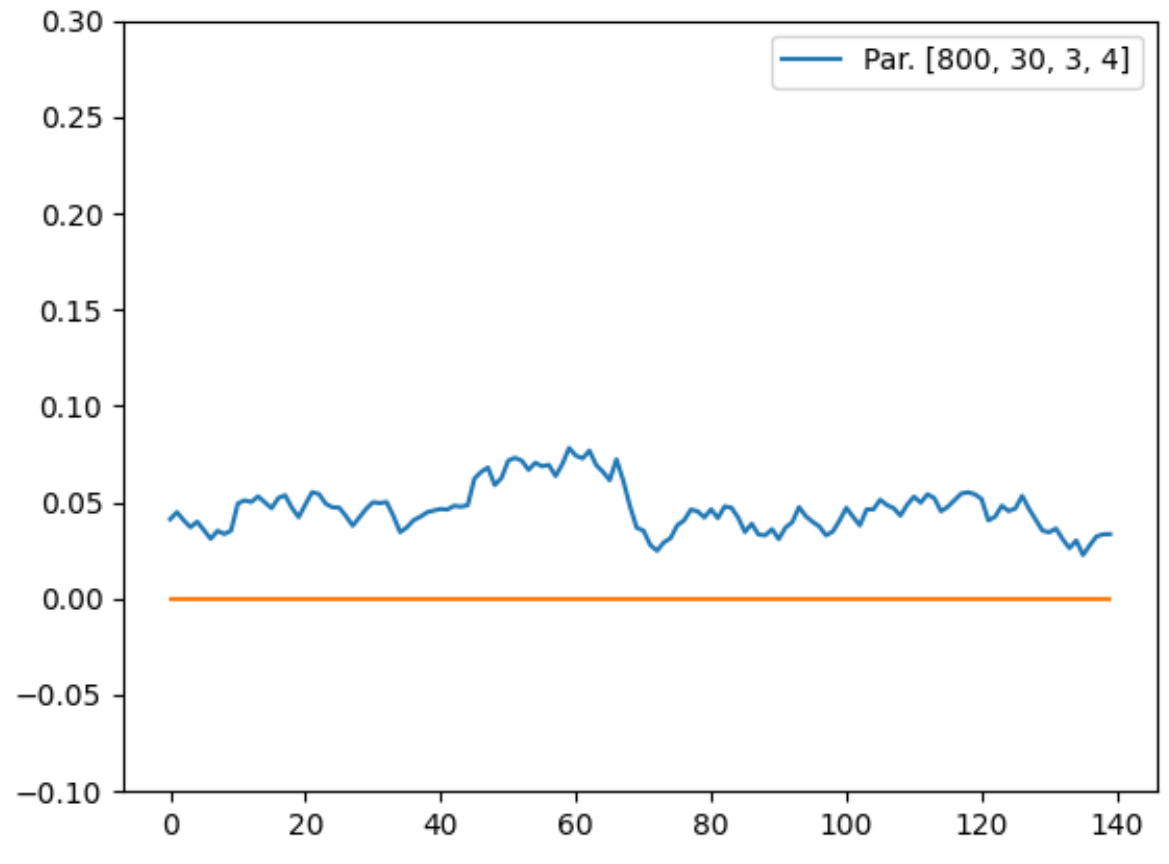}}
\subfigure[$X_4\rightarrow X_0$]{\includegraphics[width=4cm,height=3.6cm]{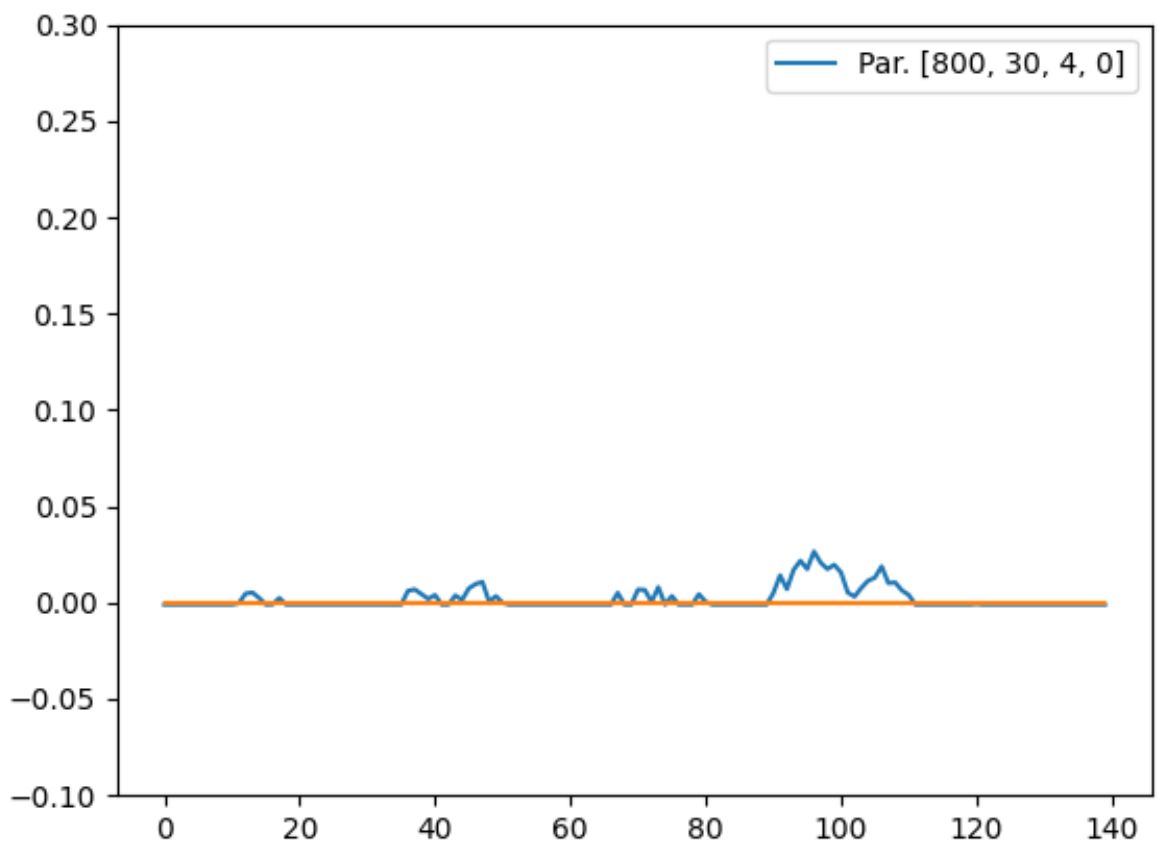}}
\subfigure[$X_4\rightarrow X_1$]{\includegraphics[width=4cm,height=3.6cm]{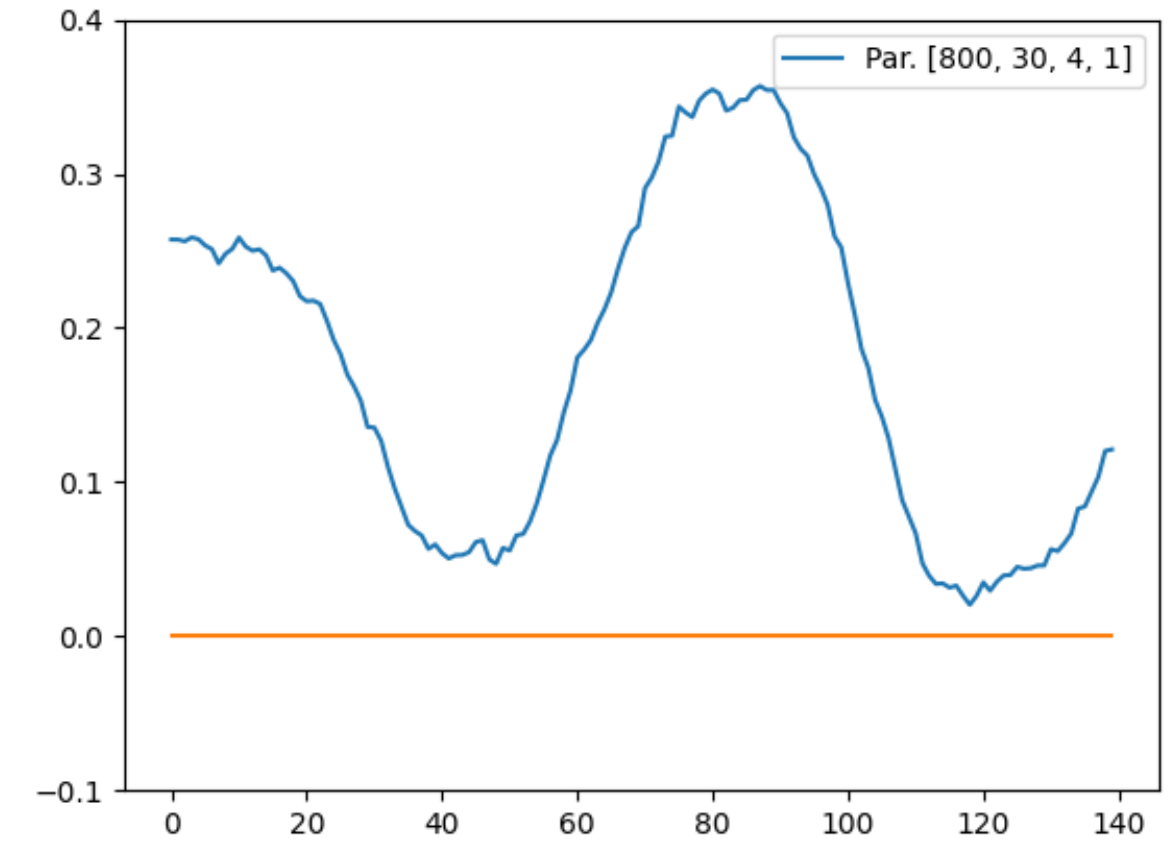}}
\subfigure[$X_4\rightarrow X_2$]{\includegraphics[width=4cm,height=3.6cm]{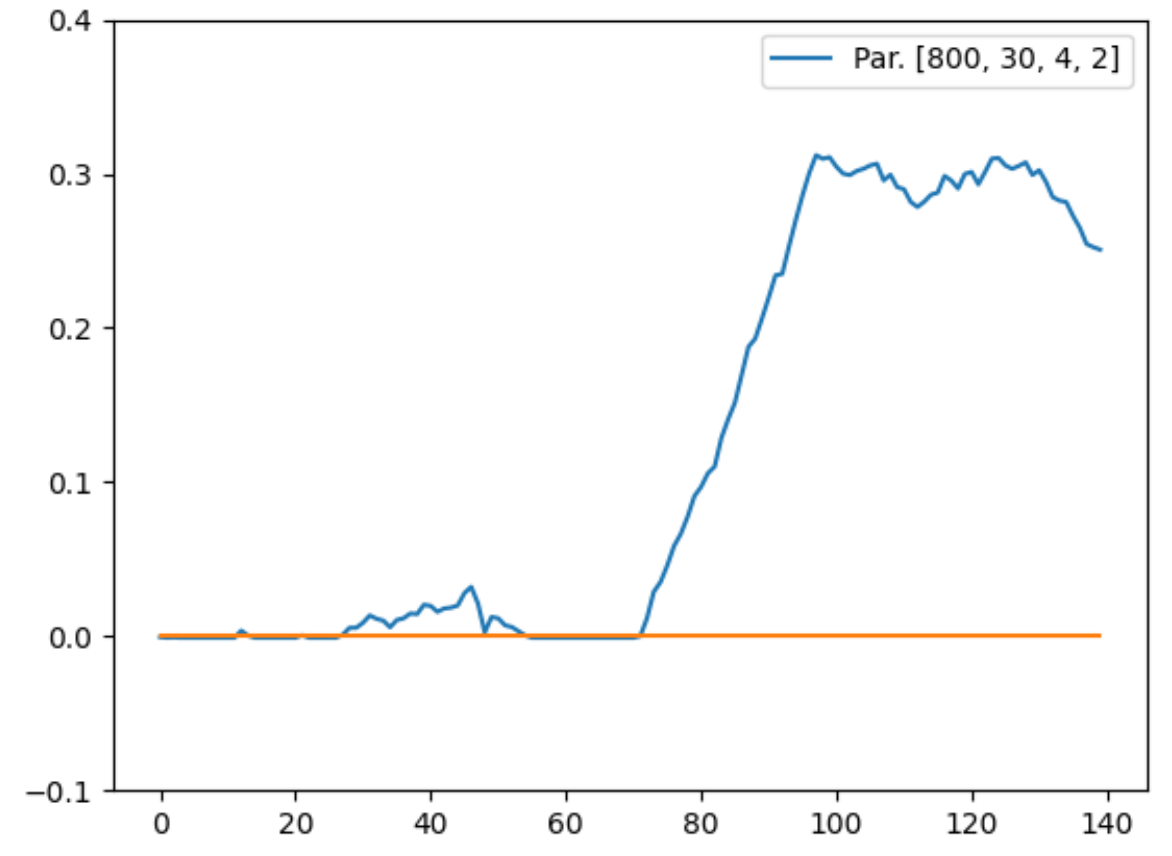}}
\subfigure[$X_4\rightarrow X_3$]{\includegraphics[width=4cm,height=3.6cm]{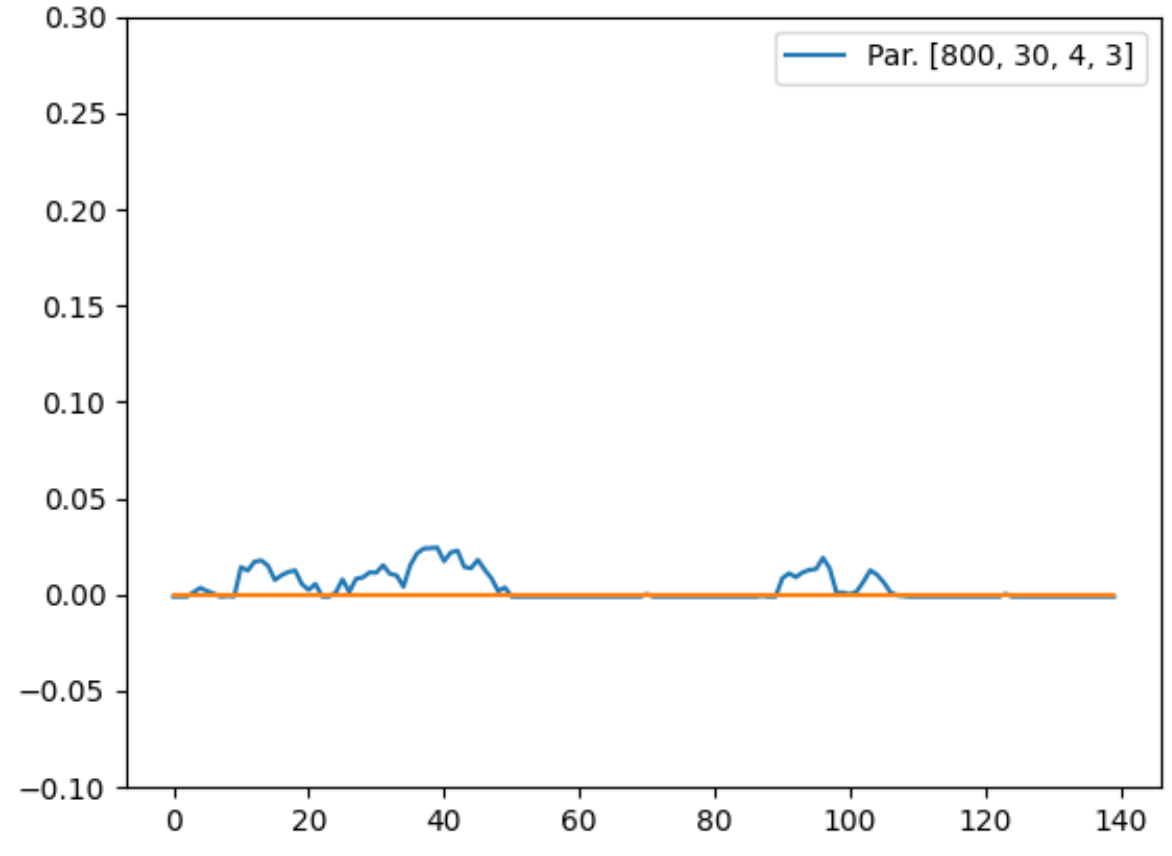}}
       \caption{Estimated DIs of the time-varying experiment.}
 \end{figure}

\subsection{Complete Results of Estimated DIs Among Industries}\label{sec:ap-emp1}
Remind that nodes with label $[0,1,2,3,4,5]$ denote industries [Banks, Diversified Financial, Insurance, Real Estate Investment, Financial Technology, Crypto], respectively.

\begin{figure}[H]
\centering
\subfigure{\includegraphics[scale=.2]{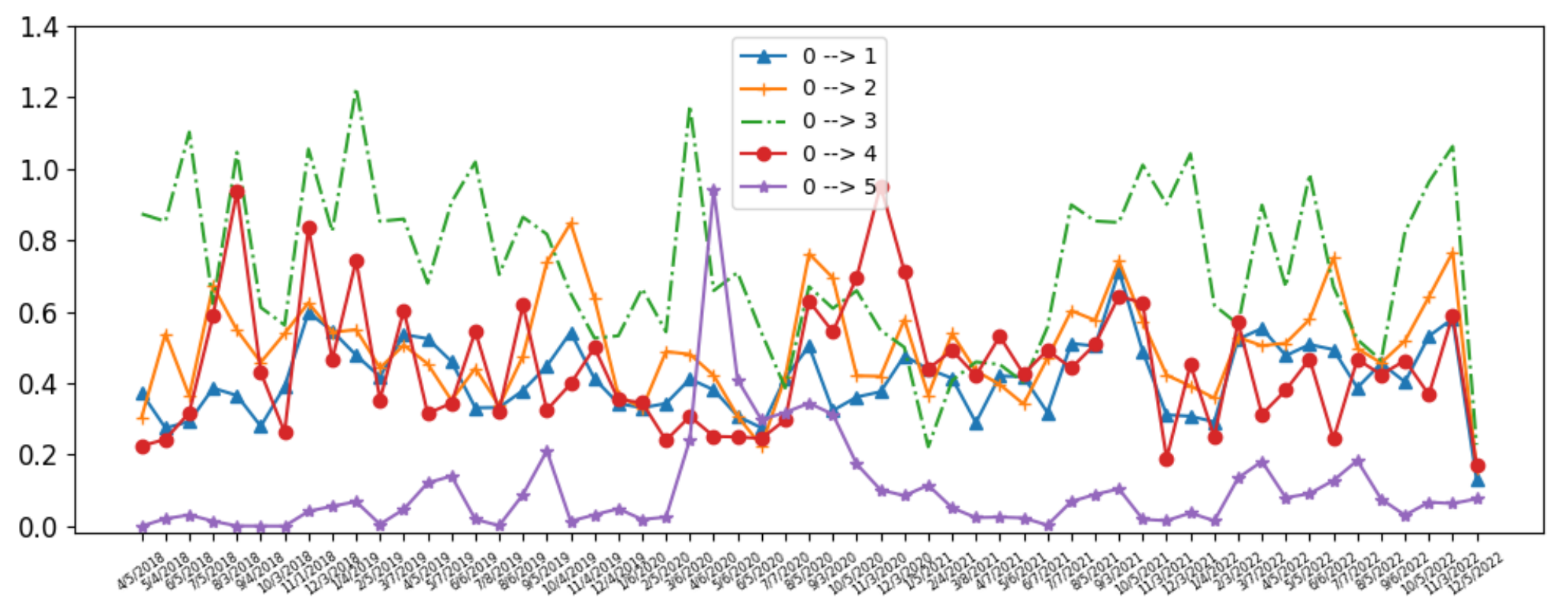}}
\subfigure{\includegraphics[scale=.2]{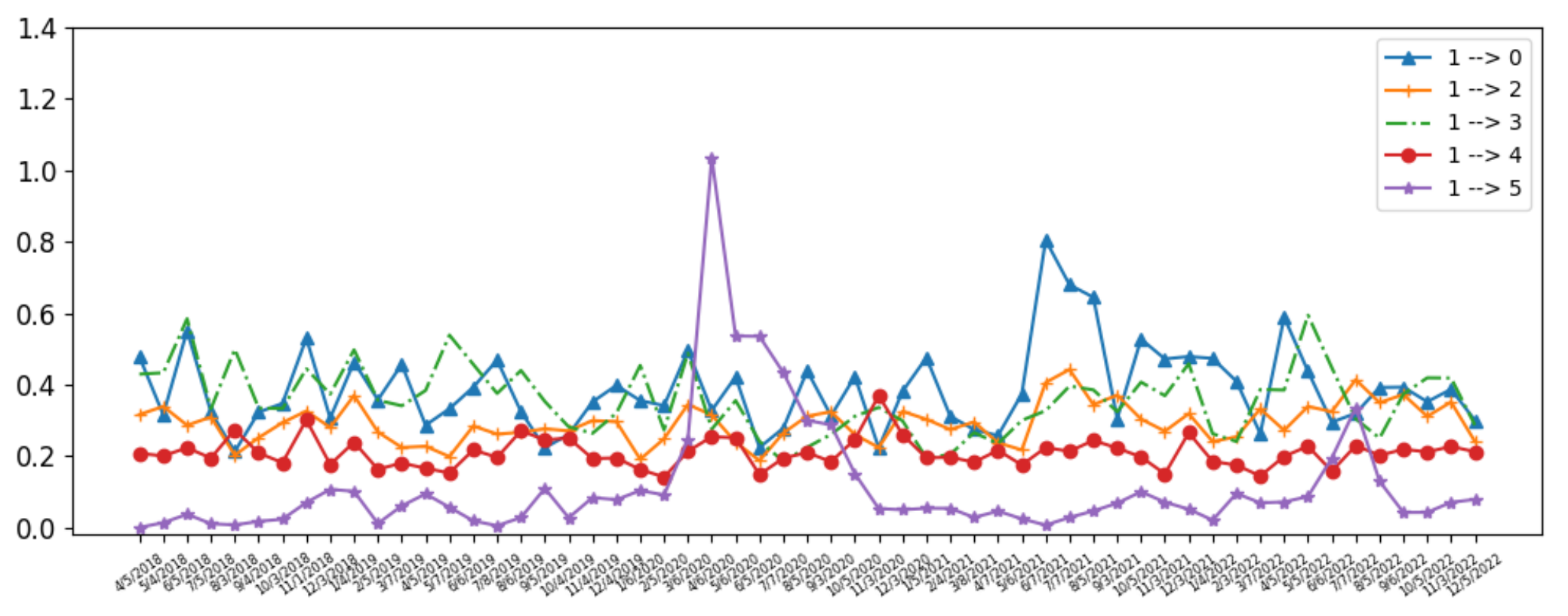}}
\subfigure{\includegraphics[scale=.2]{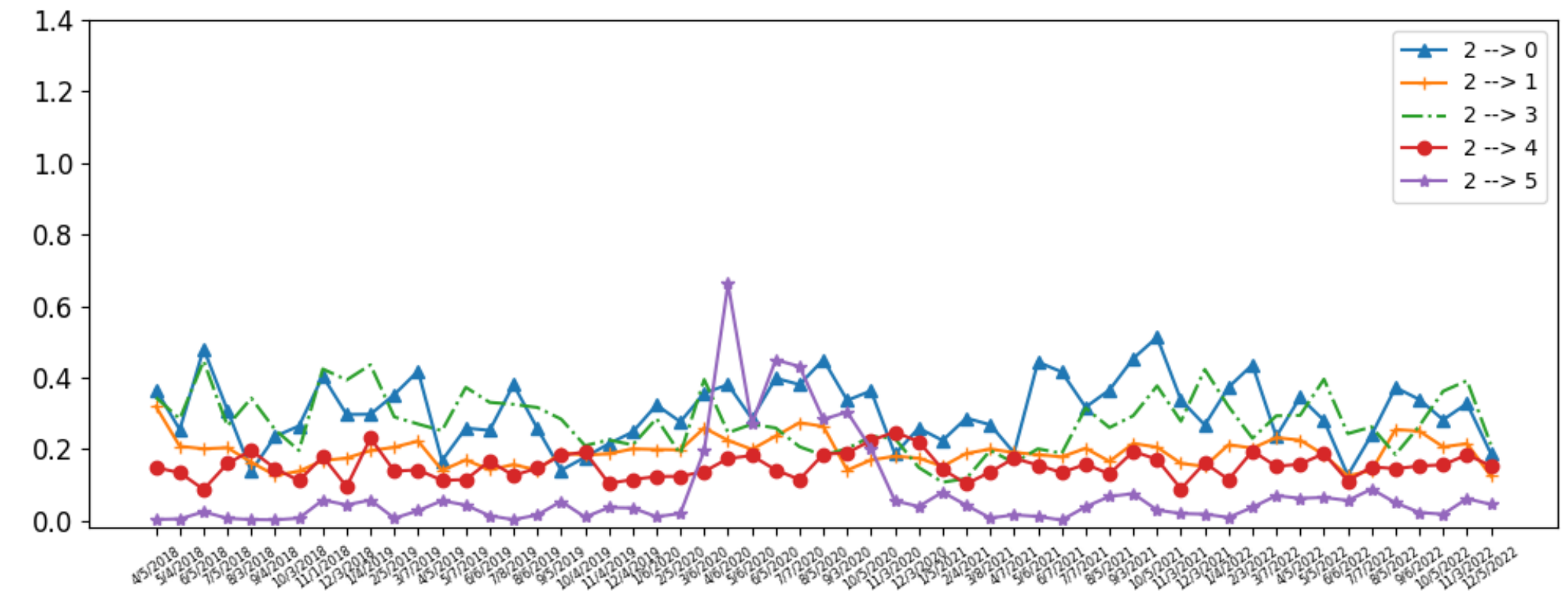}}
\subfigure{\includegraphics[scale=.2]{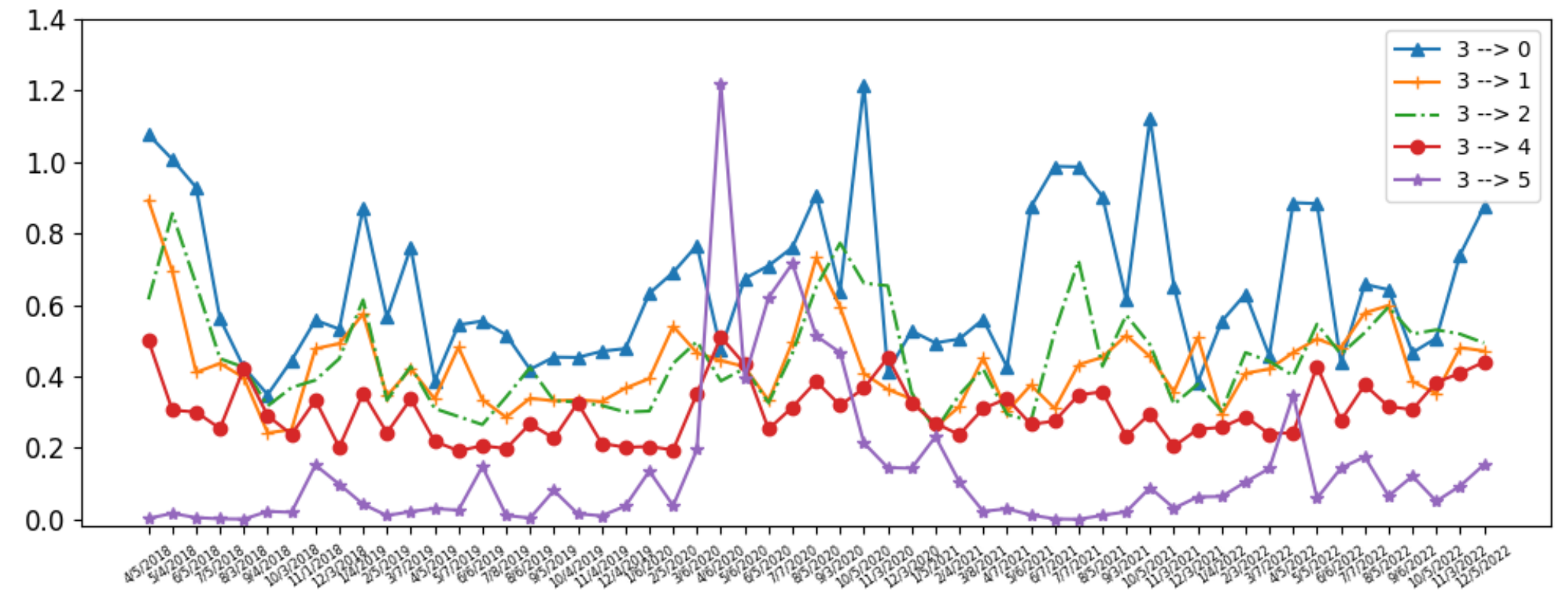}}
\subfigure{\includegraphics[scale=.2]{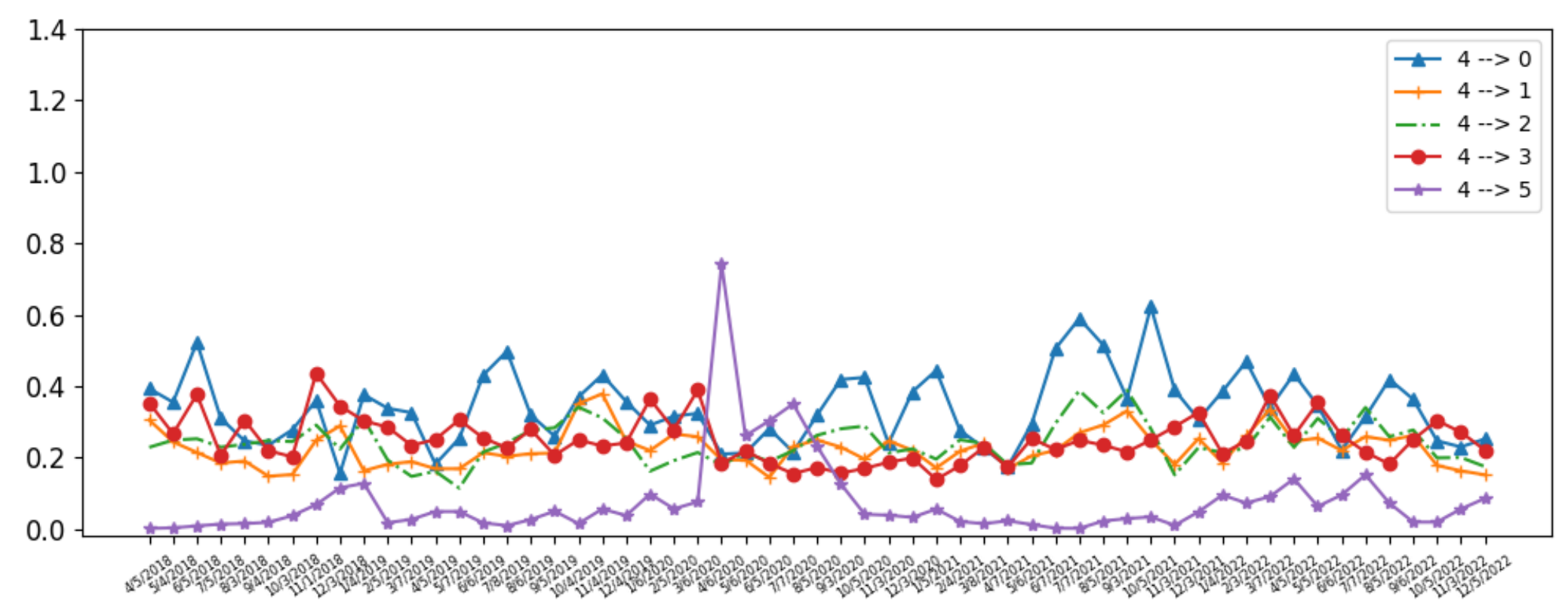}}
\subfigure{\includegraphics[scale=.2]{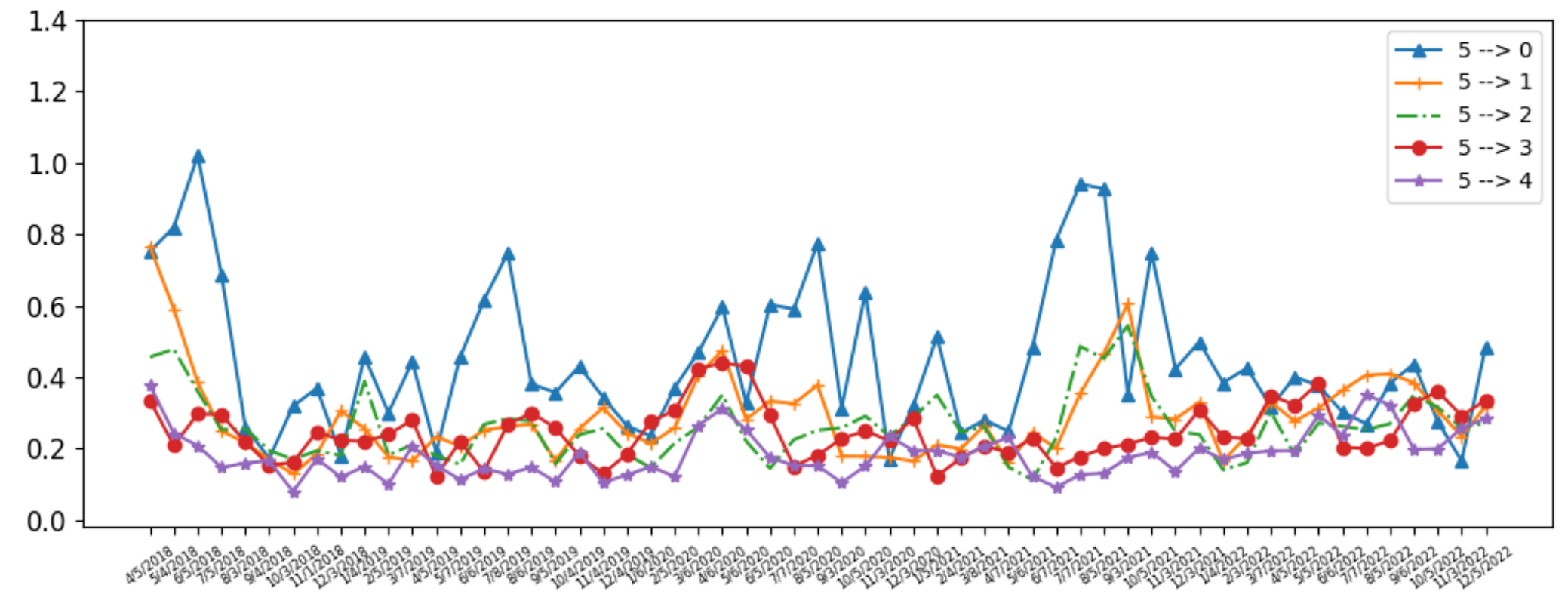}}
\caption{Estimated DIs of the time-varying influences between the industries. }
\end{figure}







\end{document}